\documentclass[12pt]{article}
\usepackage{epsfig}
\usepackage{graphics}
\usepackage{amssymb}
\usepackage{multirow}
\def\TKR{TKR}
\def\TOT{ToT}
\def\degree{${}^\circ$}

\begin{document}
\renewcommand{\thefootnote}{\fnsymbol{footnote}}


\vspace{.8cm}

{\large\bf The On--orbit calibration of the {\em Fermi} Large Area Telescope}

\vspace{.4cm}

\centerline {\it The Fermi LAT Collaboration}

\vspace{.4cm}

\pagestyle{empty}
\setlength{\oddsidemargin}{0in}
\setlength{\evensidemargin}{0in}
\setlength{\textwidth}{6.5in}
\setlength{\textheight}{9in}
\noindent
A.~A.~Abdo$^{1,2}$, 
M.~Ackermann$^{3}$, 
M.~Ajello$^{3}$, 
J.~Ampe$^{2}$, 
B.~Anderson$^{4}$, 
W.~B.~Atwood$^{4}$, 
M.~Axelsson$^{5,6}$, 
R.~Bagagli$^{7}$, 
L.~Baldini$^{7}$, 
J.~Ballet$^{8}$
% ...
\medskip
\begin{enumerate}
\item[1.] National Research Council Research Associate
\item[2.] Space Science Division, Naval Research Laboratory, Washington, DC 20375
\item[3.] W. W. Hansen Experimental Physics Laboratory, Kavli Institute for Particle Astrophysics and Cosmology, Department of Physics and SLAC National Laboratory, Stanford University, Stanford, CA 94305
\item[4.] Santa Cruz Institute for Particle Physics, Department of Physics and Department of Astronomy and Astrophysics, University of California at Santa Cruz, Santa Cruz, CA 95064
\item[5.] The Oskar Klein Centre for Cosmo Particle Physics, AlbaNova, SE-106 91 Stockholm, Sweden
\item[6.] Department of Astronomy, Stockholm University, SE-106 91 Stockholm, Sweden
\item[7.] Istituto Nazionale di Fisica Nucleare, Sezione di Pisa, I-56127 Pisa, Italy
\item[8.] Laboratoire AIM, CEA-IRFU/CNRS/Universit\'e Paris Diderot, Service d'Astrophysique, CEA Saclay, 91191 Gif sur Yvette, France
\item[9.] Istituto Nazionale di Fisica Nucleare, Sezione di Trieste, I-34127 Trieste, Italy
\item[10.] Dipartimento di Fisica, Universit\`a di Trieste, I-34127 Trieste, Italy
\item[11.] Istituto Nazionale di Fisica Nucleare, Sezione di Padova, I-35131 Padova, Italy
\item[12.] Dipartimento di Fisica ``G. Galilei", Universit\`a di Padova, I-35131 Padova, Italy
\item[13.] Department of Physics, Center for Cosmology and Astro-Particle Physics, The Ohio State University, Columbus, OH 43210
\item[14.] IRFU/Dir, CEA Saclay, 91191 Gif sur Yvette, France
\item[15.] Istituto Nazionale di Fisica Nucleare, Sezione di Roma ``Tor Vergata", I-00133 Roma, Italy
\item[16.] Dipartimento di Fisica, Universit\`a di Roma ``Tor Vergata", I-00133 Roma, Italy
\item[17.] Max-Planck Institut f\"ur extraterrestrische Physik, 85748 Garching, Germany
\item[18.] Laboratoire Leprince-Ringuet, \'Ecole polytechnique, CNRS/IN2P3, Palaiseau, France
\item[19.] Istituto Nazionale di Fisica Nucleare, Sezione di Perugia, I-06123 Perugia, Italy
\item[20.] Dipartimento di Fisica, Universit\`a degli Studi di Perugia, I-06123 Perugia, Italy
\item[21.] Dipartimento di Fisica ``M. Merlin" dell'Universit\`a e del Politecnico di Bari, I-70126 Bari, Italy
\item[22.] Istituto Nazionale di Fisica Nucleare, Sezione di Bari, 70126 Bari, Italy
\item[23.] Department of Physics, University of Washington, Seattle, WA 98195-1560
\item[24.] INAF-Istituto di Astrofisica Spaziale e Fisica Cosmica, I-20133 Milano, Italy
\item[25.] School of Pure and Applied Natural Sciences, University of Kalmar, SE-391 82 Kalmar, Sweden
\item[26.] Department of Physics, Royal Institute of Technology (KTH), AlbaNova, SE-106 91 Stockholm, Sweden
\item[27.] Agenzia Spaziale Italiana (ASI) Science Data Center, I-00044 Frascati (Roma), Italy
\item[28.] George Mason University, Fairfax, VA 22030
\item[29.] NASA Goddard Space Flight Center, Greenbelt, MD 20771
\item[30.] IRFU/SEDI, CEA Saclay, 91191 Gif sur Yvette, France
\item[31.] Laboratoire de Physique Th\'eorique et Astroparticules, Universit\'e Montpellier 2, CNRS/IN2P3, Montpellier, France
\item[32.] Department of Physics, Stockholm University, AlbaNova, SE-106 91 Stockholm, Sweden
\item[33.] University of Maryland, Baltimore County, Baltimore, MD 21250
\item[34.] Stellar Solutions Inc., 250 Cambridge Avenue, Suite 204, Palo Alto, CA 94306
\item[35.] Dipartimento di Fisica, Universit\`a di Udine and Istituto Nazionale di Fisica Nucleare, Sezione di Trieste, Gruppo Collegato di Udine, I-33100 Udine, Italy
\item[36.] ATK Space Products, Beltsville, MD 20705
\item[37.] CNRS/IN2P3, Centre d'\'Etudes Nucl\'eaires Bordeaux Gradignan, UMR 5797, Gradignan, 33175, France
\item[38.] Universit\'e de Bordeaux, Centre d'\'Etudes Nucl\'eaires Bordeaux Gradignan, UMR 5797, Gradignan, 33175, France
\item[39.] Department of Physical Science and Hiroshima Astrophysical Science Center, Hiroshima University, Higashi-Hiroshima 739-8526, Japan
\item[40.] University of Maryland, College Park, MD 20742
\item[41.] Orbital Network Engineering, 10670 North Tantau Avenue, Cupertino, CA 95014
\item[42.] Department of Physics, Tokyo Institute of Technology, Meguro City, Tokyo 152-8551, Japan
\item[43.] Centre d'\'Etude Spatiale des Rayonnements, CNRS/UPS, BP 44346, F-30128 Toulouse Cedex 4, France
\item[44.] Center for Research and Exploration in Space Science and Technology (CRESST), NASA Goddard Space Flight Center, Greenbelt, MD 20771
\item[45.] M\"atfakta i Kalmar AB, 30477 Kalmar, Sweden
\item[46.] Institute of Space and Astronautical Science, JAXA, 3-1-1 Yoshinodai, Sagamihara, Kanagawa 229-8510, Japan
\item[47.] Department of Physics and Astronomy, University of Denver, Denver, CO 80208
\item[48.] Dipartimento di Ingegneria dell'Informazione, Universit\`a di Padova, I-35131 Padova, Italy
\item[49.] Kavli Institute for Cosmological Physics, University of Chicago, Chicago, IL 60637
\item[50.] Current address: Pacific Northwest National Laboratory, Richland, WA 99352
\item[51.] Institut de Ciencies de l'Espai (IEEC-CSIC), Campus UAB, 08193 Barcelona, Spain
\item[52.] NYCB Real-Time Computing Inc., Lattingtown, NY 11560-1025
\item[53.] Department of Chemistry and Physics, Purdue University Calumet, Hammond, IN 46323-2094
\item[54.] Max-Planck-Institut f\"ur Kernphysik, D-69029 Heidelberg, Germany
\item[55.] Instituci\'o Catalana de Recerca i Estudis Avan\c{c}ats (ICREA), Barcelona, Spain
\item[56.] Consorzio Interuniversitario per la Fisica Spaziale (CIFS), I-10133 Torino, Italy
\item[57.] Current address: Yahoo! Inc., Sunnyvale, CA 94089
\end{enumerate}

\begin{abstract}

The Large Area Telescope (LAT) on--board the {\em Fermi} Gamma--ray Space Telescope 
began its on--orbit operations on June 23, 2008. 
Calibrations, defined in a generic sense, correspond to 
synchronization of trigger signals, optimization of delays for latching data, determination of detector thresholds, gains and responses, 
evaluation of the perimeter of the South Atlantic Anomaly (SAA), measurements of live time, of absolute time, and internal and spacecraft boresight alignments. 
Here we describe on--orbit calibration 
results obtained using known astrophysical sources, galactic cosmic rays, and 
charge injection into the front-end electronics of each detector. 
Instrument response functions will be described in a separate publication.
This paper demonstrates the stability of calibrations and describes minor changes observed since launch. 
These results have been used to calibrate the LAT datasets to be publicly released in August 2009.

\end{abstract}

{\small\bf Keywords:} GLAST, Fermi, LAT, gamma-ray, calibrations

{\small\bf PACS classification codes:} 07.87.+v; 95.55.Ka

\section{Introduction}
\label{sec:intro}

The Fermi Gamma--ray Space Telescope, hereafter {\em Fermi}, represents the next generation of satellite--based high-energy gamma-ray observatory.
The {\em Fermi} satellite hosts two instruments: the Large Area Telescope (LAT)~\cite{latpaper} and the Gamma-ray Burst Monitor (GBM)~\cite{gbmpaper}. 
The former employs a pair-conversion technique to measure photons from 20 MeV to energies greater than 300 GeV, while 
the latter uses NaI and BGO scintillation counters to 
record transient phenomena in the sky in the energy range from 8~keV to 40~MeV. The LAT has no consumables, 
and a very stable response unlike its predecessor, the Energetic Gamma Ray Emission Telescope (EGRET)~\cite{egret}. 
For the energy range above 10 GeV the sensitivity of the LAT is at least one order of magnitude greater than that of EGRET, allowing 
the sky to be explored at these energies essentially for the first time~\cite{latpaper}. 

The LAT consists of a tracker/converter (TKR) for direction measurements~\cite{Atwood:2007ra,ssd,Johnson98,Baldini:2006pv}, followed by a calorimeter (CAL) for energy measurements~\cite{cal}. Sixteen TKR and CAL modules are combined to form sixteen towers, which are assembled in a 4$\times$4 mechanical support structure. An anticoincidence detector (ACD), enclosed by a micrometeoroid shield, surrounds the TKRs and rejects charged cosmic-ray background~\cite{acdover,acddet}. The LAT has about one million detector readout channels. 

On-orbit calibrations relate to all aspects of LAT measurements and data analysis results, from absolute timing 
to energy and direction measurements for individual events, to fluxes and positions of gamma-ray sources.

The accurate timestamps of the LAT are obtained using the Global Positioning System (GPS) of the {\em Fermi} spacecraft, which provides timing and position information. Those are needed for phase folding pulsars and correlating gamma-ray observations with those at other wavelengths. 

As the number of photons in an observation increases, the centroid of their spatial distribution becomes better measured, and 
eventually the error is dominated by uncertainties in the alignment of the LAT, both internal and 
with respect to the {\em Fermi} spacecraft.

Source localization at GeV energies enables the LAT to resolve bright, adjacent sources 
previously labeled as unidentified~\cite{egret} and will help elucidate the origin of gamma-ray 
emissions from galactic cosmic rays accelerated in supernova remnants. 

The energy calibrations at higher energies are of utmost importance for detection of dark matter particle signals. 
Some extensions to the Standard Model predict narrow spectral lines due to the annihilations of as-yet unknown massive particles.
For detecting and characterizing these features, accurate energy determination is vital.
Even though the LAT was designed to measure gamma-rays it can also study, though not separately, the cosmic-ray electron and positron spectra.  
Cosmic-ray electron and positron spectra and intensities may also contain signatures for new physics.  
In this case, energy calibrations and position determination using extrapolated tracks into the CAL play an important role. 
At lower energies, broad features in the photon energy spectra of active galactic nuclei, or supernovae remnants originating from pion decays and bremsstrahlung may help unravel outstanding questions concerning particle acceleration in these sources. 

The purpose of this paper is to document the on-orbit calibration procedures used by the LAT; 
it begins with an overview of calibrations in Section~\ref{sec:overview}. 
Details on trigger, ACD, CAL and TKR calibrations are described in  Sections~\ref{sec:daq},~\ref{sec:acd},~\ref{sec:cal} and~\ref{sec:tkr}, respectively. 
The evaluation and updates to the perimeter of the South Atlantic Anomaly (SAA) are presented in Section~\ref{sec:saa} and 
measurements of live time are discussed in Section~\ref{sec:livetime}. The results from absolute timing follow in Section~\ref{sec:absolute}. 
Finally, internal and spacecraft boresight alignments are explained in Section~\ref{sec:ta}.  
We conclude with a table that summarizes the calibration results in Section~\ref{sec:conclusion}. Assessment of the current LAT performance is described in a separate publication~\cite{perfpaper}, while tests performed at particle accelerators are presented elsewhere~\cite{bt97,bt99,bt06}. The section describing each calibration in shown in the last column of Table~\ref{table:calib}.

\section{Overview of LAT calibrations}
\label{sec:overview}

For this paper, the word calibration represents synchronization of trigger signals, optimization of delays for latching data, determination of detector thresholds, gains and responses, evaluation of the perimeter of the South Atlantic Anomaly (SAA), measurements of live time and of absolute time and internal and spacecraft boresight alignments.

Table~\ref{table:calib} summarizes all calibration types, classified by category. 
Configuration refers to operational settings that define the state of the hardware, which are fixed before data are acquired. 
Calibrations are related to quantities that can change after data are processed and analyzed.
As shown in Table~\ref{table:calib}, for some types of calibrations, updates are not as frequent as the acquisition of the relevant data.
\begin{table}
\begin{center}
\begin{tabular}{|c|c|c|c|c|c|c|} \hline\hline
\multirow{2}{*}{Category} &   Title  & Type & Frequency to & Frequency of & Sec. \\ 
  & & & acquire data & updates & \\ \hline\hline
Trigger  &  Time coincidence window&      config   &1 year  & 1 year & \ref{sec:tsync} \\ \hline 
Trigger  &   Fast trigger delays &         config  & 1 year & 1 year & \ref{sec:tsync} \\ \hline 
Trigger  &   Delays for latching data &      config   & 1 year & 1 year & \ref{sec:tsync} \\ \hline 
ACD       &   Pedestal       &    both  & continuous & 3 months & \ref{sec:acdpeds} \\ \hline
ACD       &   Coherent noise     &     calib   & 3 months & 3 months & \ref{sec:acdpeds} \\ \hline  
ACD       &   MIP peak     &      calib  & continuous & 3 months & \ref{sec:acdmip} \\ \hline 
ACD       &   High range (CNO)   &      calib  & continuous & 3 months & \ref{sec:acdmip} \\ \hline 
ACD       &   Veto threshold &              config    & 3 months  & 3 months & \ref{sec:veto} \\ \hline
ACD       &   High level discriminator &              config    & 3 months  & 3 months & \ref{sec:veto} \\ \hline  
CAL       &   Pedestal      &     both & continuous & 3 months & \ref{sec:calpeds} \\ \hline 
CAL       &   Electronics linearity     &     calib  & 3 months & 3 months & \ref{sec:callin} \\ \hline  
CAL       &   Energy scales     &     calib  & continuous & 6 months & \ref{sec:callin} \\ \hline  
CAL       &   Light asymmetry &   calib & continuous & 6 months & \ref{sec:calpos} \\ \hline 
CAL       &   Zero-suppression threshold &       config    & continuous & 3 months & \ref{sec:calthresh} \\ \hline 
CAL       &   Low-energy threshold   &       config  & continuous & 6 months & \ref{sec:calthresh} \\ \hline 
CAL       &   High-energy threshold  &        config  & continuous & 6 months & \ref{sec:calthresh} \\ \hline 
CAL       &   Upper level discriminators  &      config   & continuous & 1 year & \ref{sec:calthresh} \\ \hline 
TKR       &   Noisy channels   &  config   & continuous & 3 months & \ref{sec:tkrch} \\ \hline  
TKR       &   Trigger threshold  &    config  & 3 months & 3 months & \ref{sec:tkrdac} \\ \hline
TKR       &   Data latching threshold     &    calib  & 3 months & 3 months & \ref{sec:tkrdac} \\ \hline
TKR       &   ToT conversion parameters &       calib  & 1 year & 1 year & \ref{sec:tkrtot} \\ \hline
TKR       &   MIP scale   &   calib  & continuous & 3 months & \ref{sec:tkrmip} \\ \hline 
SAA     &   SAA polygon &  config  & continuous & 1 year & \ref{sec:saa} \\ \hline  
Timing    &   LAT timestamps     &   calib  & continuous & continuous & \ref{sec:absolute} \\ \hline
Alignment     &   Intra tower &  calib & continuous & 1 year & \ref{sec:intra} \\ \hline 
Alignment     &   Inter tower &   calib & continuous & 1 year & \ref{sec:inter} \\ \hline
Alignment     &   LAT boresight &   calib & continuous & 1 year & \ref{sec:sc} \\ \hline 
\hline
\end{tabular}
\vspace*{+0.5cm}
\caption{List of LAT calibrations (calib) and configurations (config) that can impact LAT scientific results. Note that pedestals are used both as a configuration on-board and as a calibration on the ground. Detailed descriptions are given in the sections listed in the last column. Frequency of updates correspond to current best estimates for the period beyond the first year of operations.}
\label{table:calib}
\end{center}
\end{table}
Operationally, calibration data are acquired in two distinct modes (see Section~\ref{sec:tmodes}):
\begin{enumerate}
   \item Dedicated, meaning that the trigger, detector and software filter settings are
       incompatible with nominal science data taking.
   \item  Continuous, meaning the trigger and software can, with only
       a small penalty in live time, acquire specialized data that
       is used to calibrate, or, more generally, monitor the
       performance of the LAT during nominal science data-taking.
\end{enumerate}
Within a run the LAT acquires data with fixed  instrument configurations. 
For the most part, the dedicated calibration runs are concerned with characterizing the electronics' response to known stimuli while the continuous calibrations are aimed at calibrating the electronics with a known physics input. The stability of calibrations 
has been such that operations in dedicated-mode amount to approximately 2.5 hours every three months.

Sea-level cosmic ray muons were used to calibrate the low-energy scales and trigger thresholds, but muons do not deposit enough energy in the detector elements to calibrate the high-energy scales and high-energy trigger thresholds.  Instead, we used charge injection into the front-end electronics to calibrate the high-energy scales.
Because of rise-time slewing effects, the optimal synchronization of trigger signals and optimal delays for data latching are energy dependent. 
We used pre-launch tests to provide a best approximation of the optimal trigger timing, and verified and corrected the synchronization and delays with on-orbit data.
All these calibrations are revisited with on-orbit data selected from galactic cosmic rays and, as it will be shown later, there are only minor deviations when comparing results prior and after launch.

\section{Overview of trigger and readout}
\label{sec:daq}

The trigger and data readout system controls the composition and flow
of data from the source in the detector elements to the Solid State
Recorder (SSR) of the {\em Fermi} spacecraft. There are two distinct
stages: the latching and movement of data from the front-end detector
elements to a LAT Event Processing Unit (EPU), and the processing of
the data on the EPU and its subsequent transfer to the SSR.
The first stage is controlled by hardware, namely the trigger, and the
second stage by software. A full description of the LAT multilevel
trigger and data readout can be found
elsewhere~\cite{latpaper,tdaqpaper}.

\subsection{Acquisition modes: dedicated or continuous}
\label{sec:tmodes}

During dedicated calibration runs the trigger system and detector electronics are configured to acquire
data useful for calibrating the thresholds and responses of detectors
and synchronizing the arrival times of signals from various parts of
the detectors. We call these runs dedicated because the trigger and detector configurations
necessary to acquire these specialized data are incompatible with acquiring our primary science
data (i.e. high-energy gamma rays from celestial sources).   During continuous calibration 
acquisitions, the trigger
system and detector electronics are configured to
detect and latch not only those events thought to be gamma rays, but
also those useful for calibration. 

As used on orbit, the LAT trigger system takes inputs from the ACD, TKR, and CAL front-end electronics and from a
programmable, internal periodic trigger. Under control of LAT flight software, the periodic trigger can 
be used with a programmable charge-injection system to
calibrate the detector electronics, or it can be used simply to read out the detector front-ends at a specified cadence.

The programmed control under the charge injection system is used only in dedicated calibration 
runs.  There the trigger is configured to collect a specified number of events at a particular rate,
regardless of input from the detectors.  In most cases, the trigger is configured to instruct the
detector electronics to inject a known, programmable amount of charge at a specified time relative to each trigger.  In other
cases, no charge is injected.  For each sequence of these events, flight software sets configuration 
registers throughout the instrument that control, for example, the amount of charge injected.  By
collecting a number of events with the same injected charge, and then
varying the amount of injected charge, the electronic response can be
accurately calibrated.

Other dedicated calibration runs and all continuous calibration runs have the instrument triggered 
by signals from the detectors, not a programmed sequence,
and charge injection is not used.  For example, dedicated
calibration runs in this mode are those used to synchronize 
trigger signals from the detectors. Data are collected using a
sequence of configurations that sweep values through the various
registers that control the delay times of the trigger signals.  The data
captured are then analyzed on the ground to identify the delay settings
that best synchronize the arrival times of the trigger signals from each detector system. 

During continuous calibration runs, the trigger is configured to
detect and latch both events thought to be gamma rays and other events useful 
for calibration.  Based on the signals recieved from
the detectors, the trigger specifies the type of ACD and CAL readout. 
There are four types (see Table~\ref{table:cont}) formed by enabling
or disabling zero-suppression, in which values are committed to the
data stream only if they exceed a programmable threshold, and
selecting CAL single-range or four-range mode, in which the electronics
takes respectively only the ``best'' range or all four values (see
Section~\ref{sec:calthresh} for details). The ACD does not support a
readout mode to select both readout ranges. It does, however, respect
the zero-suppress mode.

\begin{table}[h]
\begin{center}
\begin{tabular}{|c|c|c|} \hline\hline
Purpose & Readout Mode \\ \hline\hline \multirow{2}{*}{Select photons}
& Zero-suppressed \\ & and best range \\ \hline
\multirow{2}{*}{Cross-calibration of the energy ranges (CAL)} &
Zero-suppressed \\ & and four-range \\ \hline
\multirow{2}{*}{Calibration of the MIP peak (CAL,ACD) } &
Zero-suppressed \\ & and best range \\ \hline \multirow{2}{*}{Monitor
  pedestals (CAL,ACD) and noise occupancy (TKR) } &
Non-zero-suppressed \\ & and four-range \\ \hline\hline
\end{tabular}
\vspace*{+0.5cm}
\caption{Readout modes used for continuous calibrations (nominal
  science operations).}
\label{table:cont}
\end{center}
\end{table}

After an event passes the hardware trigger it is inspected by 
on-board software filters, each configured to identify events likely to be
useful for one or more scientific or calibration purposes.  If any filter accepts an event, it
is included in the LAT data stream and forwarded to the SSR for transmission to the ground.
In addition, the filters can be configured to allow a fraction of events that would otherwise
have been rejected to be included in the data stream.  Such events are used primarily to study
filter performance.  The input trigger rate of about 2.2
kHz, averaged over many orbits, is reduced to about 450 Hz by the
software filters.  Table~\ref{table:obf} describes filter types,
purpose and average output rate. 

\begin{table}[h]
\begin{center}
\begin{tabular}{|c|c|c|c|} \hline\hline
Filter Type & Purpose & Average Rate (Hz) \\ \hline\hline
\multirow{2}{*}{Gamma} & select gamma-ray candidates & 410 \\ & and
events $>$ 20 GeV & \\ \hline Heavy ion & calibration of high-energy
scales, & 2.5 \\ \hline \multirow{2}{*}{MIP} & select non-interacting
protons & 0 (nominal) \\ & & 10 (dedicated-mode) \\ \hline
Diagnostic & filter performance, background & 22 \\ \hline \hline
\end{tabular}
\vspace*{+0.5cm}
\caption{On-board filters used to select events for calibration
  acquisitions. Rates depend on geomagnetic and other orbital
  variations.  Here we list average rates.}
\label{table:obf}
\end{center}
\end{table}

\subsection{Synchronization of trigger signals and delays for latching data}
\label{sec:tsync}

Fast trigger signals ($\sim$ few hundred ns) from the detectors must be synchronized with respect to each other for the trigger to operate 
efficiently. There are five such signals: 
veto and high level discriminator in the ACD (see Section~\ref{sec:acd}), low and high-energy in the CAL (see Section~\ref{sec:cal}) and TKR~(see Sec~\ref{sec:tkr}). 
The earliest arriving trigger signal initiates a readout. The arrival times of the other trigger signals 
relative to this signal are captured in the event data with 50 ns precision (i.e. the period of the LAT system clock), 
allowing a direct comparison of trigger signals on an event-by-event basis. 
Once the relative timing is determined, the settings of various delay registers (i.e. the instrument configuration) are modified to synchronize these trigger signals in subsequent data acquisitions.
There are two types of delay associated with the trigger: the fast trigger signal delay and the delay for latching the data.
To synchronize the detectors we use the time of arrival of trigger signals, referred to as condition arrival times. 
This is event information containing the number of 50 ns clock ticks that have passed between
the opening of the coincidence time window and the arrival of a given trigger signal. 
Since the trigger signals for TKR
and CAL have significant time walk depending on the ratio of signal size to 
threshold value, it is important to choose an appropriate dataset to optimize
the timing. We select MIP data for the relative timing of ACD and TKR, 
since the main purpose of the ACD is to reject charged particle background.
The relative
timing between TKR and CAL is optimized using photon candidates selected 
by the gamma filter. The CAL low and high-energy signals are controlled by a single delay and cannot be 
tuned independently. 

\begin{figure}
  \center
   \includegraphics[width=14.cm]{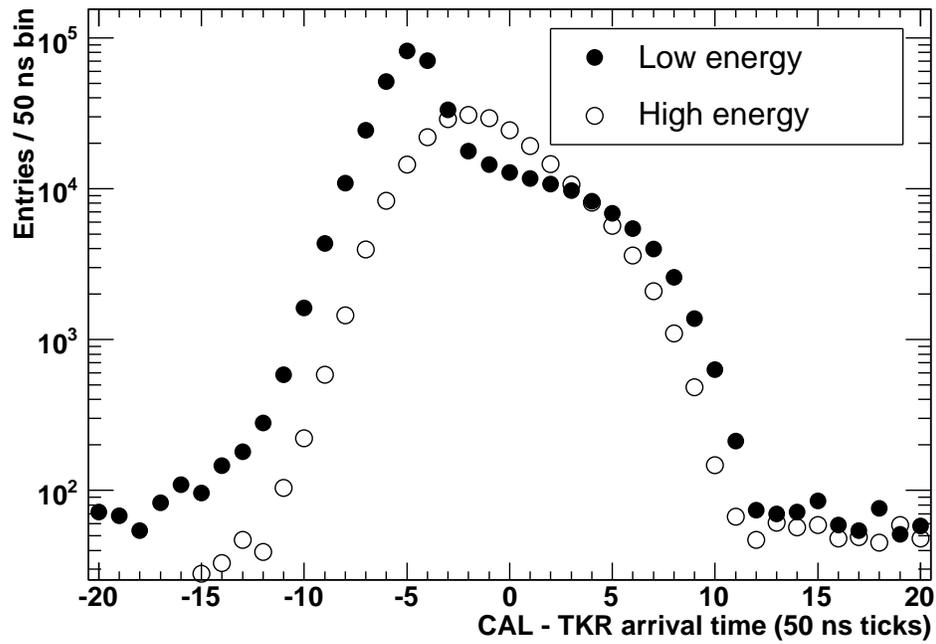}
   \caption{On-orbit arrival times of the CAL low-energy (FLE) and high-energy (FHE) trigger signals with respect to the TKR trigger. This special 
dataset was acquired with the time coincidence window of 1550 ns (31 ticks). 
Most of the entries fall within the window used for nominal science operations (700 ns or $\pm$13 ticks).}
   \label{fig:treq}
\end{figure}

Figure~\ref{fig:treq}
shows the arrival times for CAL with respect to the TKR. Negative values indicate that the TKR trigger arrived 
earlier than the CAL trigger signal. Distributions are 
fully contained inside the time coincidence window\footnote{Prior to launch the time 
coincidence window was set to 600 ns (12 ticks).}of 700 ns (14 ticks).
The arrival times are affected by the ratio of the crystal energy over the threshold, thus influencing the shape of the trigger signal curves. 
The spectrum is slightly altered from its original form due to a trigger event selection. 
There is no difference in arrival times between CAL high- and CAL low-energy triggers at energies very far above the two thresholds. 

\begin{table}
\begin{center}
\begin{tabular}{|c||c|c|c|}
\hline
Measurement &ACD Delay&TKR Delay&CAL Delay \\ \hline\hline
Ground& 800 ns (16 ticks) & 250 ns (5 ticks)&0 \\ \hline
Orbit&750 ns (15 ticks)&200 ns (4 ticks)&0\\ \hline
\end{tabular}
\end{center}
\caption{Trigger signal delays for measurements on the ground and on orbit.}
\label{table:treq}
\end{table}
On the ground, only muons were available for timing calibrations so a small change in 
parameters was observed on orbit.
The results from the synchronization of the trigger signals are shown in Table~\ref{table:treq}.

\begin{figure}
  \center
   \includegraphics[width=14.cm]{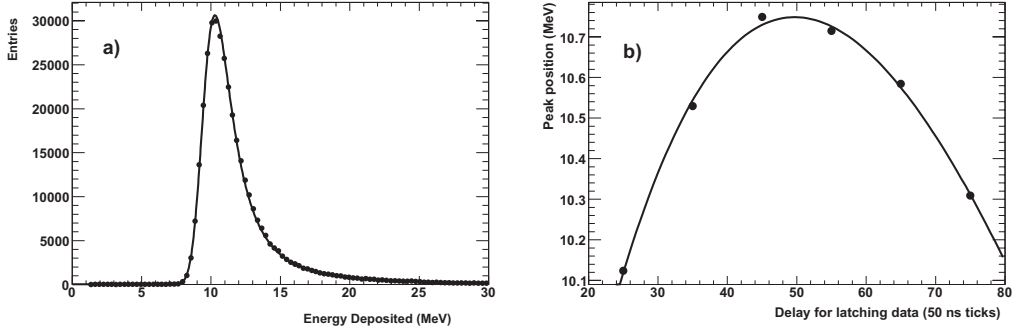}
   \caption{a) Energy deposited in a single CAL crystal (averaged over all towers) for a fixed value of the delay for latching data. The fit (curve) is used to determine the
MIP peak position in the CAL, b) MIP peak positions for different CAL delay values (all towers). The curve is a parabola fit to the data.}
   \label{fig:tack}
\end{figure}

The optimal delay is obtained after analyzing data acquired with fixed delay values.
For the CAL we fit the MIP peak 
for each dataset with a fixed delay. Figure~\ref{fig:tack}a shows an example for a delay 
setting of 750 ns (25 ticks). Figure~\ref{fig:tack}b shows CAL MIP peak positions for six delay values for latching data. The optimal 
setting is the peak position obtained from the fit to the data. There is a data point slightly off the curve 
due to changes in geomagnetic conditions along the orbit. Since data were not recorded at the same location in orbit, 
the MIP selection cuts are designed to keep variations to $<$ 0.1\% of the MIP peak value.
A similar procedure is applied to the ACD. For the TKR, the quantity of interest is not the MIP peak position but instead the detector efficiency, which is defined as the number of layer hits between the first and
the last hit of the track divided by the expected number of layers crossed by the track.
The optimal\footnote{The efficiency for latching TKR data is unchanged up to $\ge$ 750 ns (15 ticks).}settings for the on-orbit delays for latching data 
are 200 ns (4 ticks), 0 and 2450 ns (49 ticks) for the ACD, CAL and TKR delays, respectively.

\section{ACD calibrations}
\label{sec:acd}

The ACD is the LAT first-level defense against the charged particle cosmic-ray background that outnumbers the gamma-ray signals 
by 3-5 orders of magnitude. 
It consists of 97 separate plastic scintillating detectors - 89 scintillator tiles and 8 scintillator ribbons, 
each viewed by two photomultiplier tubes (PMTs) for redundancy. 
The overall ACD detection efficiency is $>$ 0.9997, which is provided by ensuring high uniformity of the detectors' response and a large 
number of photoelectrons. 
The segmentation is needed in order to minimize pulse height variations over the ACD area and to minimize unwanted self-veto due to backsplash 
of soft photons from the developing electromagnetic 
shower in the CAL. Such self-veto caused a significant 
reduction in effective area in EGRET at high energies~\cite{egret,backsplash}. 
Detailed information about the ACD can be obtained elsewhere~\cite{acdover,acddet}.

ACD calibrations include the determination of the mean values of pedestals, of the signal pulse heights 
produced by single MIP particles in each ACD scintillator, and the veto threshold settings; and the high-energy and 
coherent noise calibrations.  
All those parameters are determined for each PMT, so that each ACD tile or ribbon has two calibrated values. 
Every ACD channel has two ranges, low and high, in order to expand the dynamic range of processed signals. 
The low range covers the signals below 4-8 MIPs (depending on the channel), and the high range extends well above~1000 MIPs. 
The switching between ranges occurs automatically depending on the amplitude of the signal.

In the future there may be changes to the MIP peak positions due to degradation of
PMT photocathodes, scintillators or in an optical path between them. 
A possible way to mitigate these effects is to raise the high voltage for the PMTs.
Since each high voltage is common to groups of 16 or 17 PMTs, this requires re-calibration 
of all channels belonging to that particular PMT group.
 
\subsection{Pedestals and coherent noise}
\label{sec:acdpeds}

\begin{figure}
  \center
   \includegraphics[width=6.8cm]{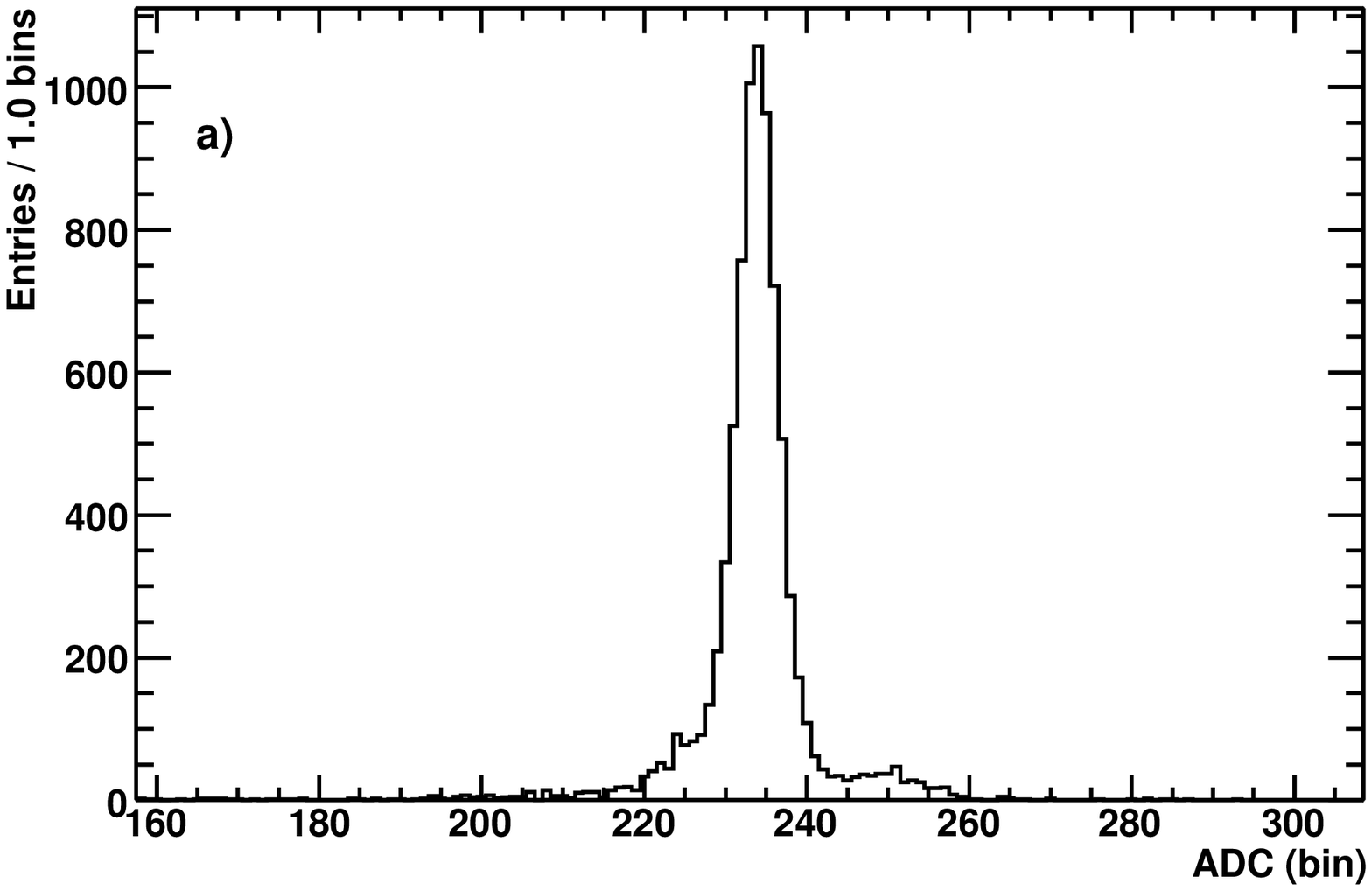}
   \includegraphics[width=6.8cm]{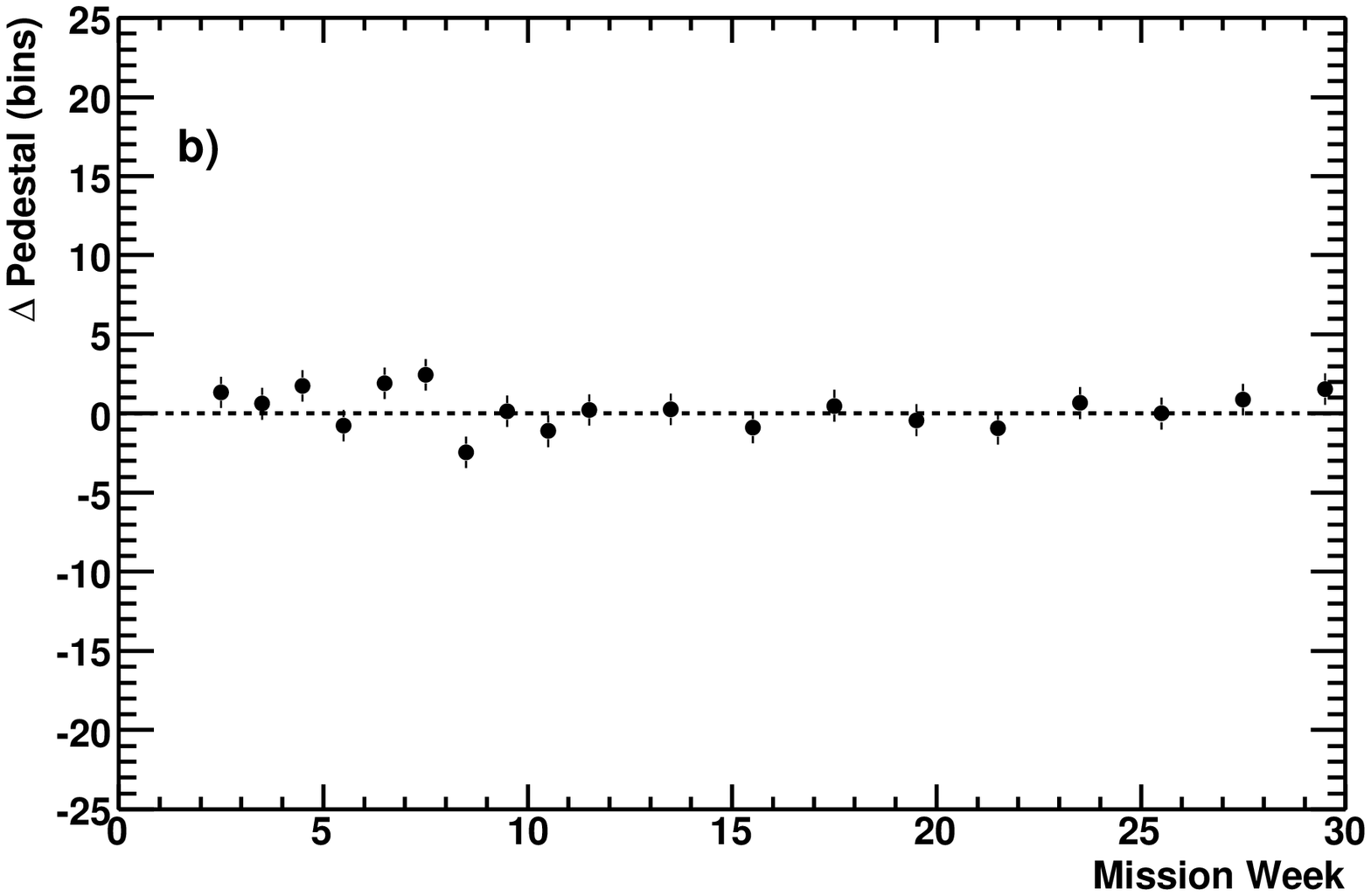}
   \caption{a) Pedestal distribution from periodic triggers for a single PMT, b) Long term pedestal trending for a single PMT.}
   \label{fig:acdpeds}
\end{figure}

Pedestals are offset voltages present at the Analog-to-Digital Converter (ADC) inputs in the low and high range readouts. 
We extract the ACD pedestals for the low range readout from the 2 Hz periodic triggers, which currently provide approximately 10,000 random samples per orbit. 
Since a small fraction of these events contains particle signals or 
electronics noise or tails from previous signals in the ACD, 
we extract the pedestal values by performing a Gaussian fit to the central 80\% of the 
pulse height distribution (truncated mean).   
Figure.~\ref{fig:acdpeds}a shows a typical pedestal distribution for a single PMT, where the width of the Gaussian (truncated) 
is about 2-4 pulse height bins ($\sim$ 0.01 MIPs). 
The narrow core of the distribution shows 
the intrinsic electronic noise; tails are residual signals from particles near in 
time to the periodic trigger.
The small peak at about 250 is dominated by coherent noise contributions. 
Extracting the pedestal for the high range readout requires a special data-taking configuration, which forces a series of 
randomly triggered events to be read out in the high range.  Since this configuration is incompatible with regular data taking 
and the pedestals are reasonably stable, these data are only acquired during the quarterly calibration periods where the LAT is in dedicated-mode. 
The data analysis is similar to that of the low range pedestals. 
Figure~\ref{fig:acdpeds}b shows the long term pedestal trending data for a single PMT.  All values are plotted relative to the calibration being used, at the time 
of writing, in the offline reconstruction software (Mission week 25). 
To reduce data volume, we use the low range pedestal values on-board. We reject all signals 
less than 25 counts above pedestal, which corresponds to approximately 8 times the electronics noise, or 0.05 times 
the MIP signal in a typical tile.
We also use pedestal values in the flight software 
data compression algorithm (for both ACD and CAL). In this case we reduce the data size 
by referencing signals against pedestal values rather than against zero. 

\begin{figure}
  \center
   \includegraphics[width=13.cm]{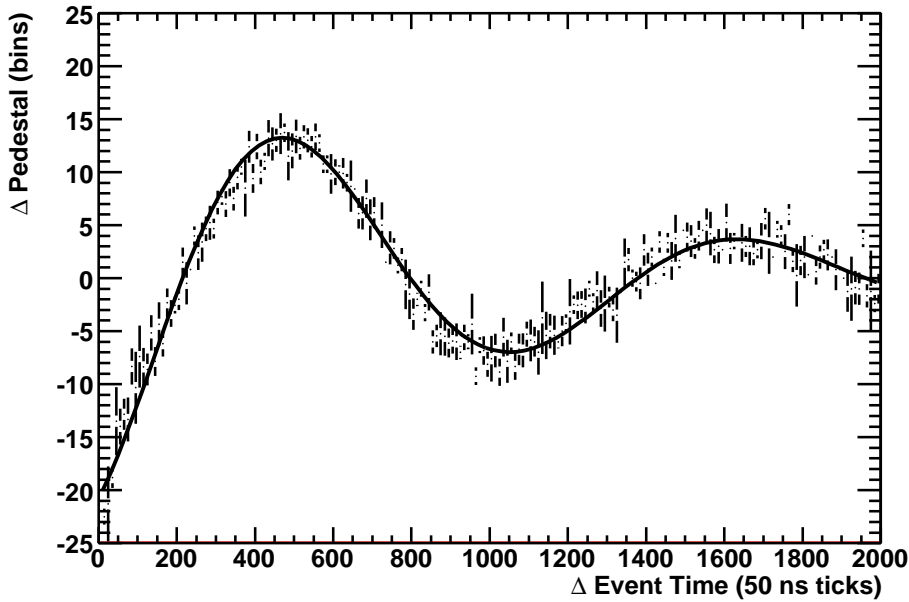}
   \caption{Readout related pedestal ringing in a single PMT.  The truncated mean and RMS ADC values obtained using periodic triggers versus the time difference between consecutive readout cycles in 50 ns clock ticks.  The curve corresponds to a fit of a sinusoidal oscillation inside a decaying exponential envelope.}
   \label{fig:acdcn}
\end{figure}

During ground testing of the ACD we discovered that the readout
process causes the electronics pedestals to ring. These oscillations occur when
elements of its internal circuitry resonate at their characteristic frequency.
This reproducible effect can be quantified. 
Figure~\ref{fig:acdcn} shows the difference ($\Delta$ Pedestal) between the coherent noise and the regular pedestal values for a single PMT.  
This truncated mean (see Section~\ref{sec:acdpeds}) is displayed versus the time difference between consecutive readout cycles in 50 ns 
clock ticks. Data were obtained using periodic triggers and the error bars correspond to the RMS ADC values. The curve represents a fit to a sinusoidal oscillation inside a decaying exponential envelope, which falls to the
electronics noise after about 200 $\mu$s.  

As this effect is present in
every channel of the ACD, events read out when the pedestal peaks at about
50 $\mu$s (1000 ticks) after the previous readout can lead to small signals in many
PMTs. This correction is applied to the offline data to avoid overestimating the
total energy in the ACD and compromising background rejection and photon selection. 
For events taken at the peak of the coherent oscillation ($\sim$ 500 ticks), this
calibration reduces the coherent noise contribution from 0.05 MIPs per PMT to less that
0.005 MIPs per PMT.
The effect is so small that we do not need to apply corrections to the on-board processing. 
Temperature dependences in ACD pedestals are negligible.

\subsection{MIP peak and high range calibrations}
\label{sec:acdmip}

\begin{figure}
  \center
   \includegraphics[width=6.8cm]{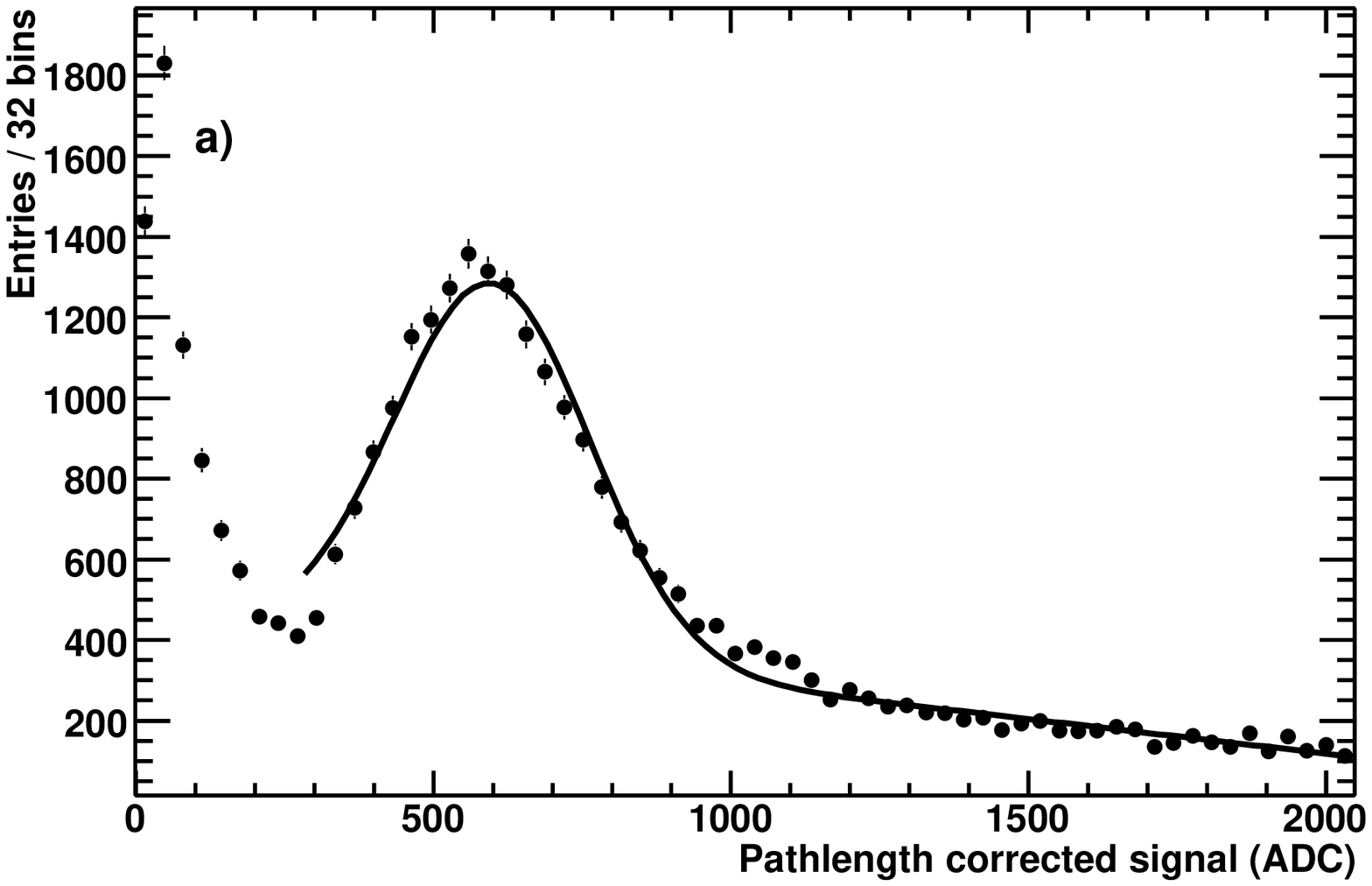}
   \includegraphics[width=6.8cm]{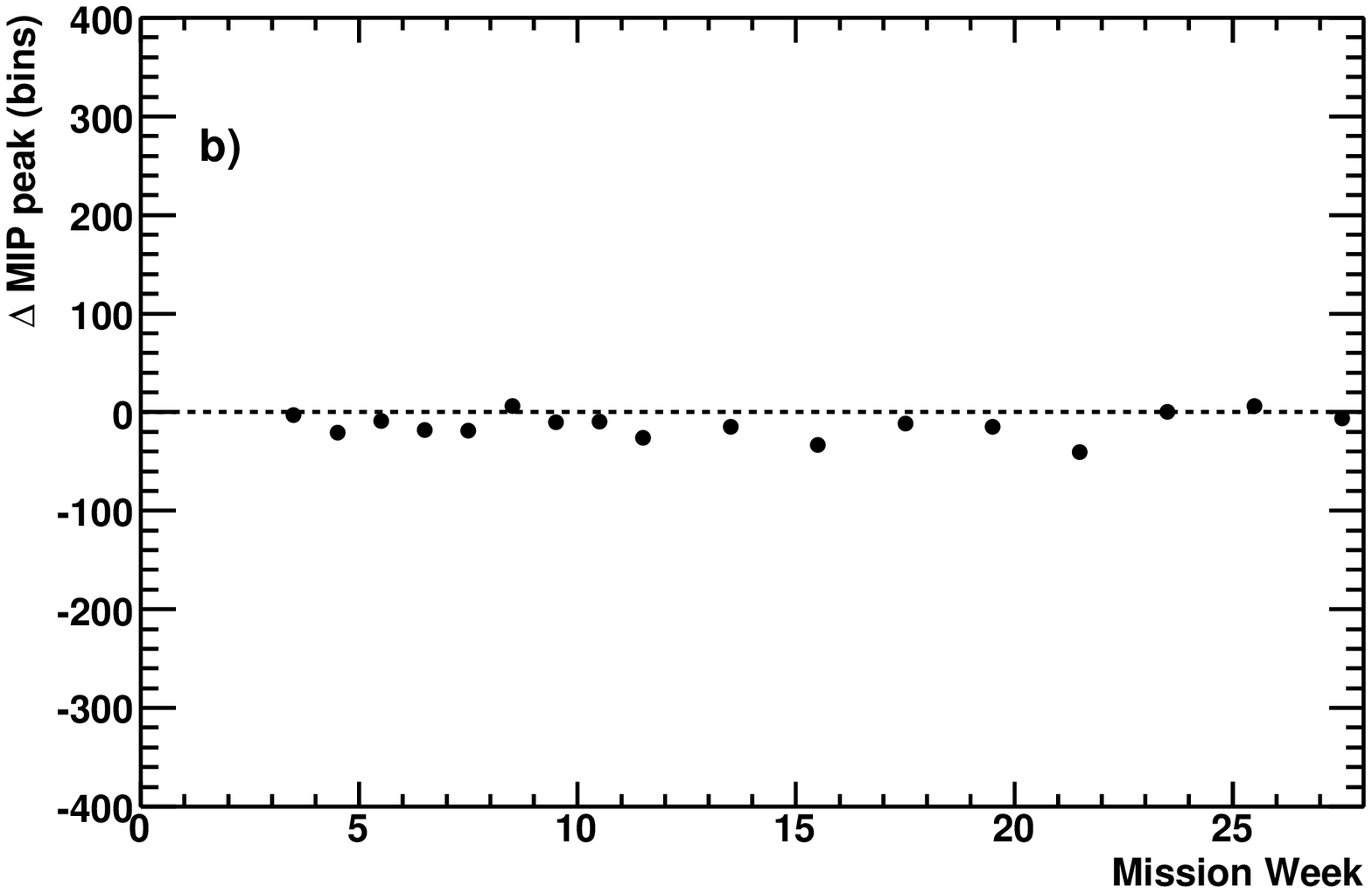}
   \caption{a) Distribution of pedestal subtracted and pathlength corrected signals for a single PMT in the ACD; 
the MIP peak is clearly visible and the curve is a fit to the data, 
b) Long term trending of the MIP peak versus mission week.}
   \label{fig:acdmip}
\end{figure}

The MIP peak values are determined using reconstructed tracks pointing at an ACD
tile or ribbon that recorded a signal. 
The peak value of the pathlength corrected pulse-height distribution
for each PMT is the MIP peak for that channel. This calibration is a heuristic
attempt to quantify the average MIP signal seen in the ACD and not a precise
determination of all details of the energy deposition in the ACD
sensors.
Figure~\ref{fig:acdmip}a shows a distribution of pedestal subtracted and pathlength corrected signals for a single PMT in the ACD. The MIP peak at about 600 is clearly seen and the peak below 100 corresponds to soft X-ray background. 
Figure~\ref{fig:acdmip}b displays the long term trending of the MIP peak for a single PMT, where 
pedestals from mission week 3 are used as a reference. The stability of the measurement is about 10\% of the 
MIP peak for a period of 20 weeks. 
During the offline reconstruction we use the MIP peak calibration values to 
express raw signal pulse height measurements in MIP equivalent values. 
These are converted into energy using the energy deposition in the 
scintillator ($\sim$2 MeV/cm). MIP peaks in the ACD
tiles occur between 400 and 1000 pulse height bins above pedestal, and are 
determined with an accuracy better than 5\%.

\begin{figure}
  \center
   \includegraphics[width=12cm]{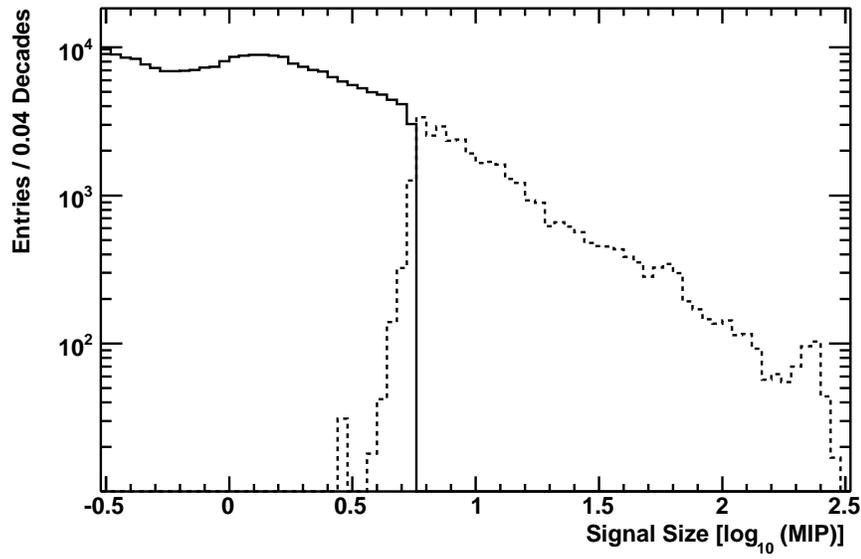}
   \caption{Deposited energy for the full dynamic range of a single PMT.  The solid line shows events read out in the low range, the dotted line shows events read out in the high range. The MIP peak occurs $\sim$ 1.3 MIPs because pathlength corrections are not applied.}
   \label{fig:acdrange}
\end{figure}

\begin{figure}
  \center
   \includegraphics[width=12cm]{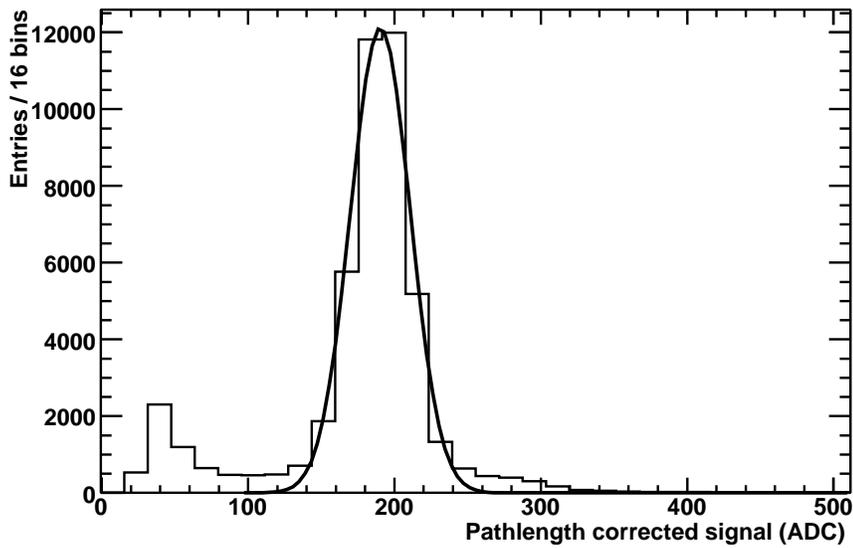}
   \caption{Distribution of pedestal subtracted and pathlength corrected carbon signals for a single PMT in the ACD in the high range. 
The carbon peak is clearly visible at $\sim$200 counts and the curve shows the fit to the data.}
   \label{fig:acdcarbon}
\end{figure}

The high and low range readout are calibrated in an analogous way.
The main difference is that for the high-range readout
we use tracks identified as carbon nuclei by the CAL, since the proton MIP-like signals are too small.
The algorithm requires the deposited energy in the CAL, which is pathlength corrected, to be consistent with that of 
carbon and the first few layers to have similar energy to avoid carbon interacting events. 
Also, in converting from raw pulse
heights to MIP equivalent values we allow for non-linearities caused
by signal saturation in the electronics and scintillators.
Figure~\ref{fig:acdrange} shows the deposited energy for the full dynamic range 
of a single PMT. 
The solid line shows events read out in the low range and the dotted line shows events read out in the high range.  
Figure~\ref{fig:acdcarbon} shows a distribution of pedestal subtracted and pathlength corrected signals for carbon in a single PMT.
\subsection{Veto threshold and high-level discriminator}
\label{sec:veto}
\begin{figure}
  \center
   \includegraphics[width=14cm]{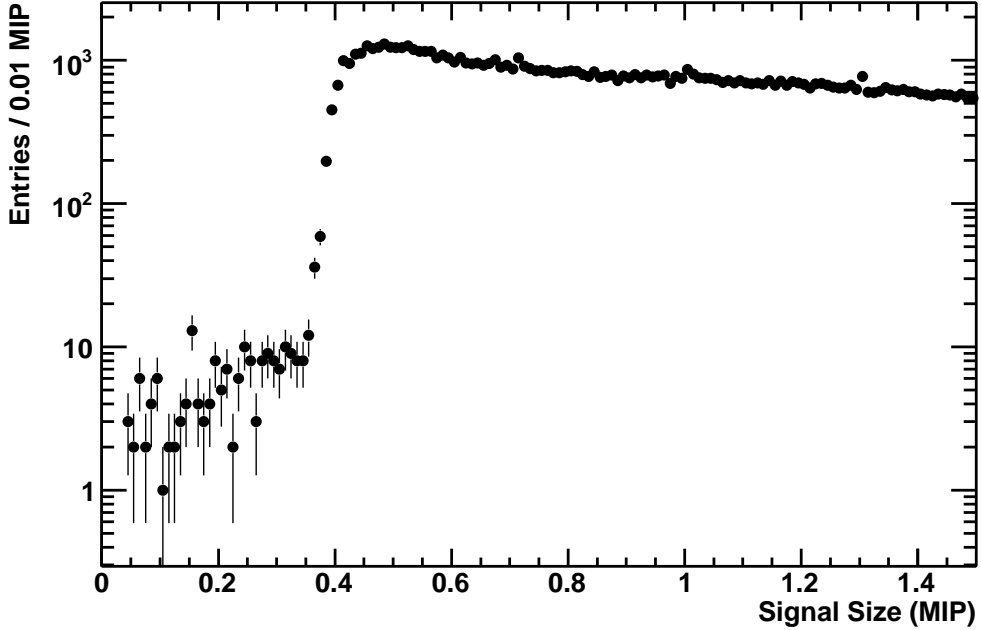}
   \caption{Veto turn-on curve for a single PMT. The small number of events (0.1\%) $<$ 0.4 MIPs correspond to overshoots in the ADC value from a previous event.}
   \label{fig:acdveto}
\end{figure}

The on-board thresholds are
controlled by DAC settings in the front-end electronics. 
These are calibrated by scanning three settings and
measuring the resulting veto threshold in pulse height bins. 
We combine that information with the MIP peak to pulse height scale, or
carbon peak for the high range, and set the veto 
and high-level discriminator thresholds as a fraction of the
MIP and carbon signals, respectively.  
The veto thresholds for each PMT are set 
separately to an accuracy of about 0.01 MIPs ($\sim$ 20 keV) relative to the calibrated MIP peak. 
As shown in Fig~\ref{fig:acdveto}, the veto turns on between 0.4-0.5 MIPs and the 50\% efficiency point is close to 0.45 MIP. 
For events below $<$ 0.4 MIPs the ADC values can be artificially small, 
because the readout electronics are relatively slow when compared to the veto electronics.
Events arriving when the electronics are overshooting the return to baseline tend to fire the veto discriminator, even
though they are not expected to have any ADC counts.
The turn-on for the high level discriminator occurs between 24-26 MIPs and 
the RMS width of the carbon peak is 20\% of the mean value. We have not 
yet monitored the stability of the carbon peak since it needs large statistical samples.
At least four months of data are required to obtain a carbon peak value with reasonable statistics.

\section{CAL calibrations}
\label{sec:cal}
The CAL is designed to measure the energy of incident photons and charged particles, and to determine the direction 
and energy of photons and charged particles for which the TKR did not provide direction information, either because
their trajectories did not cross the TKR or because they did not pair-produce in the TKR.
Its imaging properties are also a key ingredient in seeding the track reconstruction process in TKR data analysis and in the rejection of charged-particle background~\cite{latpaper}.

The CAL consists of 16 identical modules.  Each module is composed of 96 CsI(Tl) scintillation crystals arranged in a
hodoscopic configuration with eight layers each containing 12 crystals. Each 
layer is rotated 90$^\circ$ with respect
to its neighbors, forming an $x$-$y$ array.  
Crystals are read out by two dual-PIN-photodiode assemblies, one at 
each end, that measure the scintillation light produced 
in the crystal. Each photodiode assembly contains a 
large-area photodiode to measure small energy 
depositions and a small-area photodiode to 
measure large energy depositions. The active areas
of the large and small diodes have a
ratio of 6:1 with a spectral response well matched to the
scintillation spectrum of CsI(Tl).
Each of the 3072 photodiode assemblies is read out by an amplifier-discriminator ASIC, the GLAST Calorimeter
Front-End Electronics (GCFE).
To cover a large dynamic range of 5 $\times$ 10$^{5}$ in each GCFE with commercially available 12-bit ADCs for digitization,
the low and high energy photodiodes each have their own independent signal chains, the low-energy and the high-energy, 
and each chain operates with two
track and hold gains (low and high). This arrangement results in four overlapping energy ranges from 2 MeV to 70 GeV overall, as shown 
in Table~\ref{table:calranges}. Range overlap allows cross-calibration of the electronics.  Table~\ref{table:calranges} also shows the approximate factors used 
to convert ADC readout to energy units (MeV).

A detailed description of the CAL is found elsewhere~\cite{cal}.  
\begin{table}[hbt]
  \centering
  \begin{tabular}{|c|c|c|l|c|}
    \hline
    Name & Energy & Gain & Energy range & MeV/ADC  \\ \hline
   LEX8 & low & high &  2 MeV to 100 MeV & 0.033\\ \hline
   LEX1 & low & low &   2 MeV to 1 GeV & 0.30 \\ \hline
   HEX8 & high & high & 30 MeV to 7 GeV & 2.3 \\ \hline 
   HEX1 & high & low &  30 MeV to 70 GeV & 20 \\ \hline
    \hline
  \end{tabular}
  \caption{Four overlapping readout ranges for the CAL and their conversion factors.}
  \label{table:calranges}
\end{table}

Here we describe the on-orbit measurements of the following: CAL pedestals; crystal energy scale, derived from the electronics linearity and crystal light output;  crystal light asymmetry, which calibrates position measurements along the crystal; and threshold settings.
These calibrations allow determination of the location and amount of energy deposited in each crystal.  Processes for estimation of incident photon energy and the resulting overall energy resolution are discussed elsewhere~\cite{perfpaper,bt06}.

\subsection{Pedestals}
\label{sec:calpeds}

As discussed in Section~\ref{sec:acdpeds}, pedestals are offset voltages for each of the four CAL energy ranges that set the ``zero point'' for the energy scale.  We measure pedestals on orbit from the periodic triggers issued at 2 Hz by the LAT trigger system during all nominal science
data acquisitions.  
Chance coincidence energy deposits result in a small tail to the pedestal distribution, but this is suppressed both by the pedestal distribution fitting techniques used and by comparison of the various energy ranges.

\begin{figure}
  \center
   \includegraphics[width=14cm]{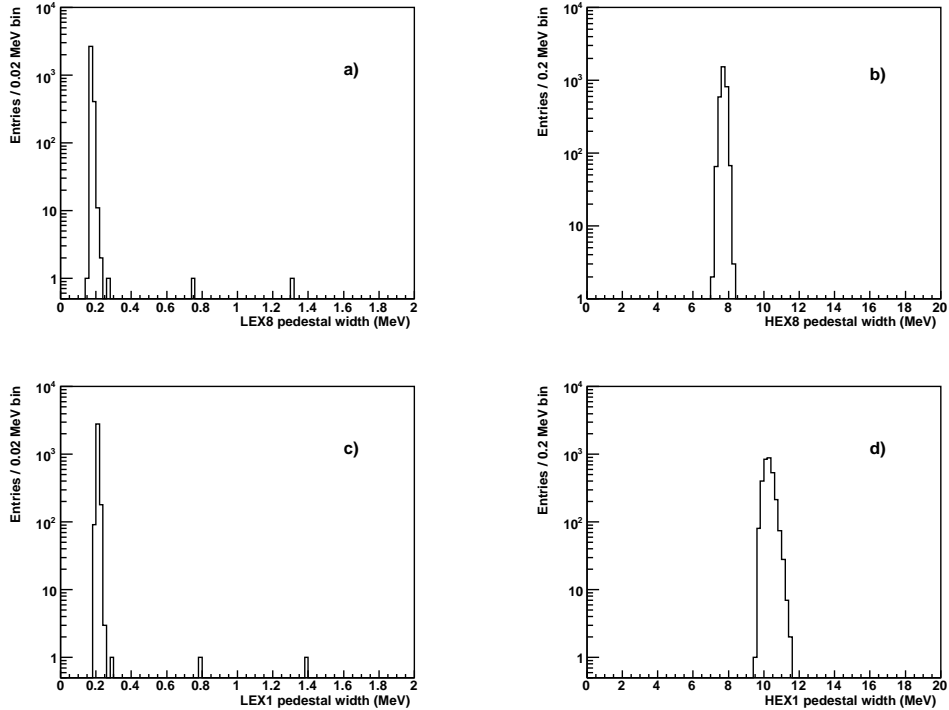}
   \caption{On-orbit pedestal widths (for all channels) for: a) LEX8, b) HEX8, c) LEX1 and d) HEX1 ranges.}
   \label{fig:calpedwidth}
\end{figure}

Figure~\ref{fig:calpedwidth} shows that typical pedestal widths (RMS) are 0.2 MeV 
for the LEX1 and LEX8 ranges and 7-10 MeV for the HEX1 and HEX8 ranges.  The pedestal values are regularly monitored
on orbit and have been extremely stable since an initial settling period.

Pedestal measurements made during thermal-vacuum tests in January 2008 indicated 
a channel-dependent linear dependence of pedestal position with temperature, where the drift magnitude varies from -3 to +3 ADC units per degree for the LEX8 and HEX8 ranges and $\sim$10 times smaller for LEX1 and HEX1, reflecting 
their smaller gains.  

When {\em Fermi} is in Pointed observing mode, the exposure of the LAT thermal control system to the warm Earth changes relative to the Sky Survey observing mode, and the LAT detector temperatures change modestly. In the CAL, the temperature changes 1--2 degrees within a few hours, leading to a temperature drift of the pedestals of a few ADC units at most.  Since Log Accept (LAC) thresholds, which determine whether a given crystal readout is included in the data stream, are set in absolute ADC unit values (i.e. not relative to the pedestal), changes in the pedestal values change the corresponding threshold energies.  A large enough pedestal change could thus influence data volume. 
However, only very large temperature changes ($\sim$ 20 degrees Celsius), which are not anticipated during the mission, could lead to significant increases in data volume.

\subsection{Individual crystal energy scales}
\label{sec:callin}

Calibration of the individual crystal energy scales involves the determination of the parameters of a transfer function between the energy deposited in the crystal and the signal output in ADC units. The transfer function consists of the response of the front-end electronics and the light emission and collection properties of the scintillation crystals.  We use charge injection calibrations to describe the electronics and ionization energy depositions from cosmic-ray muons 
or protons and various nuclei to calibrate crystal response. Ionization energy losses over a known path length are very predictable and hence make a good calibration tool.

Charge injection calibrations are used to characterize the non-linear behavior of the electronics chain.
A pulsed signal of known amplitude (controlled by a charge injection calibration DAC) is sent to the preamplifier input of 
each CAL front-end electronics channel. 
The nonlinearity of this DAC is specified by the manufacturer to be $<$0.1\% and hence negligible when compared to the nonlinearity of 
the front-end electronics.  

\begin{figure}
  \centering
   \includegraphics[width=6.7cm]{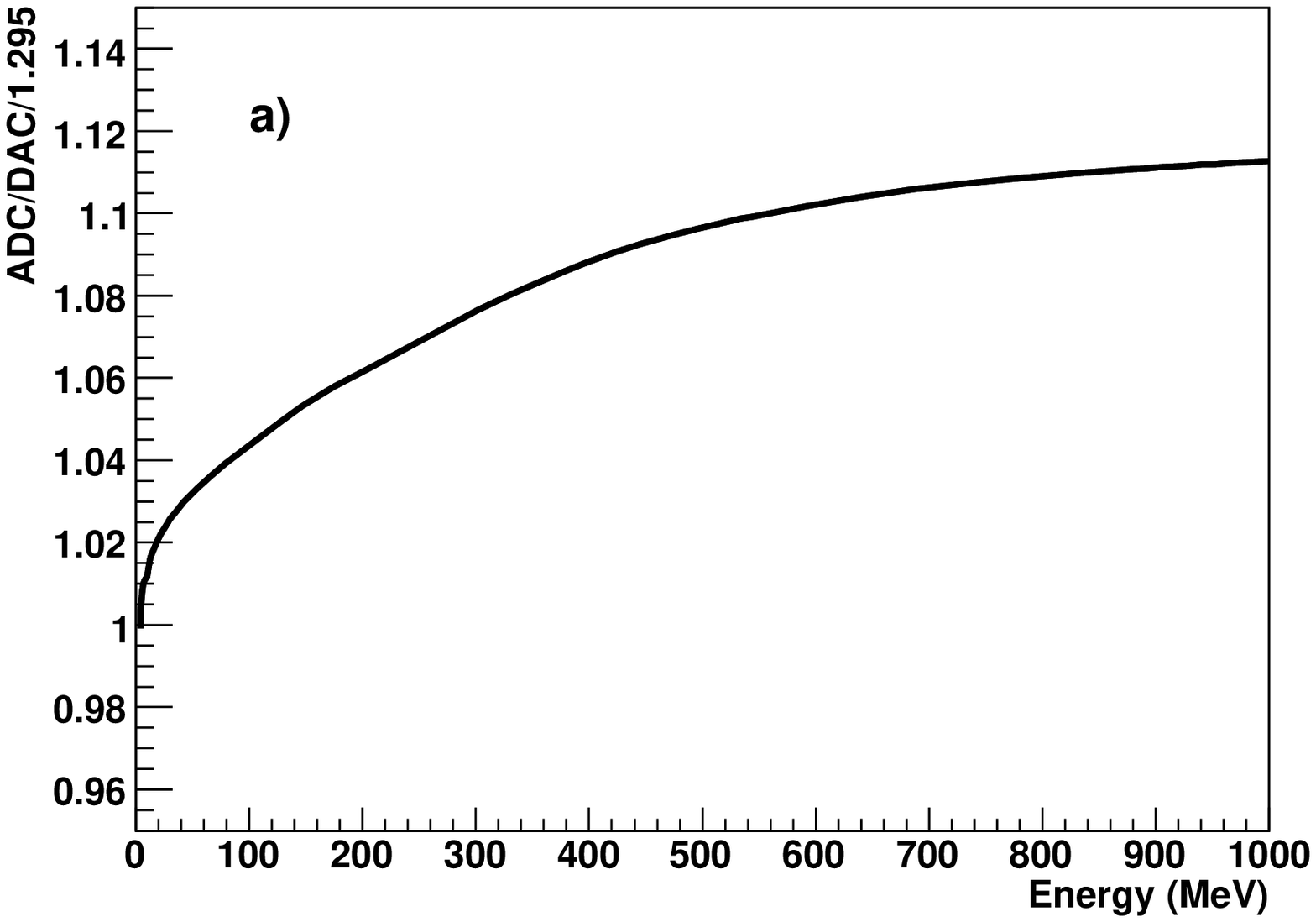}
   \includegraphics[width=6.7cm]{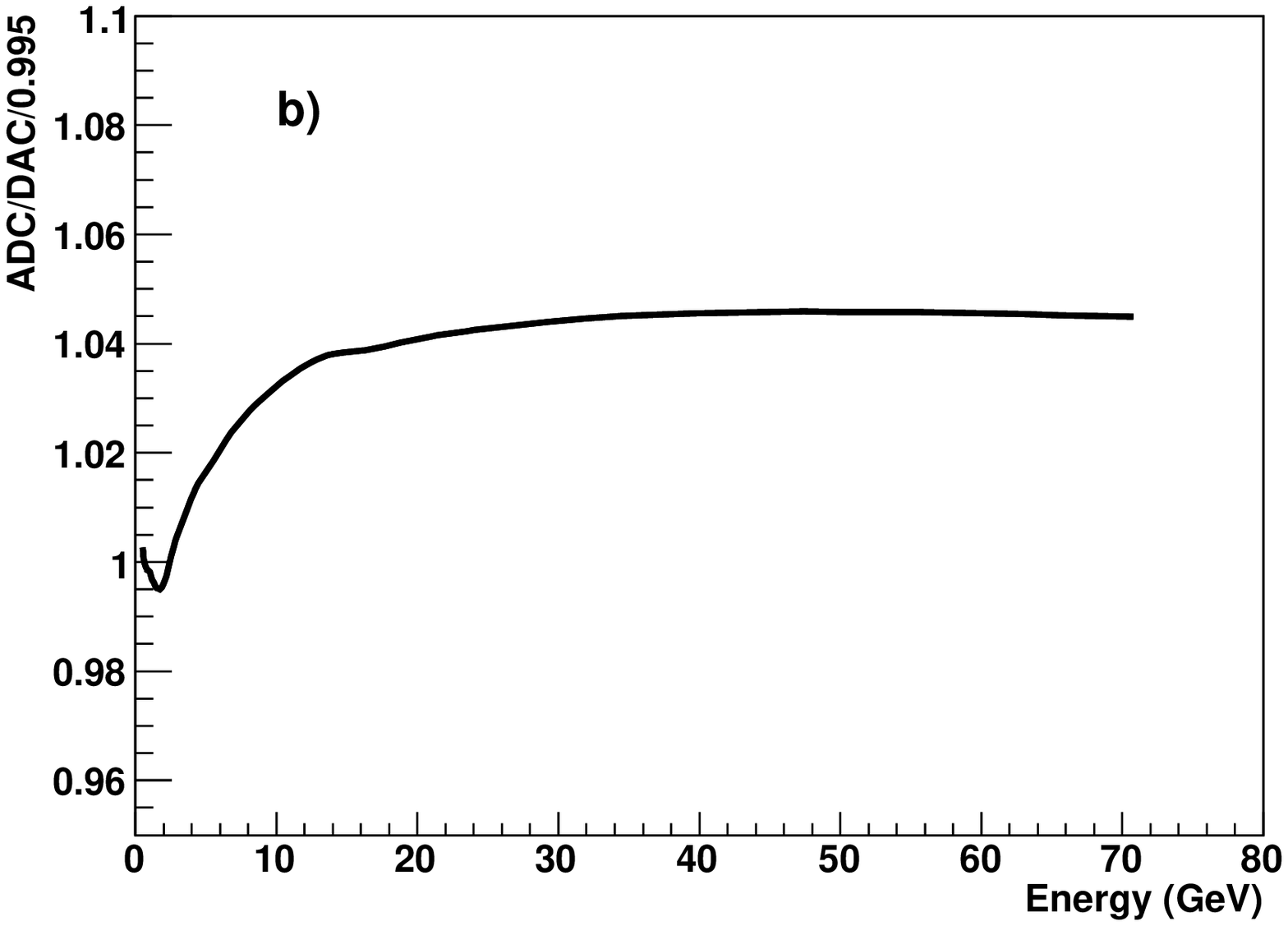}

   \caption{Characterization of electronics non-linearities. Normalized ADC/DAC versus energy for: a) LEX1 and b) HEX1 ranges.}
   \label{fig:calnonlin}
\end{figure}

For each fixed DAC setting we inject 100 pulses onto each of the electronics channels and average the resulting ADC output values.  A spline function is used to fit the resulting DAC versus ADC curve.  The functions (one for each channel), describe both electronics gain and non-linearity.  Using these functions, signals from each channel can be converted to a linear scale. 
Figure~\ref{fig:calnonlin} shows the normalized ADC/DAC versus energy, where deviations from one 
indicate non-linear behavior for measurements made at fixed energy values.
The largest deviations ($\sim$12\%) are seen in the LEX1 range displayed in Figure~\ref{fig:calnonlin}a. 
Figure~\ref{fig:calnonlin}b illustrates the case for the HEX1 range with 4\% non-linearities. The feature around 2 GeV in the HEX1 curve corresponds to 
cross-talk between LEX1, which saturates around 1 GeV, and HEX1. 
For the other ranges (LEX8 and HEX8) deviations are $<$1\%.   
After applying the measured nonlinearity calibration, residual nonlinearity is $\le$ 1\% of the measured energy, resulting in a negligible systematic effect in spectrum determination. 

The crystal response calibration, i.e. the function that relates deposited energy to the linearized signal described above, has been performed using different ionizing particles on the ground and in orbit.  In both cases, ion incident energies are, for the most part, in the slow relativistic rise region of the Bethe-Bloch curve, so the predicted energy loss per unit path length ($\frac{dE}{dx}$) is only weakly dependent on incident energy.  Using simulated incident spectra for ground and orbit environments, we determine an expected $\frac{dE}{dx}$ for each incident particle species.  We then collect spectra of the these species, correcting each event for path length, and compare the peak position in pulse height units to the predicted position in energy to yield a calibration.  Variations in incident spectrum with orbital position result in slight broadening of the energy deposit peaks, but, given integration over multiple orbits, the peak most-probable-value is well-determined and usable for calibration.  

On the ground, the low-energy scales were calibrated using sea-level cosmic ray muons while high energy ranges were calibrated using muons with the HEX1 and HEX8 channels set to a special ``muon mode" gain setting that increased the gain by a factor of $\sim$10.

On-orbit, we used a technique we refer to as ``proton inter-range calibration''. 
The low-energy scales are calibrated using protons.  Higher energies are calibrated by using energy deposits that meet two criteria:  first they must be in the overlap range between LEX1 and HEX8 and second they must result in a ``heavy ion'' trigger, the only common trigger that produces the required 4 energy range readout rather than the normal single range readout.  These events are a combination of galactic cosmic ray (GCR) primary carbon nuclei, interacting protons and other interacting or ionizing GCRs.  From events that meet these criteria, we can construct a cross calibration of the low and high energy ranges. In both ground and on-orbit cases, the ionization calibration, together with the charge injection results yield a usable energy scale that converts ADC units to deposited energy.  A week of nominal science operations data is sufficient to calibrate the energy scales using relativistic protons and heavy ion trigger events in the overlap energy range.

Protons that are accepted by either the MIP filter (when active) or the diagnostic filter (see Table~\ref{table:obf}) are required to pass a number of cuts. In order to reject all events that are not contained in a single CAL module and to eliminate ``corner clipping'' events, for which path length determination is not sufficiently accurate, we require extrapolated TKR tracks to cross the top and the bottom surfaces of a single CAL crystal, and be at least 5 mm away from the crystal edges.   In addition, we require single TKR track events with $>$ 20 TKR hits and a chi-square for the track $<$ 3. To reject low-energy re-entrant albedo protons that could broaden and bias the energy distribution, we require the multiple scattering angle, calculated by the Kalman filter used for TKR track reconstruction, to be $<$ 0.01. Finally, we reject nuclear-interacting protons by selecting events with two or fewer hits in each CAL layer and no additional crystals hit in the layer containing the crystal being calibrated.

In order to determine a calibration peak shape that represents the data well but minimizes the number of free parameters, we use a two-step process.  First, we produce a spectrum of path length corrected signals for all the crystals together.  Each peak is fit with a Landau distribution convolved with a Gaussian, for which all parameters are left free.  From this fit, we determine both the Landau and Gaussian widths, which are then fixed.

\begin{figure}
  \centering
   \includegraphics[width=15.cm]{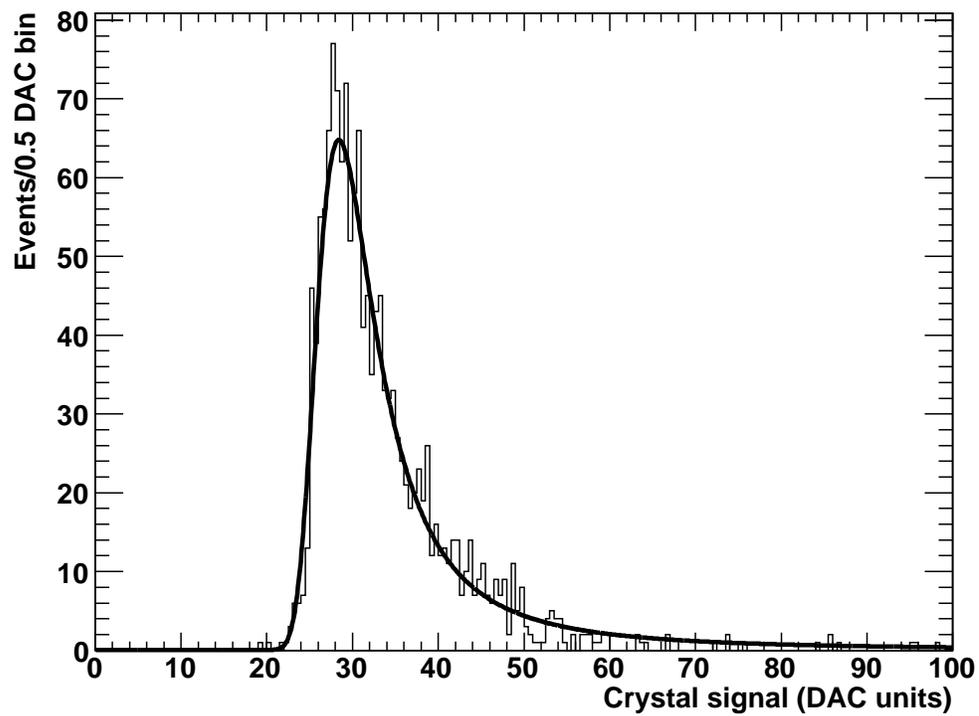}
   \caption{Energy deposited in a crystal (pathlength corrected). The position of the proton peak is given by the fit. 
The signal is corrected for electronics non-linearities.}
   \label{fig:calproton}
\end{figure}

In the second step, we fit the Gaussian-convolved Landau function to spectra from each crystal separately, allowing peak position and amplitude to vary but using the fixed widths determined above.  The peak most probable value (MPV) in energy units from the simulations (10.6 MeV for protons) is divided by the peak MPV in pulse height units determined by the fitting process just described to yield the desired calibration quantity for each crystal.   Figure~\ref{fig:calproton} shows the results, where the most probable value is the peak position.

Originally, we intended to use ``heavy'' GCR primary nuclei to calibrate the higher energy scales.  This would be desirable since $\frac{dE}{dx}$ varies as $Z^2$ where $Z$ is the atomic number of the incident ion.  The major difference between use of GCR heavy nuclei and protons or muons lies in a phenomenon known as ``quenching'', in which the crystal light output per unit deposited energy ($\frac{dL}{dE}$) is thought to be less for higher $Z$ nuclei than for protons or muons.  In addition to being $Z$ dependent, this phenomenon also depends on the ion incident energy.

Since quenching is not well studied for the combination of $Z$ and incident energies relevant to on-orbit CAL calibration, 
we measured the response of the calorimeter CsI(Tl) crystals to relativistic nuclei (from carbon to iron) at the GSI facility in 2003 and 2006.  The results of these studies indicated that for the ion energies examined at GSI, which were rather higher than those measured previously for CsI(Tl), $\frac{dL}{dE}$ was actually higher for measured nuclei with $6 \leq Z \leq 14$  than for protons~\cite{lott2006}.  Due to the lack of a physical model or understanding for this ``anti-quenching'' behavior, we felt that the systematic uncertainties introduced in using the heavy ion GCR calibration were unacceptable at this time and have used the proton inter-range technique instead.  

In order to study any possible changes in the energy scale calibration between pre and post launch measurements and during early on-orbit operations, we calculate, for each crystal, the ratio of the energy scale measured after launch to that measured 
prior to launch. 
A Gaussian fit to this distribution leads to a  mean bias, 
for low-energy diodes, of $\sim$1\% and a standard deviation which indicates crystal-to-crystal 
variations of 0.8\%. The latter characterizes the statistical precision of this calibration procedure. 

For the high-energy diodes, the bias with 
respect to ground calibration is 5\%, which is explained by the lack of any high-energy signal to 
reliably calibrate the ratio between the two diodes on the ground, as 
the muon signal is too small to be visible in the high-energy diode with flight gain settings.

\begin{figure}
  \centering
   \includegraphics[width=6.7cm]{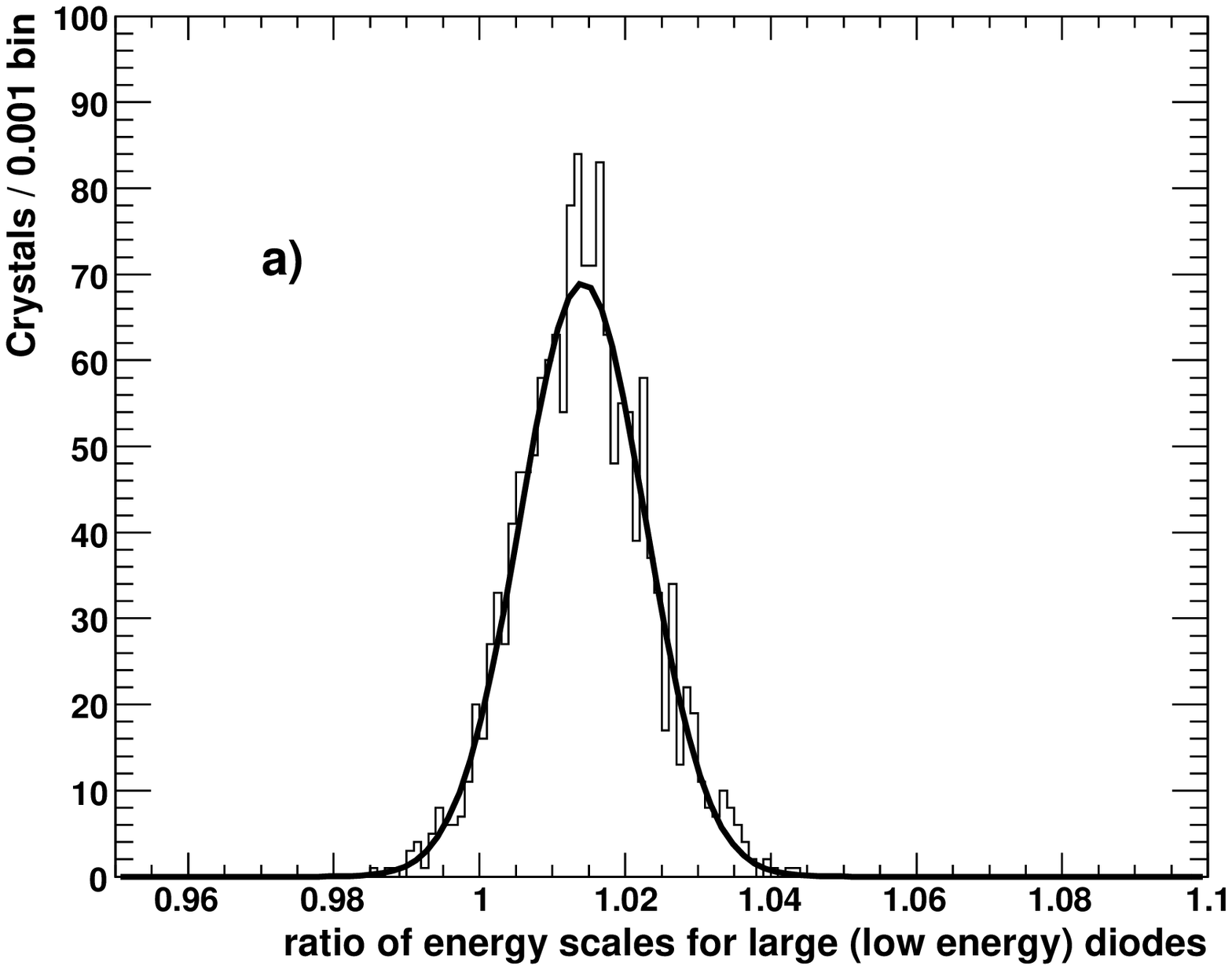}
   \includegraphics[width=6.7cm]{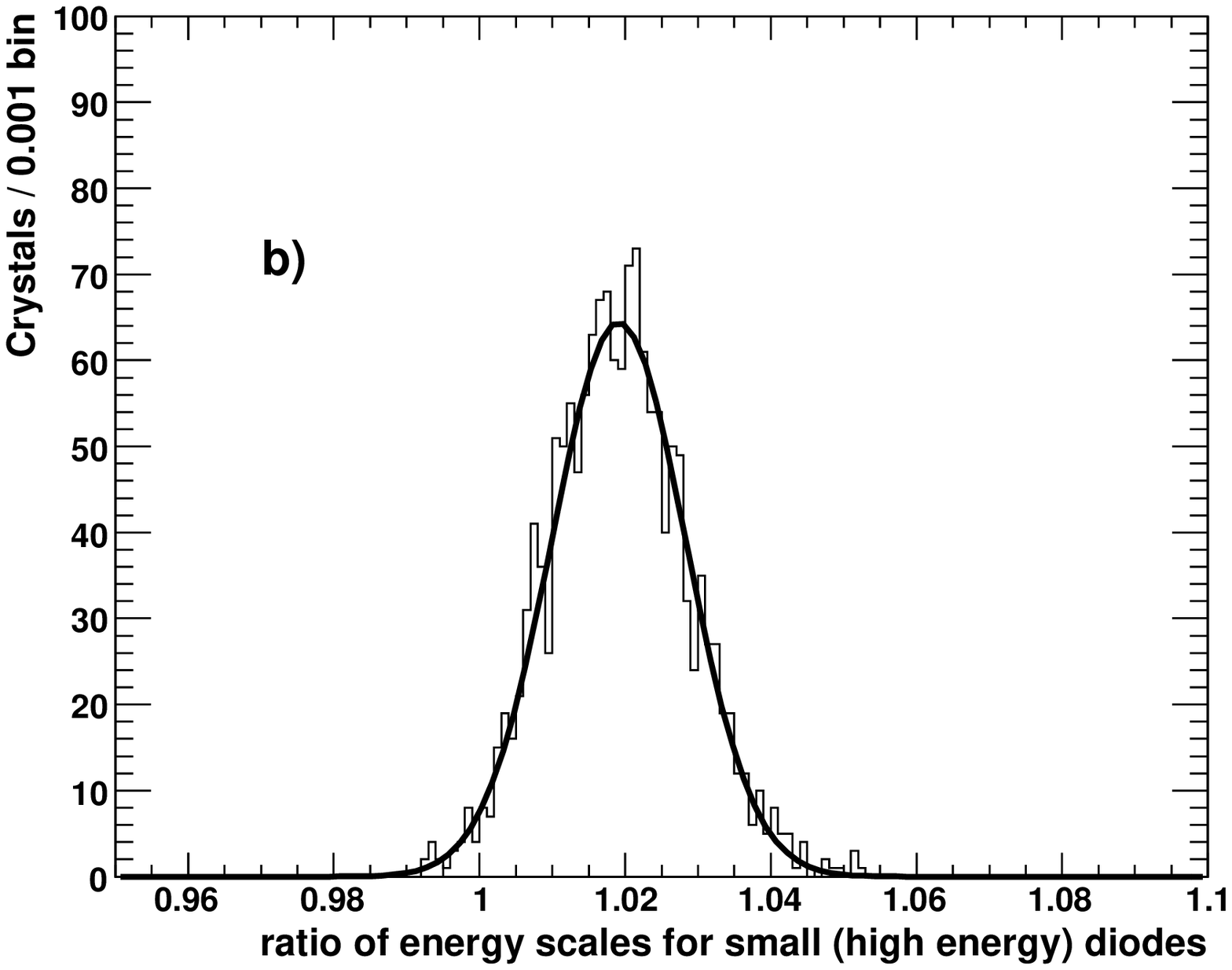}
   \caption{Ratio of the energy scales for the on-orbit calibrations performed in October and July 2008: a) low-energy diodes (mean 1.014, sigma 0.009) and b) high-energy diodes (mean 1.019, sigma 0.009). }
   \label{fig:calbias}
\end{figure}

The comparison of on-orbit results from July and October 2008 shows a shift of only 1-2\% and a spread, crystal-to-crystal, of $<$ 1\%. The 
stability of our calibrations is demonstrated in Figure~\ref{fig:calbias}a for the low-energy diode and Figure~\ref{fig:calbias}b for the high-energy diode.

\begin{figure}
  \centering
   \includegraphics[width=11.5cm]{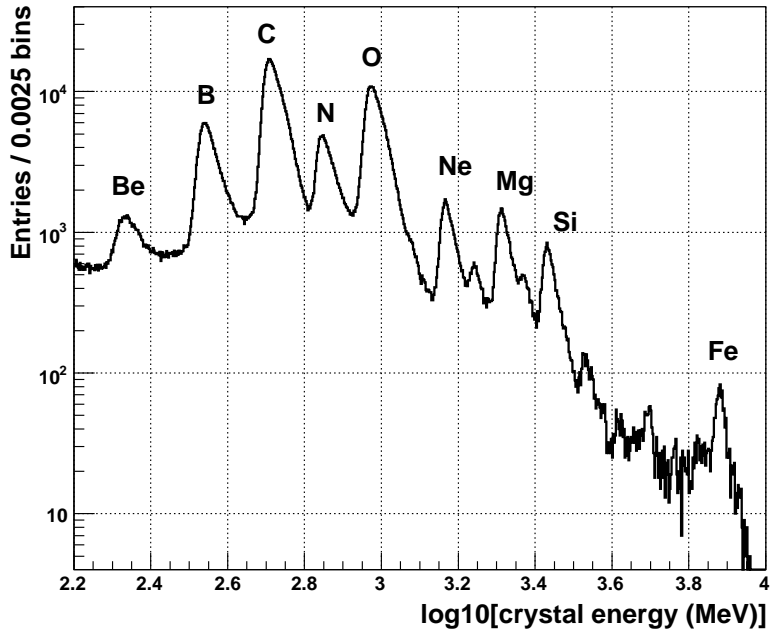}
   \caption{Energy deposited in all crystals from heavy nuclei collected during 4 days of on-orbit
operations. Pathlength corrections are applied.}
   \label{fig:heavyion}
\end{figure}

\begin{figure}
  \centering
   \includegraphics[width=11.5cm]{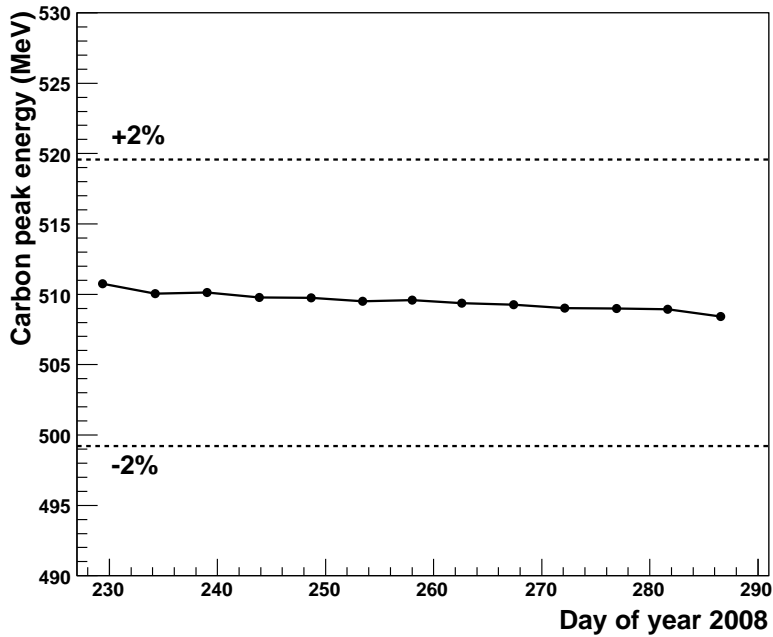}
   \caption{Position of carbon peak for 2 months of on-orbit data.}
   \label{fig:carbonpeak}
\end{figure}

Despite the fact that they are not suitably well understood for absolute calibrations, we do use energy deposits from from GCR heavy nuclei (500 MeV for carbon nuclei and 8 GeV
for iron) for independent
monitoring of the energy scale at high energies. The pathlength-corrected spectrum shown in Figure~\ref{fig:heavyion} is obtained by selecting crystal hits in low-multiplicity layers, 
thus rejecting nuclear interactions. Narrow peaks (the carbon peak has a 5\%
width) are easily identified.  It is worth noting that the measured energies 
deposited by cosmic rays are consistent with the anti-quenching effect observed in the beam test data acquired at GSI. For example, the $\frac{dE}{dx}$ for carbon is observed to be about 20\% 
higher than expected from a $Z^2$ scaling of the $\frac{dE}{dx}$ for protons. The count rate in the 
charge peaks is similar to that of the primary galactic cosmic-ray abundance, modified by loss of 
particles through charge-changing interactions above the CAL and by the decreasing efficiency of the
on-board heavy-ion filter for higher $Z$ nuclei.
For example, Figure~\ref{fig:carbonpeak} shows that, although there is a slow systematic shift, the carbon peak position is stable to within 0.5\% (for the whole CAL) after 2 months of operations.

\subsection{Light asymmetry}
\label{sec:calpos}

The design of the CAL crystals deliberately ``tapers'' the light propagation properties of the crystal so that we can determine the longitudinal position of an energy deposit by comparison of the signal at each crystal end.  We refer to the measured quantity as the signal asymmetry, 
defined as the logarithm of the ratio between the signals read out at the two ends of the same crystal. 

To calibrate light asymmetry, we select non-interacting heavy nuclei in a similar way to protons (see Section~\ref{sec:callin}), 
but with looser TKR track quality.
The TKR-determined position in a CAL crystal is obtained by extrapolating the TKR track to the center of the crystal (half-way through its thickness). 
Each track enters the crystal through one of the twelve evenly spaced bins defined along its length.  The distribution of asymmetry signals for each bin is collected by computing the asymmetry for each event for the appropriate bin.  
The first and the last bins, near the crystal ends, are not used because of known 
non-uniformities in light collection near the diodes. 
The average asymmetry is calculated for each of the ten central bins along each crystal and is fit with a {\em spline} function for purposes of interpolation. 

We use the signals from the two ends of each crystal to obtain the weighted centroid of the 
energy depositions along the crystal. The position resolution along the crystal is defined 
as the RMS of the difference of the position from light asymmetry to that from track extrapolation.

\begin{figure}
  \center
   \includegraphics[width=15cm]{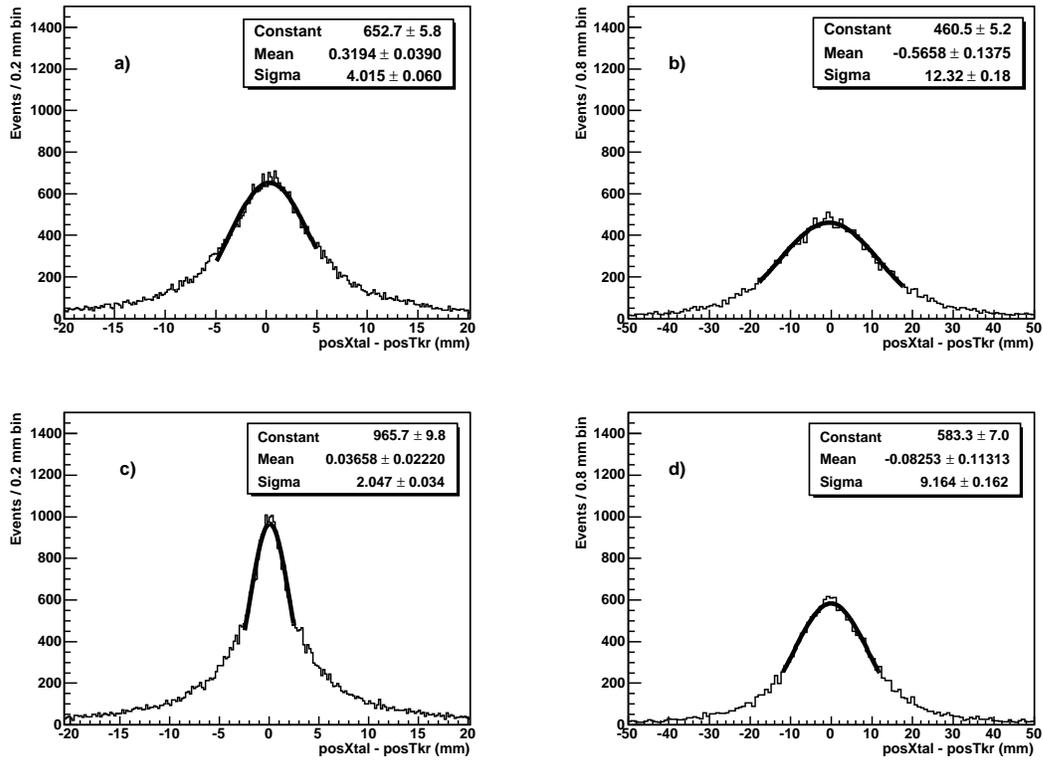}
   \caption{CAL position resolution from 200 to 900 MeV using ground calibration constants  a) LEX1 and b) HEX8, 
and using flight calibrations for: c) LEX1 and d) HEX8.}
   \label{fig:posres}
\end{figure}

We calibrated the light asymmetry of each crystal on the ground with sea-level cosmic muons, and we recalibrated
on orbit with GCRs.  The calibration constants derived on orbit are more precise than those derived on the
ground because the GCRs suffer less multiple scattering than muons and create larger scintillation signals.
The position resolution measurement for energy depositions from 200 to 900 MeV, 
for both LEX1, and HEX8 ranges, improves from 4 mm to 2 mm and from 13 mm to 9 mm, respectively. 
Figure~\ref{fig:posres} shows position resolutions with pre- and post-launch constants in the energy range 
from 200 to 900 MeV. Results are dominated by the 
carbon events which peak $\sim$500 MeV. The improvement by using flight calibration constants is clearly seen. 

\subsection{Trigger thresholds and upper level discriminators}
\label{sec:calthresh}

To reduce the data volume generated by the CAL and the additional dead time that would be created by
moving large events through the LAT data acquisition system, each GCFE has a zero-suppression discriminator
with a programmable threshold DAC.  This threshold, known as the log-accept (LAC) threshold, is nominally set
to 2 MeV, which is approximately 10 times higher than the average electronic noise.  The zero-suppression for 
an entire crystal is performed on
the logical OR of the LAC discriminator states at the two ends; thus data from {\em both} ends 
of a crystal are included in the CAL data stream if the LAC discriminator on either end fires.  

\begin{figure}
  \centering
   \includegraphics[width=14cm]{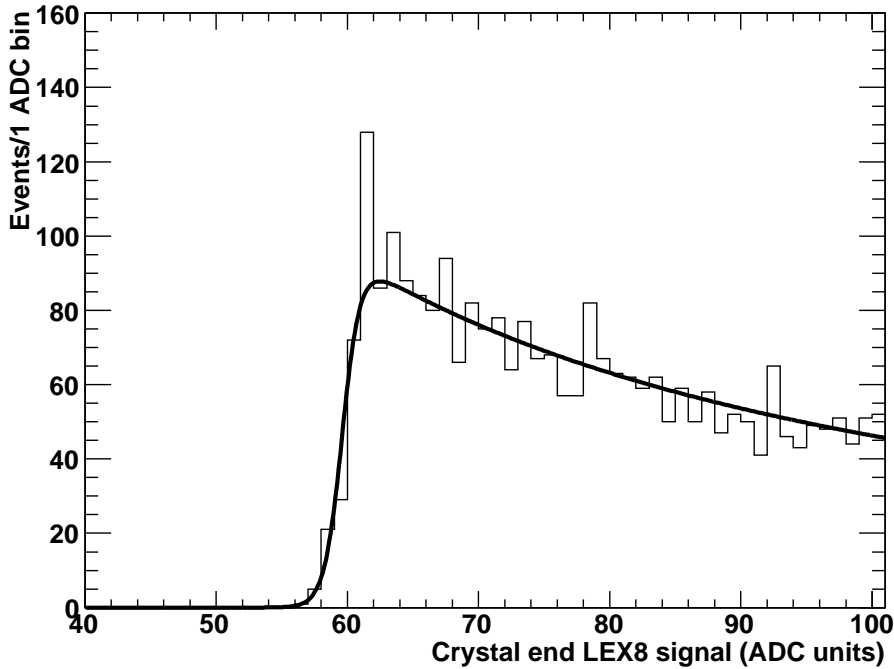}
   \caption{Signals from one end of a crystal (LEX8 range) and a fit used to determine the LAC threshold.}
   \label{fig:lac}
\end{figure}

Extensive ground testing with the LAT charge injection system and sea-level cosmic ray muons established the
functionality, linearity, and energy scale for each LAC discriminator.  We launched with LAC settings derived from
the ground calibrations, but the LAC settings are temperature dependent, primarily because the pedestal values
are temperature dependent.  Thus we revised the LAC calibration constants on orbit once the LAT had achieved
its stable operating temperature.  For each discriminator, 
we characterize the
relationship between DAC setting and LAC value (in MeV) with a linear model that is derived from calibration
data acquired with the LAC set at two values near the nominal setting.
We calibrate one end of all crystals at a time using a set of four configurations such that
the LAC threshold at the crystal end {\em not} being calibrated 
is set to its maximum possible value, preventing it from initiating 
the readout, and the LAC threshold at the end being calibrated is set to the test value. 
This process give a LAC measurement for each channel with 
a statistical precision of 5\% ($\sim$ 0.1 MeV). 
Figure~\ref{fig:lac} shows an example of the signals measured at one end of a crystal (LEX8 range) and the fit that determines the 
LAC threshold. The values can be easily converted into energy by using results from Table~\ref{table:calranges}.
We are monitoring four GCFEs with out-of-family electronic noise (out of 3072 in the CAL).  In February 2009, we 
inhibited one of the channels from participating in 
zero-suppression decisions because its noise level reached 1.5 MeV. This has 
no impact on the scientific performance of the CAL.

We monitor the stability of each LAC threshold in all nominal science operations data acquisitions.  To measure
the threshold value, we select events for which we are certain which discriminator qualified the crystal for inclusion
in the data stream.  Because the LAC value at each crystal end is within 10\% of the setting at the opposite end,
we achieve this certainty by selecting events where the signal from the two ends differs by more than 10\%.

The CAL provides two fast signals that participate in the formation of LAT trigger, the low energy CAL\_LO and 
high energy CAL\_HI triggers.  The CAL\_LO and CAL\_HI trigger requests are formed as the logical OR of the
outputs of the programmable fast low-energy and fast high-energy trigger discriminators, respectively FLE and FHE, 
at each end of a crystal.  The nominal values for the FLE and FHE thresholds are 100 MeV and 1000 MeV per
crystal, respectively, as measured at the center of the crystal by each GCFE.

Extensive testing on the ground with the LAT charge injection system clearly demonstrated the functionality and
linearity of each FLE and FHE discriminator; however it gave only an approximate absolute calibration (i.e. in MeV deposited)
of the threshold DACs.

The FLE and FHE thresholds are calibrated on orbit using background events recorded in dedicated-mode with 
additional information provided by the tower electronics module.  However, this additional trigger 
diagnostic information only provides the logical OR of the combination of 
all 12 FLE or all 12 FHE discriminators for each CAL layer-end; thus it does not clearly identify which 
crystal end produced the trigger signal.  Distinct procedures for FLE and FHE thresholds 
are necessary to resolve this 12-fold ambiguity.  To calibrate the
FLE discriminators near the nominal 100 MeV setting, we separately enable the trigger for each of 
two groups of six crystals in a layer (the six odd-numbered
and the six even-numbered crystals) and require that five of the six enabled crystals have a signal below 50 MeV.
Most showers share their energy between adjacent crystals in a layer, so the separation into two sets of six 
non-adjacent crystals readily resolves that ambiguity.  The 50 MeV energy cut ensures that only one of the six is
the source of the trigger signal.  To calibrate the FHE discriminators, we allow only the CAL\_LO signal to initiate a LAT trigger,
and we enable the FHE discriminators in two groups of six crystals per layer while we read the
state of the FHE trigger diagnostic information.

\begin{figure}
  \center
   \includegraphics[width=6.7cm]{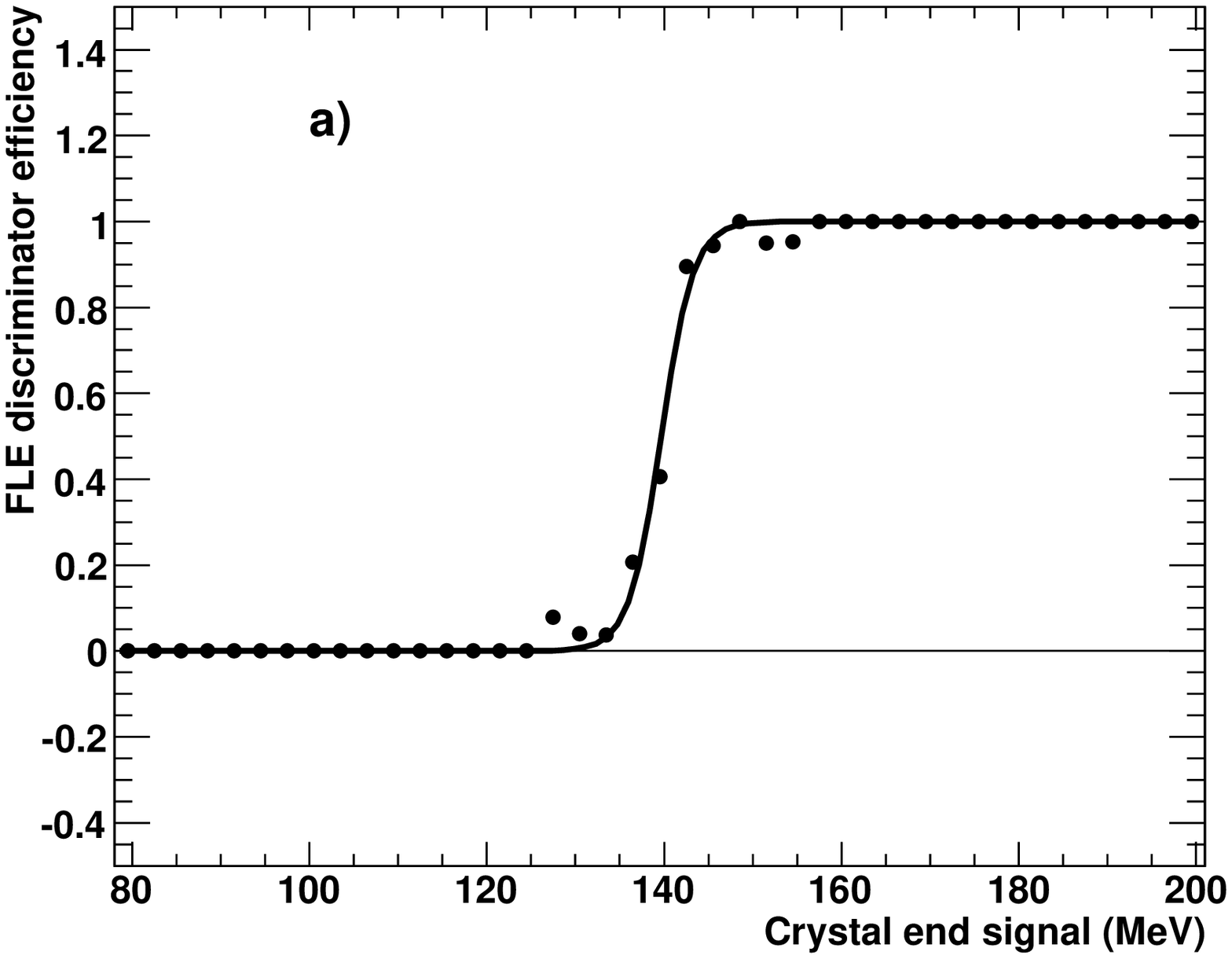}
   \includegraphics[width=6.7cm]{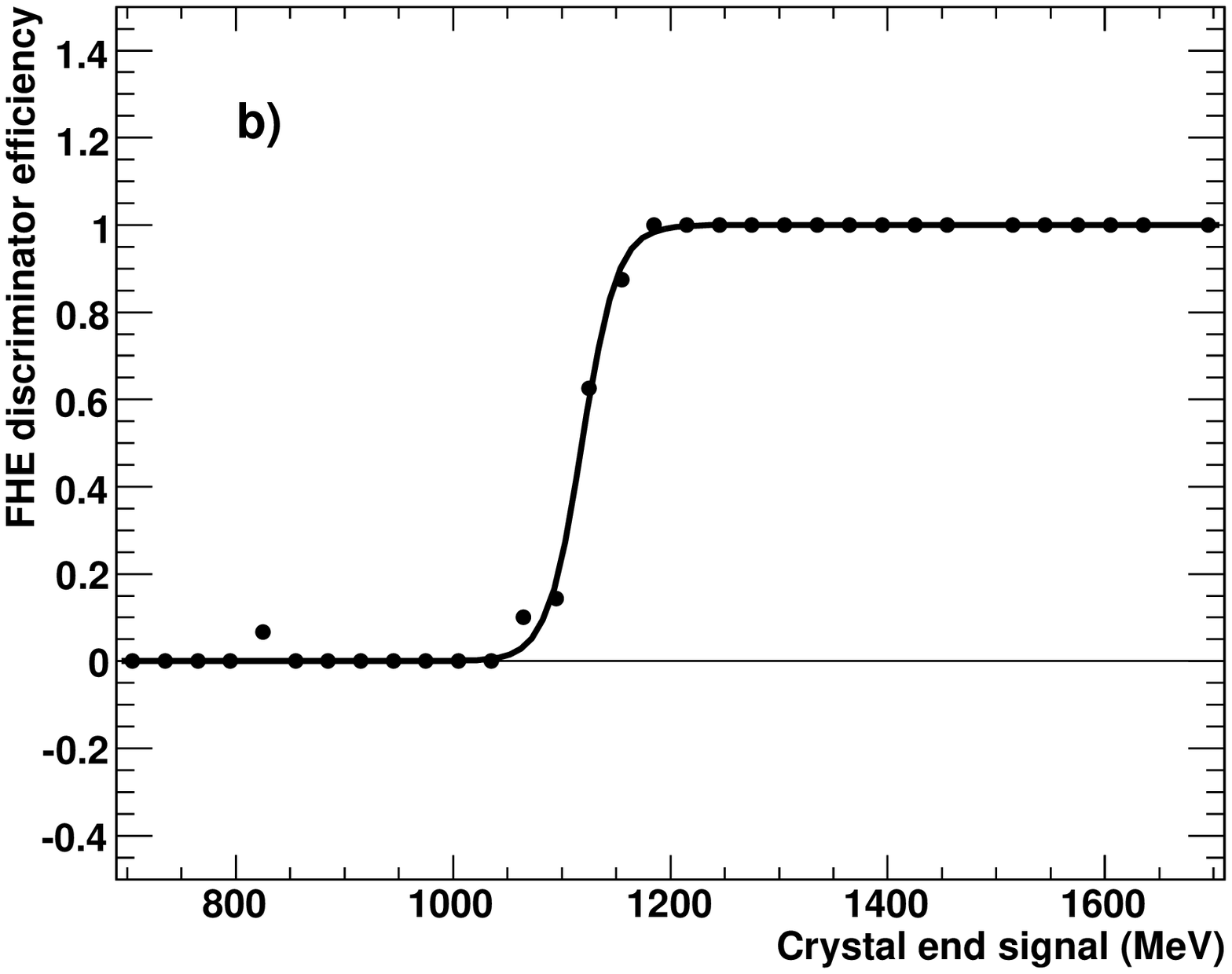}
   \caption{Efficiency versus energy for the FLE and FHE thresholds.}
   \label{fig:caleff}
\end{figure}

The efficiency of a discriminator as a function of signal (in MeV) is 
determined by calculating the ratio of the spectrum 
from events for which the discriminator fires to the spectrum for all events.
As shown in Figure~\ref{fig:caleff}, the value of the threshold is obtained by fitting a step function to this ratio.  Having established that the FLE and FHE discriminators are linear in threshold DAC setting, we calibrate
each discriminator at two settings near nominal value and fit the measurements with a linear model.  We calibrate FLE at
100 MeV and 150 MeV, and we calibrate FHE at 1000 MeV and 1500 MeV.
The statistical error in determining the threshold values is $<$ 1\% for FLE and $<$ 2\% for FHE thresholds.
Figure~\ref{fig:caleff} illustrates how these efficiencies are obtained for the FLE and FHE thresholds for one GCFE.
These data are from dedicated calibration runs taken during early operations (July 2008), 
when thresholds were set using calibration coefficients derived from ground tests.  As it happens, the ground calibration
gave threshold values for FLE and FHE somewhat higher than intended, viz. $\sim$140 MeV and $\sim$1200 MeV,
respectively.  The ground FLE and FHE calibration relied on the charge-injection system, which gives
pulse shapes that differ from those produced by CsI(Tl) scintillation signals.  Since the trigger signal is fast ($\sim$ 250 ns) and the 
energy measurement depends on the slow shaper ($\sim$ 4000 ns), the difference in the shape of the pulses is important and
creates the 40\% and 20\% bias we observed.  We then adjusted the FLE and FHE settings using the calibration constants
derived from the on-orbit calibration.  We continue to monitor the FLE and FHE threshold values with data from the
nominal science acquisitions.

The programmable Upper Level Discriminator (ULD) in each GCFE is responsible for switching 
between CAL energy ranges.  To select the best range for digitization, three ULDs in each GCFE compare the outputs 
of three ranges (LEX8, LEX1 and HEX8) with corresponding ULD threshold (one setting per GCFE).
The output of these discriminators is analyzed by the range selection logic which selects 
the best range as the highest range without an ULD signal~\cite{cal}. 
All three ULD thresholds of each crystal end are defined by one DAC and are set to $\sim$5\% below the saturation level 
of the ADC.  We calibrated the ULD threshold DACs on the ground with charge injection and
verified those settings and the linear calibration model on orbit during nominal science continuous acquisitions.
We measured the ULDs on orbit 
by finding, for each energy range, the highest observed signal for each individual crystal end.  Most on-orbit
data-taking configurations read out only one of the four available ranges, namely 
the one providing largest signal below the ADC saturation, but the nominal science data explore
all four ranges fully.

\begin{figure}
  \center
   \includegraphics[width=6.7cm]{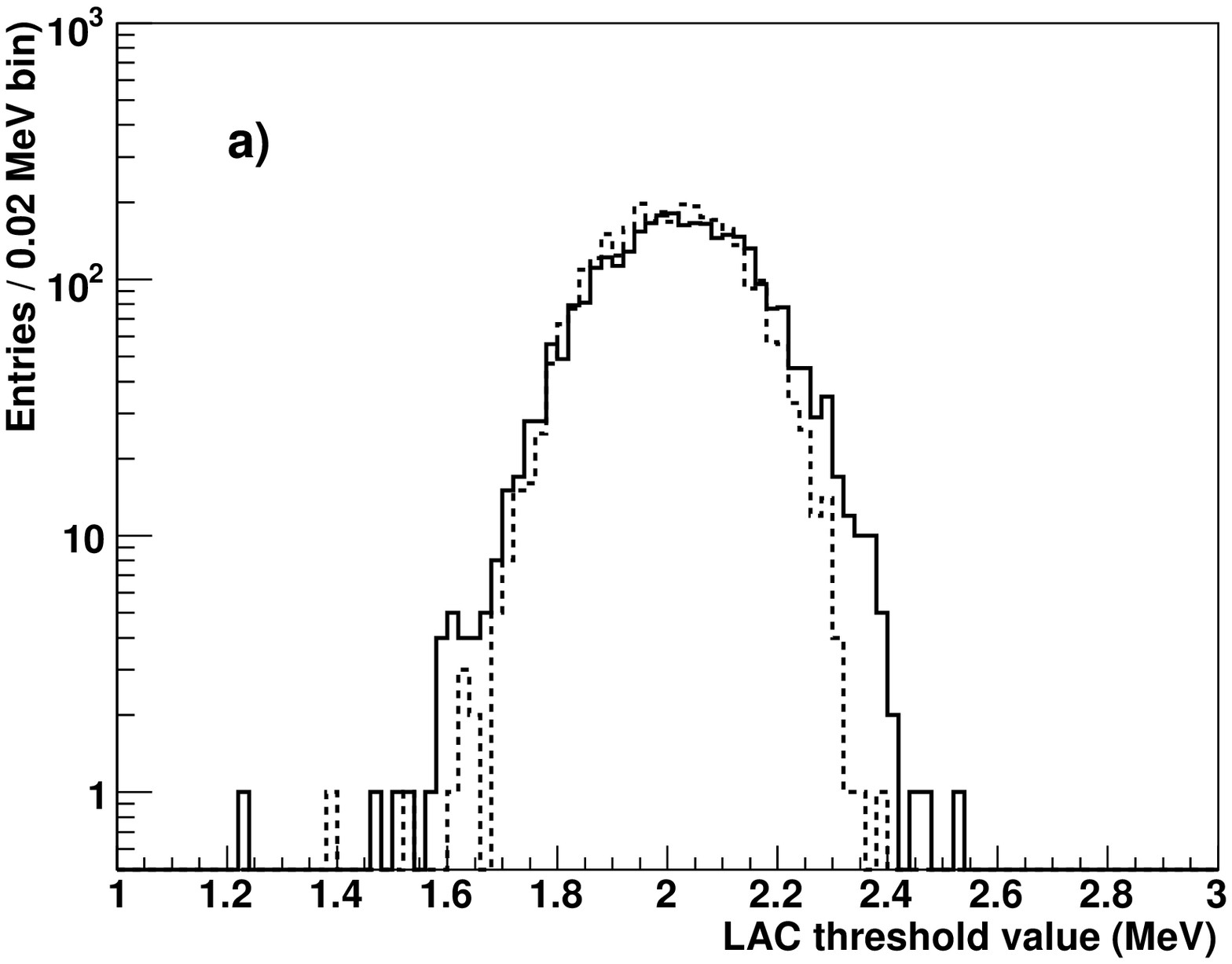}
   \includegraphics[width=6.7cm]{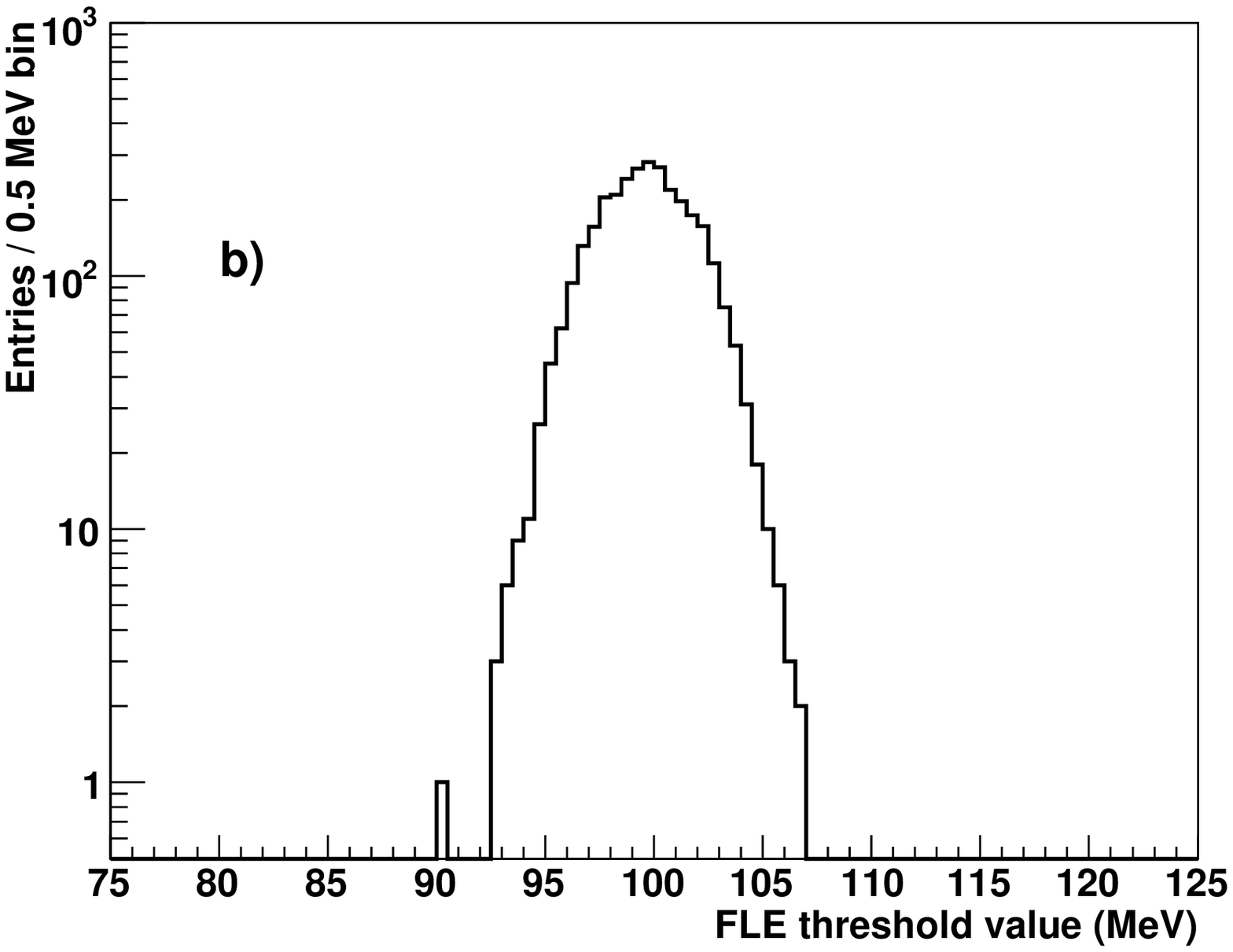}
   \includegraphics[width=6.7cm]{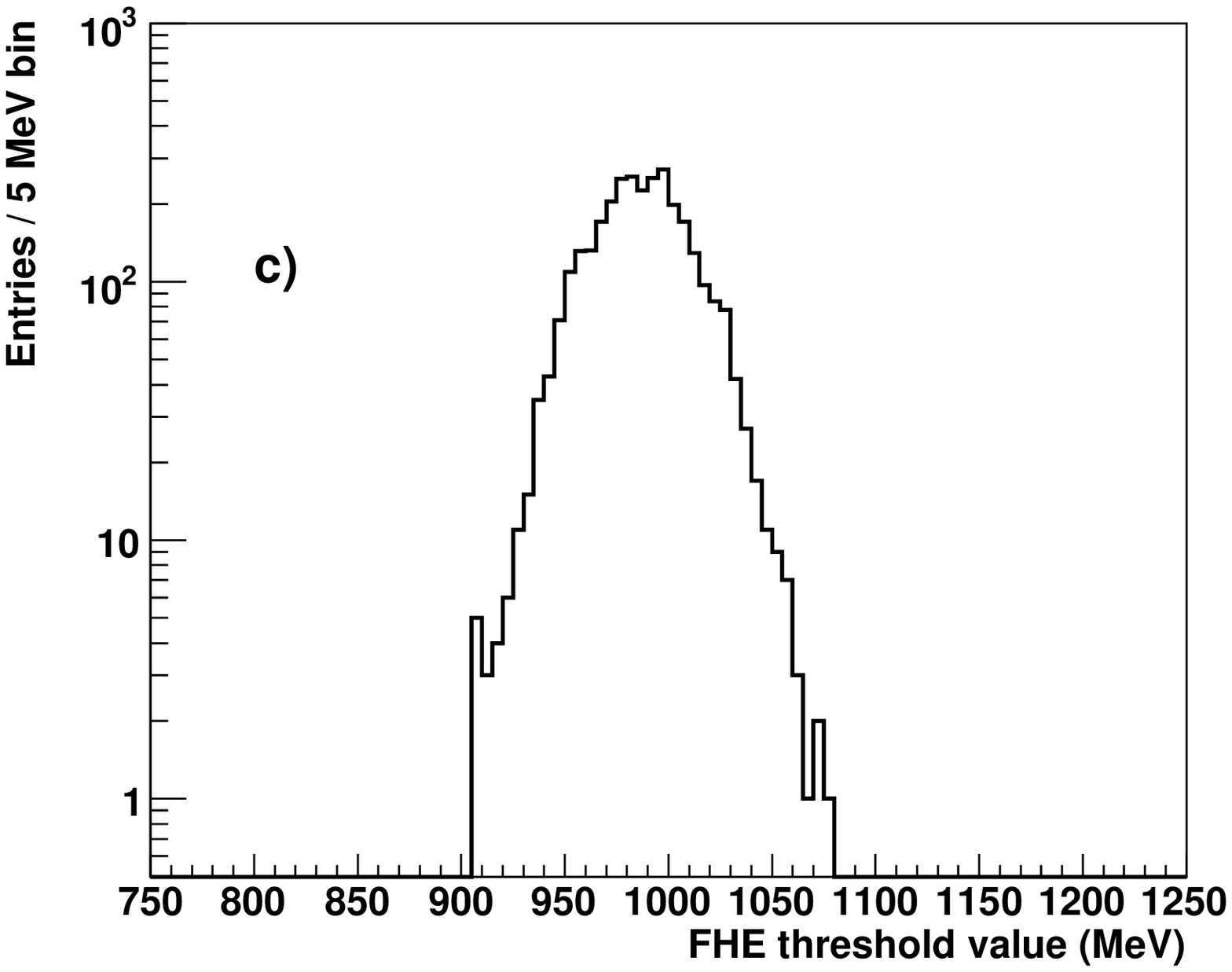}
   \includegraphics[width=6.7cm]{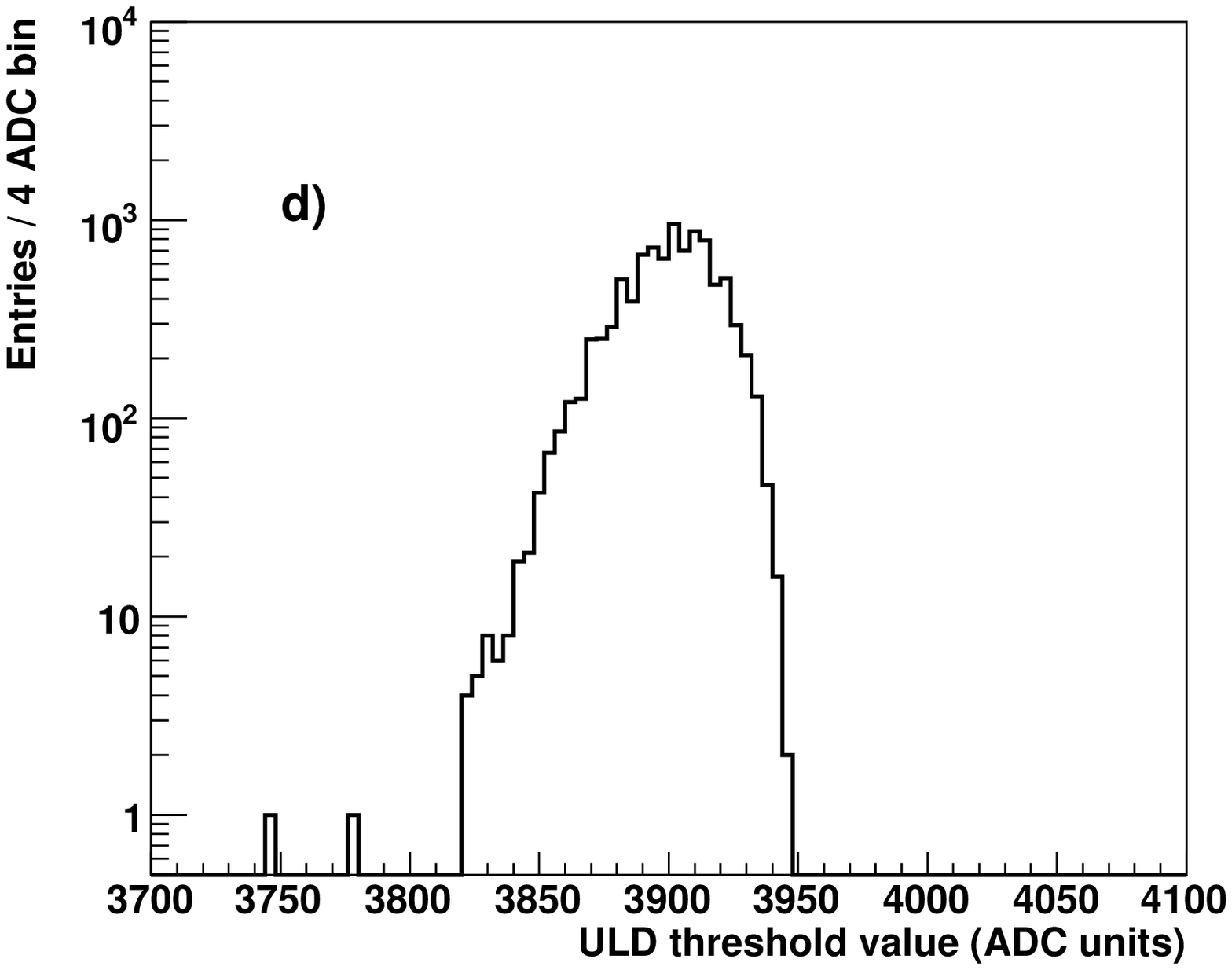}
   \caption{On-orbit measurements of threshold values for November 2008: a) LAC, b) FLE, c) FHE and d) ULD. The dashed line 
in the LAC histogram shows the effects of pedestal drifts seen in August 2008, prior to stabilization. The ULD thresholds are expressed in 
non-pedestal-subtracted ADC units. This makes it easier to judge how thresholds are set. The saturation limit corresponds to 4095 ACD units.}
   \label{fig:calthresholds}
\end{figure}

We updated the LAC, FLE, FHE, and ULD threshold settings after their initial on-orbit calibrations and later adjusted
them to accommodate the small settling drift in pedestal values (see  Section~\ref{sec:calpeds} for details). 
Figure~\ref{fig:calthresholds} shows the distribution of all thresholds for all channels.
The LAC threshold data from August (dashed) and November 2008 (solid) are used in Figure~\ref{fig:calthresholds}, 
where the effect of pedestal evolution since launch is seen as a slight broadening of the distribution.  
This effect is negligible for the other thresholds. 
The slight asymmetry in the distribution of ULD values (Figure~\ref{fig:calthresholds}d) has absolutely no effect on CAL
performance; it means only that for a small fraction of channels the range switching will happen at a slightly lower energy than
for the majority of channels.

\section{\TKR\ calibrations}
\label{sec:tkr}

The TKR is used to convert the photon to an e$^+$/e$^-$ pair and to 
determine the incoming photon direction. It is also the main contributor to the LAT trigger.
It consists of sixteen modules each composed of a stack of 19 trays. 
A tray is a stiff, lightweight carbon-composite panel with silicon-strip detectors (SSDs) mounted on both sides 
with strips oriented along the same direction. All but the three bottommost trays in each TKR module, contain an array of 
tungsten foils, which matches the active area of each SSD. These foils act as photon converters. 
Depending on their location within the tower (front or back), foils 
are 3\% and 18\% of a radiation length. 
Each tray is rotated 90\degree\ with respect to the one above and the one below. 
Therefore, two consecutive trays are needed to provide orthogonal measurements of the $x$,$y$ coordinates.
Details of the \TKR\ design are described elsewhere~\cite{Atwood:2007ra}.

Each side (top or bottom) of the tray consists of 1536 silicon strips read out by 
twenty four 64-channel amplifier-discriminator ASICs, GLAST Tracker Front-end Electronics (GTFE), which are controlled by 
two digital readout-controller ASICs, GLAST Tracker Readout Controller (GTRC).
Each channel in the GTFE has a preamplifier, shaping amplifier, and discriminator similar, although not identical, to 
the prototype circuits described in \cite{Johnson98}.
The amplified detector signals are discriminated by a single threshold per GTFE chip; no other measurement of the signal size is made within the GTFE.
The \TKR\ electronics is discussed in detail elsewhere~\cite{Baldini:2006pv}.

The trigger information is formed within each GTFE chip from a logical OR of the 64 channels. 
Any latched, noisy or inoperable channel can be masked.
The OR signal is passed to the left or to the right, 
depending on how the chip is configured, and combined with the OR of the neighbor.  
This procedure continues down the line, until the GTRC receives a logical OR of all non-masked channels it controls.
This ``layer-OR" initiates a one-shot pulse of adjustable length in the GTRC, which is sent as a fast trigger signal input for the trigger decision.
In addition, a counter in the GTRC measures the length of the layer-OR signal, i.e. the time-over-threshold (ToT), and buffers the result for inclusion in the event data stream.
Upon receipt of a signal that acknowledges the trigger decision, each GTFE chip latches the status of all 64 channels into one of the four internal event buffers. 
Another 64-bit mask, which is separate from the trigger mask mentioned above, can be used to mask any subset of channels from contributing data, as may be necessary in case of noisy channels.

TKR calibrations include the determination of the noisy channels that form the trigger and data masks, of the trigger threshold 
settings, and ToT calibrations.

\subsection{Noisy channels}
\label{sec:tkrch}

Since noisy channels can increase the false trigger rate, and can affect instrumental dead time, \TOT\ measurements and data volume,
they are 
disabled in trigger and/or data masks. 

The trigger mask determines the active channels that can participate in the formation of the layer-OR trigger signals.
The noise occupancy for each strip is measured using not only periodic triggers but all available events, thus 
increasing statistics by almost 2 orders of magnitude.
The computation of noise occupancy for non-periodic triggered events excludes consecutive (2 or more) layers with at least one hit in each. 
The occupancy measured by this method is consistent with that obtained from periodic triggers.

Noisy channels produce off-timing trigger signals resulting in a dead time of $\sim 1~\mu$s for the layer involved.
Noisy channels also lead to incorrect \TOT\ measurements if the noise hit occurs at the tail of the main pulse.
To minimize this effect, we mask any channel with occupancy greater than 0.7\%\footnote{0.7\% corresponds to the value in which the loss in efficiency due to dead time is comparable to that from masking noisy channels.}.
Furthermore, if the occupancy of layer-OR is greater than 8\%, we mask the highest-occupancy channels until we reduce it 
to this fraction.

The data mask determines the active channels whose data can be transmitted to the ground. 
Since the offline track reconstruction software is tolerant of high-occupancy channels, data mask is driven 
by the constraint on the data rate given by noisy channels. 
We mask any channel with occupancy greater than 50\% since it does not carry any useful information.
We limit the data size due to noisy channels to be less than 10\% of TKR total data size, by requiring the 
average strip occupancy to be less than $5\times10^{-5}$, which corresponds to 44 strip hits per event.
The typical strip occupancy of $\sim2-3\times10^{-6}$ is dominated by accidental hits due to off-timing cosmic-ray tracks. The strip occupancy due to electronics noise is $10^{-7}$ or less.
We mask highest-occupancy channels until the TKR average occupancy is reduced to $<$ $5\times10^{-5}$. 

The number of masked channels for trigger and data purposes was 203 before launch and changed to 206 in July 2008, to 
220 in August 2008, to 284 in Octobter 2008 
and finally to 316 in January 2009. These 
additional 113 channels are distributed across 
seven SSDs, while 60 of these channels are concentrated in a region of one SSD. The total number of disabled channels 
corresponds to only 0.04\% of the total number of TKR channels.

\subsection{Trigger and data latching thresholds}
\label{sec:tkrdac}

In order to minimize the noise occupancy while maximizing the hit efficiency, the nominal threshold level 
is set to 1.4~fC ($\sim$0.28 MIP). The threshold DAC value for each GTFE is calibrated using charge injection. 
The charge injection DAC is set to the value corresponding to 1.4~fC and the threshold DAC is scanned (see Section\ref{sec:tkrmip}).
The best threshold for each channel is determined by a fit to the occupancy versus threshold using the error function (integral of a Gaussian).
The average threshold for each GTFE is obtained by calculating the mean value of the threshold DAC values after removing all dead and 
masked channels, and 5\% of the channels with the largest and the smallest values.

The calibration of the threshold DAC performed after launch yields identical DAC values to those before launch for 86\% of GTFEs. 
Only 0.04\% of the total number of GFTEs (6 out of 14 000) exhibits a difference of more than one DAC value. This was already known 
from pre-launch measurements and it corresponded to additional noise in the system.

The GTFE data is latched following a trigger acknowledge signal, which is 0.8~$\mu$s later than the typical peak of the \TKR\ pulse shape.
This value was determined using an external trigger during pre-launch tests\footnote{Although this implies that the TKR delay 
should be negative, the smallest allowed value is zero. This has no effect on the TKR performance.}. 
Due to the delayed data latch timing, the effective threshold for the data is different from the trigger threshold.
Once the threshold DAC value is determined for all GTFEs, we measure the effective thresholds at the time of the data capture 
by scanning charge injection calibration DAC values. The best threshold for 
each channel is determined by a fit to the occupancy versus threshold data using the error function.
The ratio of data latching threshold measurements before and after launch yields a 2\% shift in RMS and 0.5\% shift in the mean value, 
which implies no significant changes from values measured prior to launch. Therefore, trigger thresholds were assumed not to have 
changed after launch and were not recalibrated.

\begin{figure}[htbp]
\begin{center}
\includegraphics[width=14.0cm]{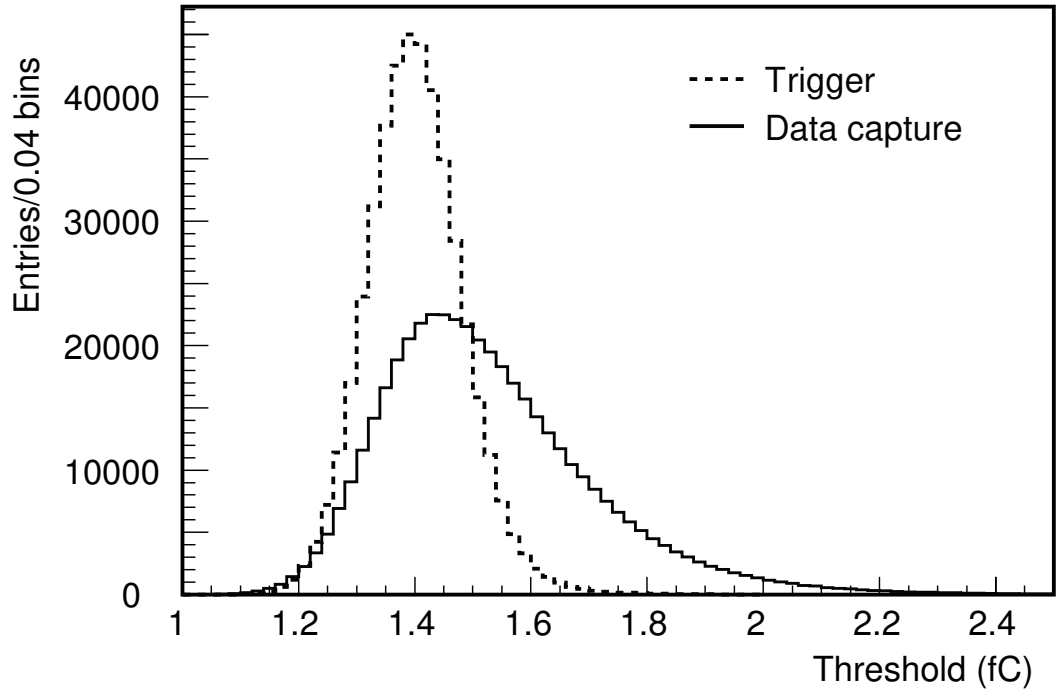}
\caption{Effective threshold for data capture (solid) and trigger threshold (dashed).}
\label{fig:tkrthresh}
\end{center}
\end{figure}

Figure~\ref{fig:tkrthresh} shows the trigger thresholds obtained prior to launch and the data latching thresholds measured on-orbit.
Because of the delay in latching the data, the threshold for the data capture is slightly higher and has a broader peak than that of 
the trigger threshold. The RMS dispersion is $\sim$5\% and $\sim$12\% for the trigger and data latching thresholds, respectively.

\subsection{Time-over-threshold conversion parameters}
\label{sec:tkrtot}
\begin{figure}[htbp]
\begin{center}
\includegraphics[width=11.0cm]{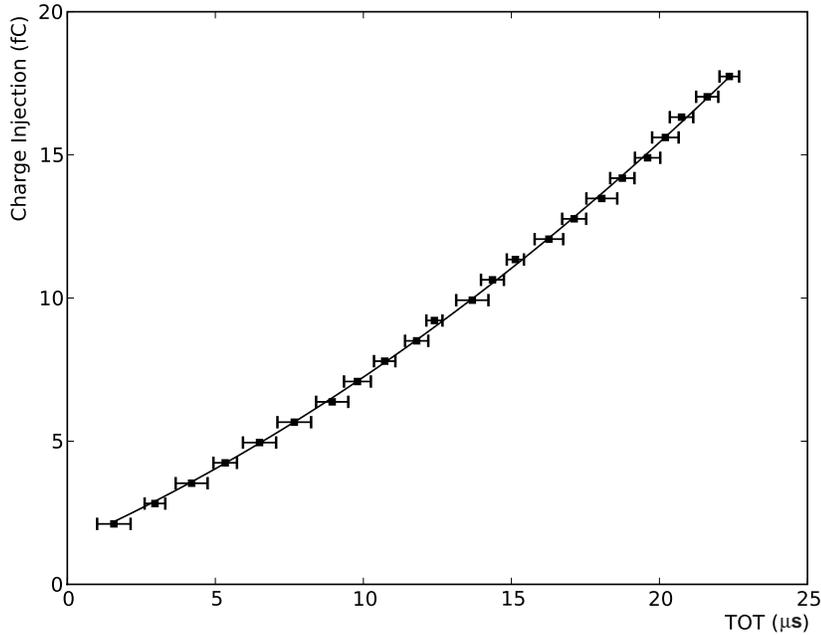}
\caption{Amplitude of charge injected versus ToT values. The curve corresponds to a second order polynomial fit to the data.}
\label{fig:totconvfit}
\end{center}
\end{figure}

To determine the conversion parameters from \TOT\ (ns) to charge deposit (fC), we measure for each channel, the \TOT\ value for several settings of the 
charge injection calibration DAC.
Since by definition the \TOT\ values cannot be negative, the \TOT\ measurements near threshold are biased 
toward positive values and result in slightly biased conversion parameters.
Figure~\ref{fig:totconvfit} shows the amplitude of charge injected versus ToT values, where the values for charge injection have 
not been corrected by the MIP scale calibration described in Section~\ref{sec:tkrmip}.
In the fit shown in Figure~\ref{fig:totconvfit}, the \TOT\ is described as a second order polynomial 
of injected charge, whose offset corresponds to the threshold value. To avoid biases near the threshold, the fit uses a fixed value for the intercept that
corresponds to the calibrated threshold.
The statistical error in the fit of order 8\% and 
is estimated by comparing two measurements of the same curve done prior to launch.
This calibration is important since the \TOT\ gain may vary by as much as a factor of three within a GTFE.
We have not performed this calibration after launch, 
since other \TKR\ results (see Section~\ref{sec:tkrdac}) indicated little change in the response of \TKR\ pulses 
and this calibration requires more than 10 hours of data-taking in dedicated-mode. 
We expect to recalibrate these values annually.

\subsection{MIP scale calibration}
\label{sec:tkrmip}

\begin{figure}
\begin{center}
\includegraphics[width=13cm]{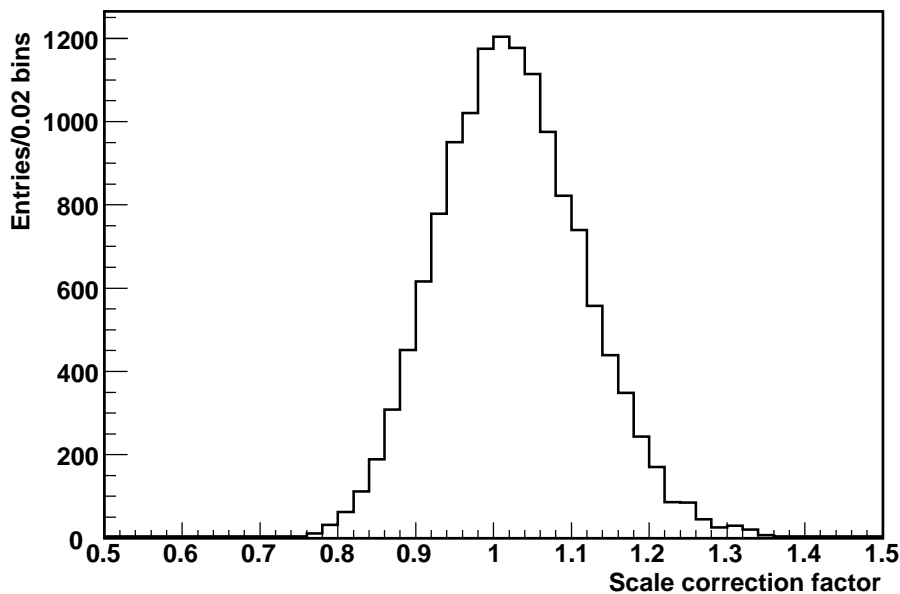}
\caption{Measured MIP peak divided by simulated MIP peak obtained prior to launch. This charge scale correction factor is obtained for all GTFEs. }
\label{fig:totratio}
\end{center}
\end{figure}

The absolute calibration of the charge injection DAC was performed prior to launch using the charge deposited by surface cosmic rays.
A correction factor was defined by the ratio of MIP peaks between data and Monte Carlo simulations. 
The distribution of this ratio shown in Figure~\ref{fig:totratio} exhibits an RMS dispersion of $\sim$9\%. 
The MIP deposited in each channel was calibrated using these correction factors.
On-orbit, we select single-track events, close to normal incidence ($\cos\theta<-0.85$) and with CAL energy 
consistent with that of a MIP. We require  
the RMS values of TKR hit positions with respect to the reconstructed track position to be consistent with the resolution needed 
to reject low-energy tracks. 
To avoid confusion with charge sharing between adjacent strips, we only consider layers with single hit strips.
The charge deposited by a particle is corrected by taking into account its path length in the silicon. 
The measured \TOT\ of the hit associated with the track is converted to charge using the conversion parameters.
Following that, we fit the data for each GTFE with a Landau distribution convolved with a Gaussian. 

\begin{figure}
\begin{center}
\includegraphics[width=13cm]{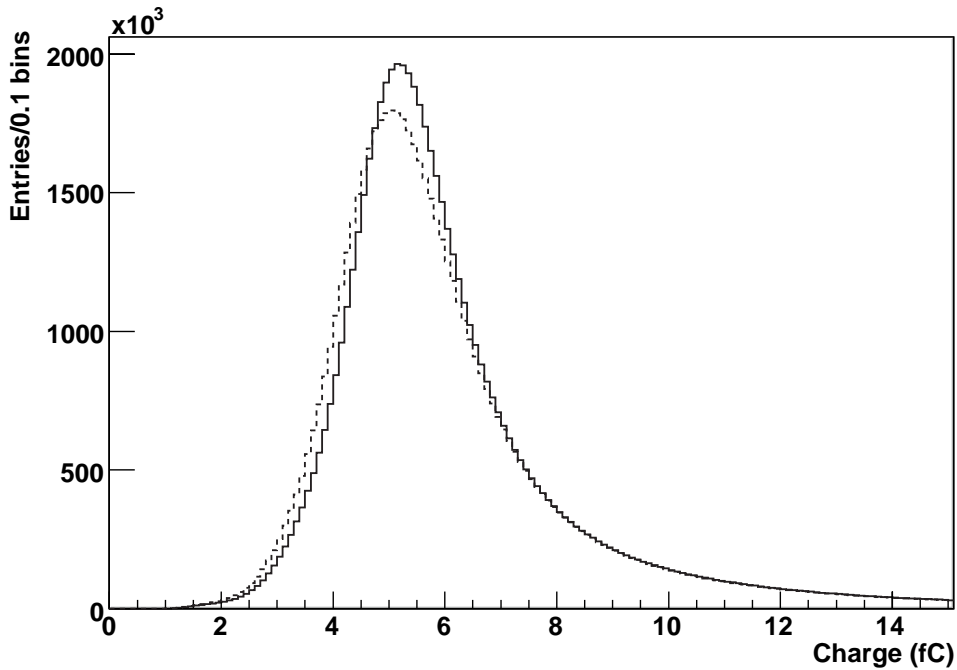}
\caption{MIP signal (pathlength corrected) before (dotted) and after (solid) calibrations.}
\label{fig:tot}
\end{center}
\end{figure}

More than 10 million MIP tracks are required to accumulate 
sufficient entries for all GTFEs, which takes 5 days during 
nominal science operations. Figure~\ref{fig:tot} 
shows the MIP charge deposit distribution for all channels 
before (dotted histogram) and after (solid histogram) each GTFE is calibrated. The dispersion 
correction factor of 9\% is included in the calibrated results.

\section{Determination of SAA polygon}
\label{sec:saa}
The orbit of {\em Fermi} intersects the Earth's inner radiation belt
in a region which is known as the South Atlantic Anomaly (SAA). This region features geomagnetically 
trapped protons with energies up to hundreds of MeV and electrons with energies up to tens of MeV. The flux of 
protons and electrons in the LAT energy range
reach levels which are several orders of magnitudes above those of primary cosmic rays.
This extreme particle flux imposes constraints on LAT operations. The TKR electronics 
saturate due to the increase in the charge deposited per live time, leading to large dead time fractions, thereby hampering 
scientific observations. The continuous influx of particles generates high currents in the ACD photomultiplier 
tubes (PMT), thus exceeding safe operating limits, which leads to slow deterioration.
Therefore, during SAA passages, triggering, recording and transmission of science data are stopped and the bias voltages of the PMTs are lowered from~900V to 
$\sim$400V. Only LAT housekeeping data are recorded and transmitted to the ground.  

The position along the orbit defined by the GPS receiver aboard the {\em Fermi} spacecraft determines the transition between nominal science operations and the SAA transit mode.
The latitude and longitude 
of the {\em Fermi} position are compared to the bounds of a polygon defined by 12
latitude-longitude vertices stored in the spacecraft memory. As the spacecraft position crosses this polygonal 
boundary it triggers the SAA transit mode. To avoid multiple entries and exits during a single orbit, a convex polygon 
is used to define the SAA region. 

\begin{figure}
  \center
  \includegraphics[width=15cm]{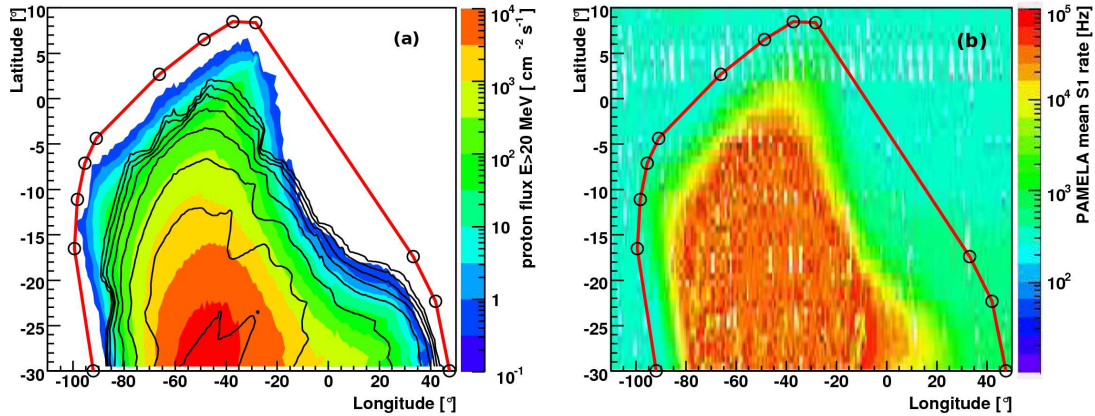}
   \caption{Trapped radiation models versus geographic latitude and longitude: a) trapped proton flux above 20 MeV predicted by the AP-8~\cite{ap8} (black contour lines) and PSB97~\cite{psb97} (color contours), b) top scintillator count rates reported by the PAMELA experiment~\cite{pamela_icrc}. In both of the plots, the red polygon shows the locations of the SAA boundary edges determined before launch and used during the initial phase of the LAT operations.}
   \label{fig:saa1}
\end{figure}

We chose a conservative definition for this initial SAA boundary, with the expectation that we would update 
the boundary based on particle rate measurements made with the LAT once it was on orbit.
The first version of the polygon, or SAA boundary, was determined before launch based on models of the Earth 
radiation belts and data from other spacecraft. The inner radiation belt was modeled using trapped radiation models: AP-8~\cite{ap8} and PSB97~\cite{psb97} in 
conjunction with the current version of the International Geomagnetic Reference Field (IGRF-10)~\cite{igrf}. 
The 12-edge SAA boundary polygon was calculated from these models based on the contour in latitude and longitude, where the $E>20$ MeV 
trapped proton flux reached 1 cm$^{-2}$s$^{-1}$. 
For regions where two models predictions disagreed, we chose the larger flux. 
The smallest convex polygon circumscribing this contour was selected and padded by a 4$^\circ$ margin. Figure~\ref{fig:saa1}a shows the trapped proton flux profiles above 20 MeV predicted by trapped radiation models versus geographic latitude and longitude. Figure.~\ref{fig:saa1}b shows top scintillator count rates in the altitude range between 532 km and 575 km, reported by the PAMELA experiment~\cite{pamela_icrc}, which is similar to that of the {\em Fermi} orbit. The red polygon shows the locations of the SAA boundary edges determined before launch and used during the initial phase of the {\em Fermi} LAT mission.
This definition of the SAA boundary resulted in a loss of observation time of about 17\%.

\begin{figure}
  \center
   \includegraphics[width=15cm]{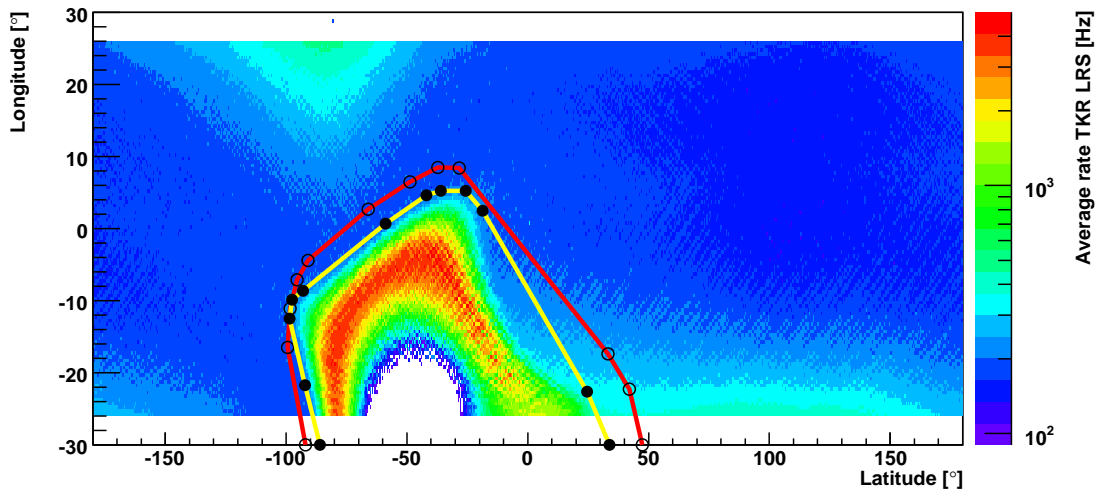}
   \caption{Average rate of TKR counters obtained during 26.6 days of LAT nominal science operations versus geographic latitude and longitude. Superimposed are the pre-launch SAA boundary (red) used during the initial phase of the mission, and the updated SAA boundary (yellow) derived from measurements of the TKR counter data (see text)}
   \label{fig:saa2}
\end{figure}

After launch, diagnostic data of the LAT were used to refine 
the size of the polygon. 
Even though science triggers are disabled during SAA passages, fast trigger signals remain operational.
Special TKR and ACD counters can sample the rate of fast trigger signals to determine position-dependent rates 
of the LAT along the orbit. Figure \ref{fig:saa2} shows the rates recorded in the TKR counters versus spacecraft 
position. A rate increase is visible at the edges of the SAA before the TKR electronics saturates and suppress fast trigger signals, thus bringing the 
count rates to zero. 

To define the SAA boundary using these data one has to account for particle rates and rate variations from primary and secondary
 cosmic rays, where both depend on the local geomagnetic cutoff rigidity at the {\em Fermi} location. Therefore, we used data from the region outside the 
pre-launch SAA boundary to determine these rates and set an upper limit on the expected number of cosmic-ray counts per second as a 
function of the local geomagnetic rigidity cutoff.
The optimized SAA boundary polygon is calculated including the points that exceeded this limit. A padding of 1$^\circ$ is applied to account
for the limited resolution and sensitivity of the measurement.

Figure \ref{fig:saa2} shows the average rate of TKR counters obtained during 26.6 days of LAT nominal science operations versus geographic latitude and longitude. Superimposed are the pre-launch SAA boundary (red) used during the initial phase of the mission, and the refined polygon (yellow) uploaded to spacecraft memory. The updated polygon reduced the loss in observation time to approximately 13\% of the total on-orbit time. This polygon has been the default for the LAT operations since July 28, 2008. A cross-check during nominal science operations is performed with 
the ACD trigger signal counters. These are more sensitive to the low-energy component (E $\approx$ 10-60 MeV) of the trapped particle flux than those from the TKR. 
There is no significant increase in the rate of ACD fast trigger signals as {\em Fermi} approaches the SAA boundary, thus validating the optimized polygon.  

Since the SAA moves at a rate of a few tenths of a degree per year and its size and particle fluxes vary with the solar cycle, we expect annual updates to 
the SAA boundary.

\section{Live time}
\label{sec:livetime}
The live time is accumulated taking into account
the variety of dead time effects. Science data taking is disabled during SAA
passages (see Section~\ref{sec:saa}). Instrumental dead time, during event
latching and readout, is about 8\% on average outside the SAA, although this
fraction depends on the trigger rate (primarily background). The dependence on the geomagnetic latitude of {\em Fermi} is secondary.  
Other losses are caused by failures in
transmission and ground processing and dedicated-mode calibrations.

Accurate accounting for live time is essential for obtaining calibrated fluxes and spectra for 
astrophysical sources of gamma rays.  The live time relates the effective collecting area of the LAT to the overall exposure.  Owing to the very large field of view of the LAT ($>$2 sr) and the relatively slow scanning rate ($\sim$4 deg min$^{-1}$) the accumulated live time typically is needed only coarsely ($\sim$30 s intervals) for accurate exposures to be calculated.  For very bright transient sources, finer accounting for live time is used, owing to the high and variable rates of triggers.  For example, a bright GRB in the field of view of the LAT, such as GRB08019C~\cite{grb0801916C}, can induce a dead time fraction of about 16\% during the impulsive phase of the burst.  

The dead time for event latching and readout is tracked on-board the LAT every 50 ns using 
the 20 MHz LAT system clock. 
The stability of the 20 MHz clock is closely
tracked with a 1 pulse-per-second (PPS) signal from the GPS system of the spacecraft.

During nominal science operations, the instrumental dead time is dominated by the 
fixed time to read out the event timing and trigger information, 
which imposes a minimum dead time of 26.5 $\mu$s per event. 
Every time the LAT triggers, further data taking is disabled until the data from the event are read out. 
After the end of the time coincidence window the latency of the trigger system is 100 ns before the new time coincidence window is available, even if the previous event was not read out.
Single front-end electronic channels can be dead for several microseconds while the pulse is above threshold. 
In special cases, such as when the periodic trigger is enabled (2 Hz), the entire CAL is read out with no zero suppression and in this case, the dead time can be as large as 620 $\mu$s.
 During nominal science operations one of the trigger combinations is configured to read the CAL with all four energy ranges and zero suppression for which the dead time is about 65 $\mu$s.

Losses of data in transmission can effectively remove events from the data
stream both directly and indirectly.  The indirect effects relate to how the
events are assembled in the telemetry stream.  In each packet, the event
times are encoded as times relative to the start of the packet.  In addition,
the 20 MHz counters mentioned above roll over every 1.6 s and the times of
the roll overs are also encoded in the packets.  Loss of certain parts of a
packet can cause times and live time accumulations to be lost for up to
$\sim$200 events.  At the 450 Hz nominal event rate in telemetry, these losses
can be as large as 0.4 s.  The mission specification for data loss is
less than 2\%, and in practice loss due to transmission errors has been much
less than this, in part because on-board the LAT data are retained for
approximately 24 hours and retransmissions can be requested. At the 
ground station, {\em Fermi} telemetry is buffered for 1 week and losses
in ground transmission generally can be recovered.  The overall average loss
has been much less than 1\% to date.

Losses in ground processing of event data are also rare and are significantly less than 1\%.

\section{Overall timing accuracy}
\label{sec:absolute}

Recording accurate arrival times of LAT photons is essential for studies of gamma-ray bursts and pulsar timing. Absolute timing tests were performed during pre- and post-launch activities. A discussion of LAT pulsar timing can be found elsewhere~\cite{pulsartiming}.

\begin{figure}[h]
\begin{center}
\includegraphics[width=9cm]{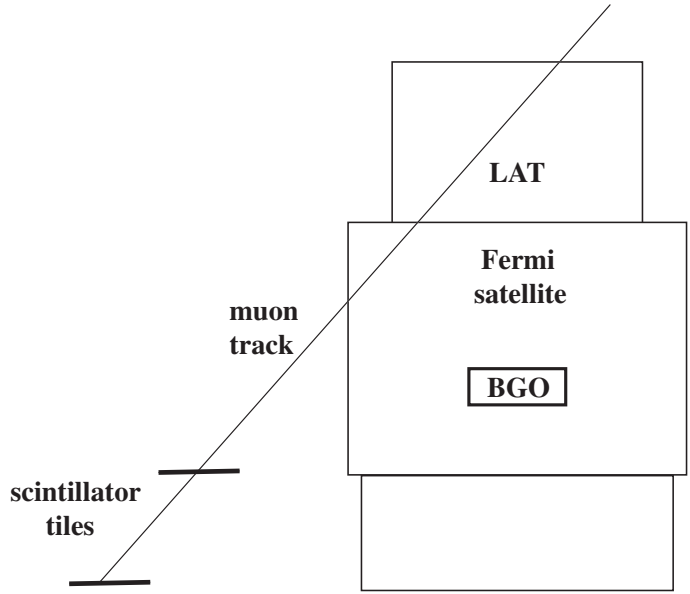}
\caption{Diagram of the muon scintillator telescope placed next to the {\em Fermi} satellite
during pre-launch tests. 
\label{fig:MuTimeTestSketch}}
\end{center}
\end{figure}

\begin{figure}
\begin{center}
\includegraphics[width=15cm]{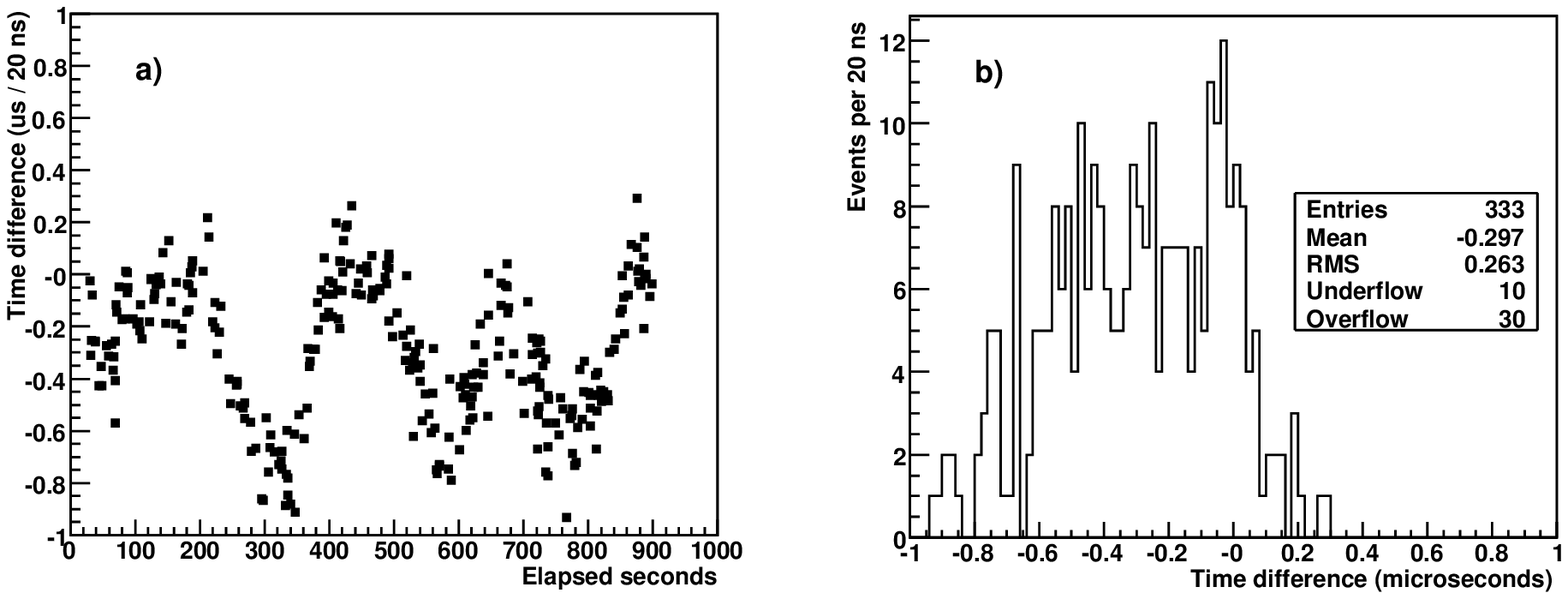}
\caption{Results from pre-launch tests: a) differences in the times recorded with the two acquisition systems, versus elapsed time; b) histogram of time differences indicating mean and RMS values around 0.3 $\mu$s.
\label{fig:MuTimeTestResults}}
\end{center}
\end{figure}

During pre-launch tests we recorded cosmic rays to measure the time difference between two GPS systems.
As shown in Figure~\ref{fig:MuTimeTestSketch} a pair of scintillator tiles provided a reference for the LAT timestamps. 
The coincidence signal from these tiles triggered 
a VME-based GPS time system previously used by the ground gamma-ray telescope CELESTE. Its absolute time accuracy was previously demonstrated by 
measuring the Crab optical pulsar~\cite{celeste}.
Reconstructed muon tracks traversing the LAT detector were extrapolated to their impact point on the 
laboratory floor and their timestamps were measured with respect to the
GPS of the {\em Fermi} satellite. If a muon 
passed through the pair of scintillators placed next to {\em Fermi}, a GPS timestamp
from a standalone VME data acquisition system was also recorded. Figure~\ref{fig:MuTimeTestResults} shows that the LAT timestamps agreed with the reference GPS to within 0.3 $\mu$s.

GPS receivers use the arrival times of reference signals from 
other GPS satellites to calculate their time and position and
transmit that 
information to processors on {\em Fermi}.
This is accompanied by
an electronic ``Pulse Per Second'' (PPS) at the moment of 
validity of the timestamp word. The processors, using precision
oscillators on-board the spacecraft, maintain the PPS accuracy in the case of occasional
short losses of GPS signal reception.

Bright gamma-ray pulsars were used to verify
that the integer seconds of absolute time from the GPS receiver conform to Coordinated Universal Time (UTC), since
an integer offset in the {\em Fermi} clocks would make a large shift in observed gamma-ray
phase, different for each pulsar. The rotational phase of the gamma-rays peaks 
of the Vela pulsar relative to the radio peak agree with that measured by previous 
experiments~\cite{vela}. The first gamma-ray peak of the Crab pulsar is aligned
with the radio reference, since Crab radio and gamma-ray beams appear to come
from the same part of the neutron star magnetosphere~\cite{celeste}.
 
The fractional part of LAT event timestamps is obtained from the counts of a 20 MHz
oscillator recorded by scalers latched at the reception of
a GPS PPS signal and at the reception of an event trigger.
The behavior of the
oscillator was extensively characterized during the ground tests, and its frequency 
is recalibrated each second using the
PPS-latched scaler values. On-orbit telemetry monitoring shows that the internal spacecraft timing signals 
behave as before launch, from which we conclude that LAT timestamps are still
well within 1 $\mu$s of the GPS times used by the spacecraft.  GPS times are
maintained within 20 ns (1 sigma) of UTC~\cite{gpsStandard}.

At present the best observational validation
of the on-orbit clock performance comes from the pulsars PSR J0030+0451 and PSR J1939+2134.  The
peak width of $<1$20 $\mu$s reported for PSR J0030+0451 in~\cite{0451} demonstrates the stability of the LAT 
event times over six months of data-taking, but not their absolute accuracy. However, the 1.56 ms pulsar 
PSR J1939+2134 appears to have a gamma peak aligned with the radio peak to better than 1/20 of a rotation 
of the neutron star, that is, 80 $\mu$s~\cite{2134}.  This translates to an absolute time 
accuracy if one assumes that both the gamma and radio emission come from the same region in the pulsar's magnetosphere.  

\section{Internal and spacecraft boresight alignments}
\label{sec:ta}
The accuracy of the directions of reconstructed events dependes on the knowledge of the exact position of each element of the TKR, i.e individual SSD or planes.
Misalignments of any element can broaden the instrument response function,
thus lowering the LAT sensitivity.
We perform the following alignment procedures:
\begin{enumerate}
  \item intra-tower alignment to determine the position and orientation of {\em each element} of a single tower with respect to an ideal coordinate system;
  \item inter-tower alignment to determine the position and orientation of {\em each tower} with respect to an ideal coordinate system;
  \item spacecraft alignment to determine the rotation of the LAT with respect to the {\em Fermi} on-board guidance, navigation and control system.
\end{enumerate}

\subsection{Intra-tower alignment}
\label{sec:intra}

As described in Section~\ref{sec:tkr}, each TKR tower consists of 36 silicon planes  
each instrumented with sixteen silicon microstrip detectors.
All strips in a plane are oriented along the same direction. 
The intra-tower procedure aligns the planes 
horizontally (along the measured coordinate) and vertically (perpendicular to the silicon plane),
and determines all rotations of the planes within the TKR tower.

Due to the procedures for construction and assembly of towers, 
we expect only minor deviations from the original positions. This simplifies the formalism 
used for the alignment, and we assume first-order approximations to deviations.

\begin{figure}
  \center
   \includegraphics[height=6cm]{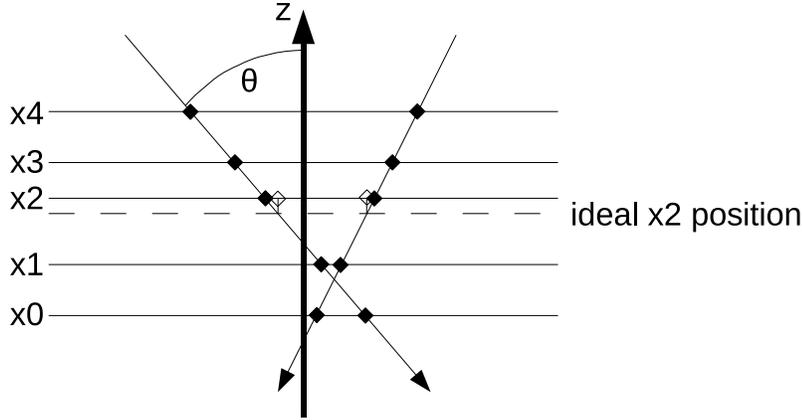}
   \caption{The diagram shows five silicon layers (x0 to x4) and the ideal x2 layer, crossed by two tracks.
     Measured hits are denoted as filled diamonds,
     extrapolated hits (assuming the ideal position of the x2 plane) as open diamonds.
     The residual is composed of two contributions: $\Delta x$ due to horizontal misalignment,
     and $\Delta z\cdot\tan(\theta)$ due to vertical misalignment.}
   \label{fig:align}
\end{figure}

\begin{figure}
  \center
  \includegraphics[height=9cm,angle=0]{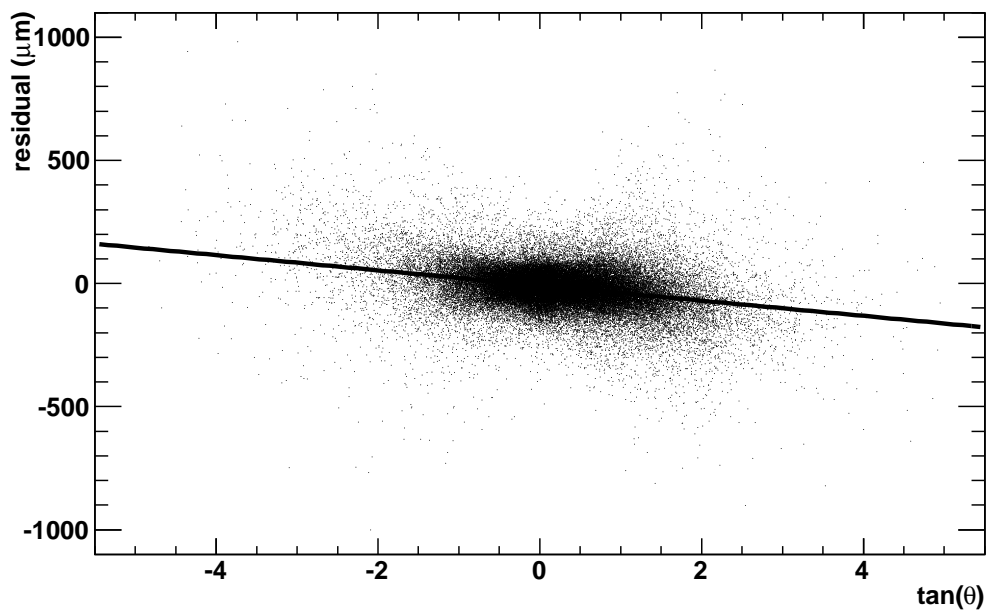}
  \caption{Residual versus $\tan(\theta)$ of the track, for a silicon plane arbitrarily chosen.
    A fit with a straight line provides an offset and a slope,
    which correspond to deviations of the real position to the assumed one.}
  \label{fig:hz}
\end{figure}

The trajectory of a particle traversing the TKR is characterized by a reconstructed track, 
which consists of a list of associated silicon strip hits joined by a straight line.
As the particle encounters material in its path its direction is affected by multiple scattering.
The effect is more pronounced in the presence of dense materials such as the tungsten converter foils.
Thus, a fit to an otherwise straight track results in deviations of the real hit positions
from an ideal straight track referred to as residuals.
The diagram in Figure~\ref{fig:align} shows two 
tracks in the $xz$ plane and illustrates how residuals relate to measured and ideal positions.
A distribution of residuals should be centered at 0, and 
any deviation from zero is an indication that the element studied is not at its assumed position.
Figure~\ref{fig:hz} illustrates how straight tracks are used to determine the horizontal and vertical misalignments.
It shows the hit residuals versus $\tan(\theta)$ (i.e. inverse of the slope) of the track, for one silicon plane (arbitrarily chosen). 
A straight line fit through these points 
produces an offset and a slope, which correspond to deviations of the real position from the assumed one. 
Results from the fit determine horizontal ($\Delta x$) and vertical ($\Delta z$) deviations, corresponding to shifts along and perpendicular 
to the strip coordinates, respectively.

Intra- and inter-tower TKR alignments are iterative processes that use reconstructed events recorded 
during nominal science operations with no requirement on any on-board filter. 
As a result, each TKR element is subject to a wide range of event types, 
thus minimizing the possibility of selection bias.

Here we briefly describe a method common to both intra- and inter-tower alignments.
The event reconstruction associates hits to tracks and classifies them according to their track length and straightness.
The internal TKR alignment procedures benefits from track reconstruction by using only 
the first (best) track in the event, i.e. all other tracks are discarded.
The list of the hits is then input to the alignment algorithm. 

The alignment algorithms rely on two important pieces: the raw position information, i.e. the plane and strip containing a hit,
and the {\em smoothed} position, which results from track finding and optimization through the Kalman filter algorithms~\cite{latpaper}.
Since the raw position cannot be used to determine the location of the hit along the strip, the algorithm uses the 
corresponding {\em smoothed} position, instead.
Finally the algorithm fits the hit positions with a straight line, computes the $\chi^2$ of the fit, and evaluates 
the residuals, and the angle $\theta$ of the track with respect to the tower axis.

Correlations similar to those depicted in Figure~\ref{fig:hz} are used to determine 
all alignment parameters listed in Table~\ref{table:corr}.
Note that the angle $\theta$ is always measured in the coordinate of the silicon plane (e.g. $\theta_x$ for $x$-planes).
The sign of the correlation for rot$z$ is different for the $x$ and the $y$-planes.
There are six parameters to determine, but only four correlations.
Each correlation gives two parameters: the horizontal position, either $\Delta x$ (for $x$ planes) or $\Delta y$ (for $y$ planes),
and $\Delta z$ or any of the three rotations.
\begin{table}[hbt]
  \centering
  \begin{tabular}{|c|c|c|c|}
    \hline
    Correlation of track parameters & Alignment Parameters \\
    \hline
residual versus $\tan(\theta)$                 & $\Delta h$ + $\Delta z$ \\ \hline
residual versus $x\cdot\tan(\theta)$           & $\Delta h$ + rot$x$ \\ \hline
residual versus $-y\cdot\tan(\theta)$          & $\Delta h$ + rot$y$ \\ \hline
residual versus $\mp$ position along the strip & $\Delta h$ + rot$z$ \\ \hline
    \hline
  \end{tabular}
  \caption{Correlations of track parameters, and resulting alignment parameters. $\Delta h$ is the measured horizontal coordinate of a plane,      i.e. $\Delta x$ for $x$-planes and $\Delta y$ for $y$-planes. See text for more details.}
  \label{table:corr}
\end{table}
For each TKR element the iterative algorithm determines five parameters that are measured relative 
to other elements whose positions are also unknown. The knowledge of the precise positions and rotations increases with successive iterations,
and each parameter converges to its correct value.
The decision for convergence comes from comparing the largest deviation of any positional (rotational) parameter between two successive iterations 
and by requiring the maximum deviations to be 0.01\,$\mu$m (1\,$\mu$rad). Convergence typically occurs after 100 iterations. 
Currently, the smallest elements considered in intra-tower alignment are single silicon planes (top and bottom of trays as defined in Section~\ref{sec:tkr}).
In the future, we will extend the method to align SSD's.

Table~\ref{table:intrafinal} shows the results 
from the on-orbit intra-tower alignment of 576 TKR planes. 
Offsets and slopes (as in Figure~\ref{fig:hz}) are entered in a histogram and the standard deviation ($\sigma$) of each distribution 
is shown in Table~\ref{table:intrafinal}. 
Since planes are aligned with respect to an ideal frame, by construction, the sum of all shifts (for all parameters) 
average to the mean value of 0. Results clearly demonstrate the 
quality of the assembly of individual towers since standard deviations are within $\pm$ 61 $\mu$m and $\pm$ 220 $\mu$rad.
\begin{table}[hbt]
  \centering
  \begin{tabular}{|c|c|}
    \hline
    Parameter &  Standard deviation ($\sigma$)  \\
    \hline
    $\Delta$x &                  $\pm$ 43 $\mu$m \\
    $\Delta$y &             $\pm$ 59 $\mu$m \\
    $\Delta$z &                  $\pm$ 61 $\mu$m \\
    rot$_x$   &                $\pm$ 220 $\mu$rad \\
    rot$_y$   &              $\pm$ 220 $\mu$rad \\
    rot$_z$   &               $\pm$ 210 $\mu$rad \\
    \hline
  \end{tabular}
  \caption{On-orbit intra-tower alignment constants, for 576 planes (averaged for all towers).}
  \label{table:intrafinal}
\end{table}

\subsection{Inter-tower alignment}
\label{sec:inter}
The inter-tower alignment procedure aligns towers spatially and determines their rotations 
with respect to the LAT reference frame, and 
is performed only after the intra-tower alignment constants have been determined.
While the latter uses the residuals of each hit to determine the alignment constants,
the former relies on track segments of events that cross tower boundaries.
We determine the orientation of each tower by evaluating the angles between pairs of track segments.
The rotation in one axis is given by the scalar product of the track segments,
projected on the plane spanned by the two other axes,
and averaged over all selected events:
\begin{equation}
  \label{TA:scalar}
  cos(\mbox{rot}_i) = \left< \sum \frac{\vec{t} \,\vec{t}'}{|t| |t'|} \right>,
\end{equation}
where $\vec{t}$ and $\vec{t}'$ are the track segments in vectorial form and $i$ corresponds to $x$, $y$, or $z$.
For convenience, Eq.(\ref{TA:scalar}) can be rewritten in terms of track slopes $cot(\theta)$,
yielding
\begin{equation}
  \label{TA:rot}
  \mbox{rot}_i = \left< \sum \frac{\Delta\cot(\theta_i)}{1 + \cot(\theta_i)^2} \right>.
\end{equation}

\begin{figure}
  \center
 \includegraphics[height=8cm,angle=0]{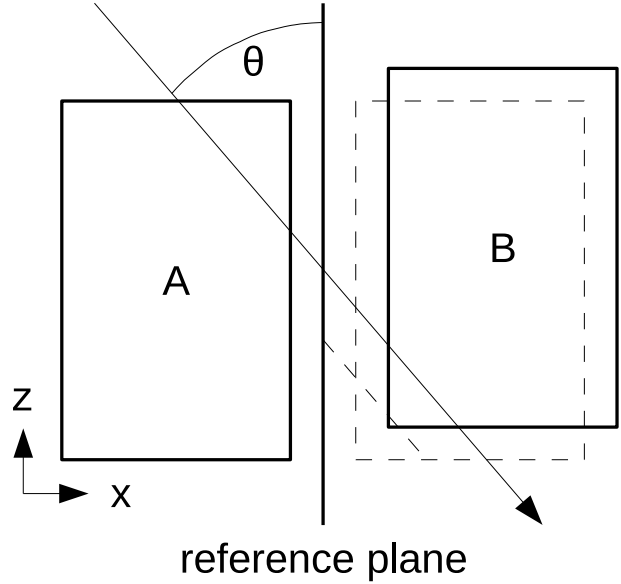}
   \caption{
     Horizontal and vertical alignment for a pair of towers.
     the real positions are drawn with solid lines,
     the assumed position of tower B with dashed ones.
     The black line between the towers marks the reference plane.
     The real particle track is shown with solid line,
     the extrapolated track based on the assumed position dashed.
   }
   \label{fig:inter}
\end{figure}

After fixing rotations, each pair of track segments is projected onto a reference plane between both towers,
as depicted in Figure~\ref{fig:inter}.
Each intersection defines a position on this plane. 
For perfect alignment, these positions are identical.
The differences between both positions are analyzed as in Section~\ref{sec:intra}, 
i.e. by searching for correlations between the position differences and $\cot(\theta)$ of the track.
In the example of Figure~\ref{fig:inter},
the residual has a $\Delta$z and a $\Delta$x$\cdot\cot(\theta)$
contribution due to vertical and horizontal misalignments. 

\begin{figure}
  \center
   \includegraphics[height=7cm,angle=0]{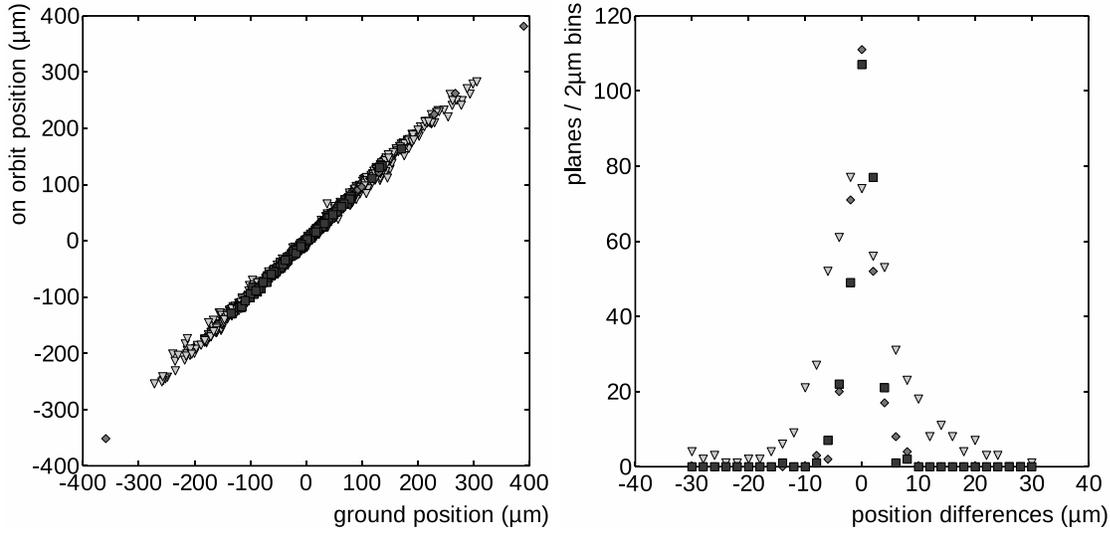}
   \caption{On-orbit data: a) correlation between ground and on-orbit inter-tower positional 
alignment constants, b) the histogram of the differences ($\Delta\sigma$), where  $x$, $y$ and  $z$ correspond to squares, 
diamonds and triangles, respectively.}
   \label{fig:data}
\end{figure}

Table~\ref{table:inter} summarizes the results 
from the on-orbit inter-tower alignment of 16 TKR towers. 
Contrary to the intra-tower alignment, only a small number of values (16 towers instead of 576 planes) are available  
to determine the $\sigma$ of the 
distribution for each of the alignment parameters. Because of that, 
Table~\ref{table:inter} has an additional column that shows the largest and smallest value for each parameter.
Results clearly demonstrate the 
quality of the assembly of towers in the LAT since standard deviations are within $\pm$ 120 $\mu$m and $\pm$ 260 $\mu$rad.
\begin{table}[hbt]
  \centering
  \begin{tabular}{|c|c|c|}
    \hline
    Parameter &  Values (mix,max) & Standard deviation ($\sigma$)  \\
    \hline
    $\Delta$x &  (-250,+190) $\mu$m & $\pm$ 119 $\mu$m  \\
    $\Delta$y &   (-90,+130) $\mu$m & $\pm$ 68 $\mu$m\\
    $\Delta$z &  (-150,+150) $\mu$m & $\pm$ 87 $\mu$m\\
    rot$x$    &  (-450,+400) $\mu$rad &$\pm$ 260   $\mu$rad\\
    rot$y$    &  (-360,+480) $\mu$rad & $\pm$ 250  $\mu$rad\\
    rot$z$    & (-360,+360) $\mu$rad & $\pm$ 230  $\mu$rad\\
    \hline
  \end{tabular}
  \caption{On-orbit inter-tower alignment constants for 16 towers.}
  \label{table:inter}
\end{table}
Alignment procedures were validated using a simulated dataset of cosmic protons,
generated with randomly misaligned geometry for the LAT. 
Figure~\ref{fig:data} shows the results when the alignment procedure is applied to data acquired before and after launch.
Here the standard deviations ($\sigma_{\Delta}$) are calculated from a histogram of differences between ground and on-orbit measurements.
Figure~\ref{fig:data}a shows the correlation for intra-tower positional alignment constants,
while Figure~\ref{fig:data}b displays the histogram of these residual differences.
Table~\ref{table:data} compares unbiased datasets obtained on ground ($\sim$8 million surface muons)
with orbit data ($\sim$20 million events, mostly cosmic protons) and
\begin{table}[hbt]
  \centering
  \begin{tabular}{|c|c|c|}
    \hline
    Parameter &  $\sigma_{\Delta}$ (intra-tower) & $\sigma_{\Delta}$ (inter-tower)  \\
    \hline
    $\Delta$x & 2.4 $\mu$m &  30 $\mu$m \\ \hline
    $\Delta$y & 2.5 $\mu$m & 32 $\mu$m \\ \hline
    $\Delta$z & 8.8 $\mu$m &  17 $\mu$m \\    \hline
    rot$x$    & 51 $\mu$rad & 53 $\mu$rad\\\hline
    rot$y$    & 53 $\mu$rad & 42 $\mu$rad \\\hline
    rot$z$    & 36 $\mu$rad & 24 $\mu$rad\\ \hline
  \end{tabular}
  \caption{Standard deviations ($\sigma_{\Delta}$) of the differences between ground and on-orbit alignments.
    Inter-tower values are averaged over all three positions and rotations.}
  \label{table:data}
\end{table}
lists the average differences for both intra- and inter-alignment, for all parameters.
Results from Figure~\ref{fig:data} and Table~\ref{table:data} demonstrate that misalignments 
due to launch or temperature variations on-orbit are small and limited to within $\pm$ 35 $\mu$m and  $\pm$ 55 $\mu$rad. 

\begin{figure}[h]
  \center
   \includegraphics[height=14cm,angle=0]{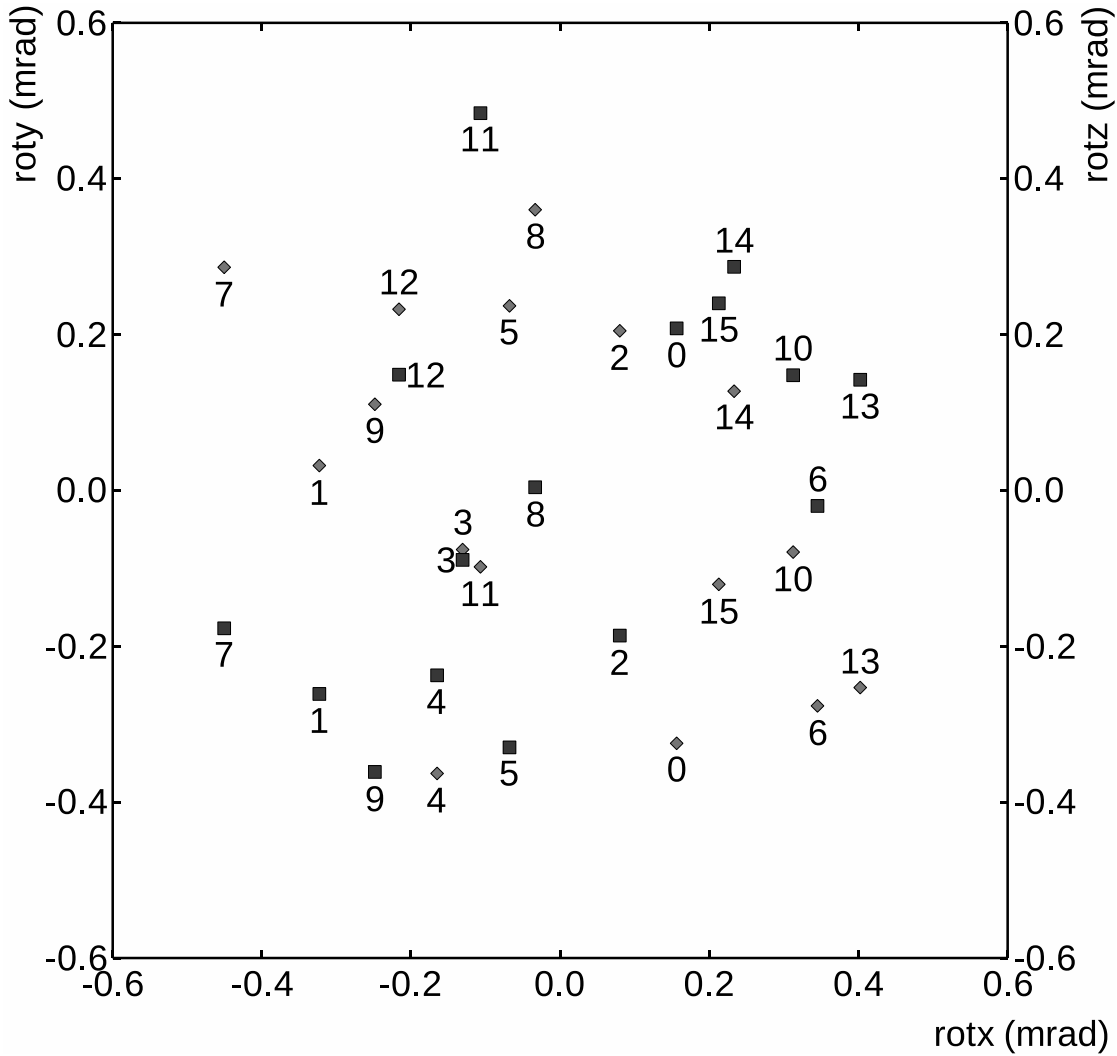}
   \caption{Correlation between the different rotation angles of each tower (the number of each tower is denoted in the diagram).
     Squares denote the correlation between rot$x$ ($x$-axis) and roty (left $y$-axis label),
     and diamonds between rot$x$ and rot$z$ (right $y$-axis label).}
   \label{fig:pointing}
\end{figure}

Finally, Figure~\ref{fig:pointing}
shows all correlations between the rotation angles obtained for each tower using on-orbit data acquired by the end of October 2008. 
Tower numbers are labeled from 0 to 15.
Squares denote the correlation between rot$x$ ($x$-axis) and rot$y$ (left $y$-axis label),
and diamonds between rot$x$ and rot$z$ (right $y$-axis label).
The spread in the correlation between rot$x$ and rot$y$ translates directly into a pointing inaccuracy on the celestial sphere, e.g. 
a misalignment of 0.3 mrad in rot$x$ or rot$y$ for one tower would cause a source position error of the same size.
The absence of clustering in this distribution indicates that there are no obvious systematic effects.

The alignment of the TKR has been 
checked twice before launch (since October 2006) and four times after launch (since June 2008).
No significant changes of the positions and rotations were observed.
We expect updates to be made annually.

\subsection{LAT alignment with respect to the spacecraft}
\label{sec:sc}

The LAT reconstructs the direction of each photon in the LAT reference system. 
To convert this to celestial coordinates, we require the celestial orientation of the LAT. 
The spacecraft orientation is provided by the Guidance, Navigation and Control system, or star-tracker. 
This system involves optical telescopes on an optical bench, and a star field pattern recognition system, which 
was calibrated before launch. 
The LAT system is nominally the same as the spacecraft system, but small deviations are 
expected due to uncertainty in the ground alignment process, thermal variations, 
launch vibrations, relaxation in 0g, or the inter-tower alignment process. 
Measuring these deviations, by comparing the LAT view of the sky with that seen by the star-tracker is 
called ``boresight alignment".

The boresight alignment is determined via likelihood maximization for gamma rays near bright, identified 
celestial point sources of gamma-ray emission. The likelihood analysis is ``binned'' in the sense that the gamma 
rays are not considered individually but binned into maps in celestial coordinates. Separate maps are used 
for different energy ranges because the point-spread function (PSF) depends strongly on energy.
The likelihood is defined in terms 
of the Point Spread Function (PSF), the background (sum of celestial sources and residual charged particle bacgrounds, 
estimated independently for each source), and the boresight alignment angles. 
We measure the boresight alignment by maximizing the total binned likelihood of the point source model (background is included) 
with respect to three angles characterizing an arbitrary rotation. We chose the convention of performing consecutive rotations about the $x$, $y$, and $z$ axes. 
These are defined by the spacecraft: $z$ is along the symmetry axis (pointing direction), and $y$ is along the solar panels. 
Since the angles are small, less than a degree, the order of the rotations is irrelevant. 
Rotations about the $x$ and $y$ axes thus correspond to angular deviations of the same size, 
while the rotation about the $z$ axis affects only tracks that traverse the detector at off-axis angles. 
The procedure for calculating the likelihood transforms both the source position (i.e. celestial coordinates) and the photon position 
into spacecraft coordinates, rotates the photon's position, and determines the 
likelihood that the photon is consistent with the source location. Details of the likelihood method are discussed elsewhere~\cite{like}.

We first define a set of reference bright point sources using the likelihood itself to select the most significant. 
We select photons within the 99\% PSF containment of the sources. 
Since the angular containment is energy dependent, photons are divided into energy bands, using an 
average PSF for each band to define the expected deviation distribution. 
The PSF is non-Gaussian, and has been found to be well modeled by the following power-law expression~\cite{perfpaper},

\begin{equation}
PSF(\delta,E_i)=\left( 1- \frac{1}{\gamma(E_i)} \right)\left( 1+\frac{\delta^{2}}{2\sigma(E_i)^{2}\gamma(E_i)}\right)^{-\gamma(E_i)}, 
\end{equation}

where $E_i$ is the energy bin of the photon, $\delta$ is the angular separation between reconstructed and true direction, and $\gamma(E_i)$ and $\sigma(E_i)$ are energy dependent parameters. The parameter $\gamma(E_i)$ determines the tails of the distribution and at low energues 
has a value of $\sim$2.25, which decreases at high energies, creating 
longer PSF tails. The parameter $\sigma(E_i)$ has a power-law dependence at lower energies and reaches an asymptotic value at high energies determined by 
the silicon detector readout pitch. 
We compute $\delta$ for all spatial bins and weight the likelihood by the number of counts per bin.
The parameters of the PSF were estimated by extensive Monte Carlo simulations, but have been since refined with on-orbit data.
We select photons from energy bands above 500 MeV, which contain little or no background. 
We calculate a likelihood for each source by modeling it as point source in a uniform background.
Treating the remaining background on the angular scale of the PSF as uniform is a good approximation, and would be 
significant only in the tails where the contribution to the likelihood is small.

\begin{figure}
  \center
   \includegraphics[width=17cm]{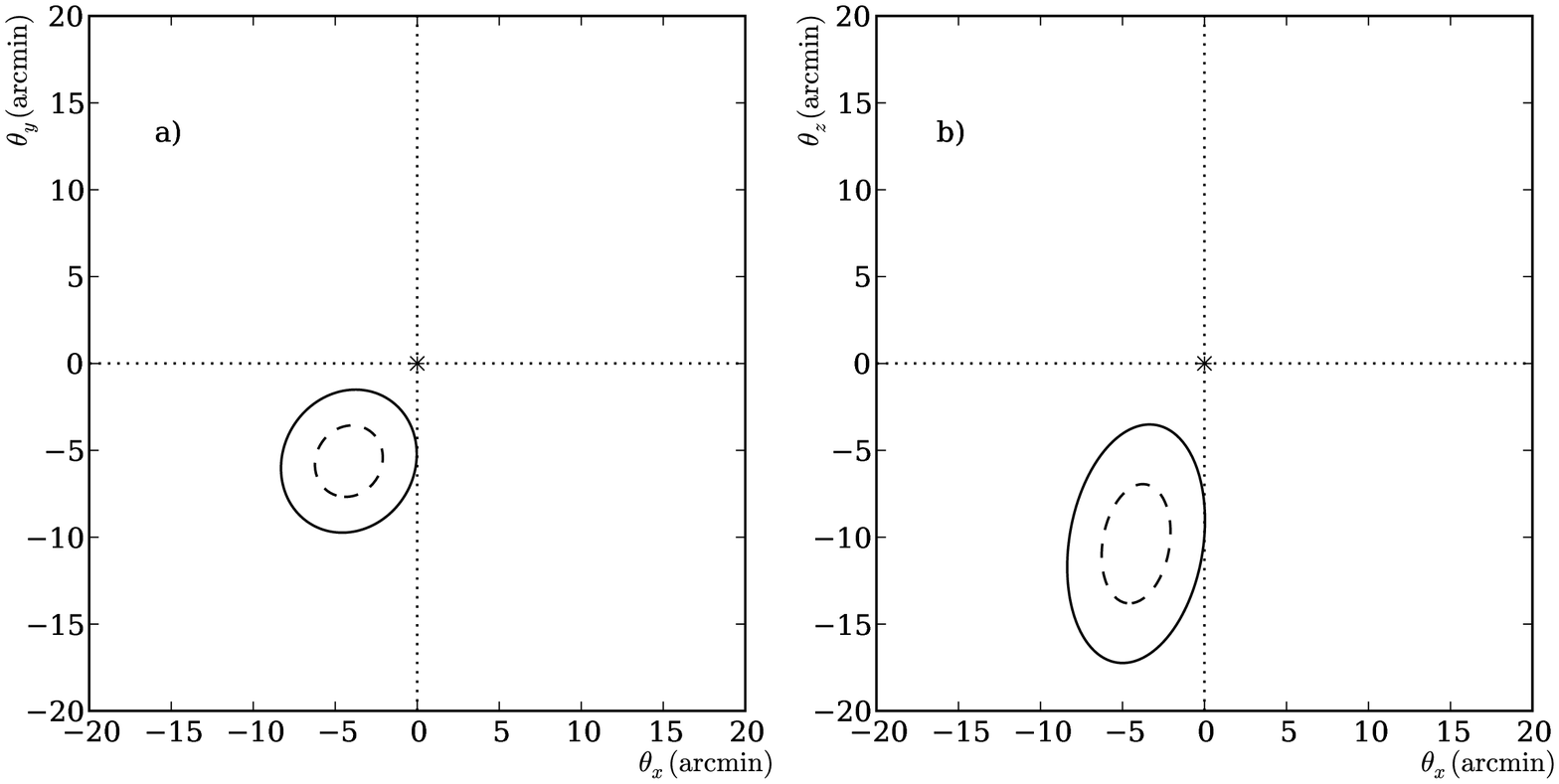}
   \includegraphics[width=8.5cm]{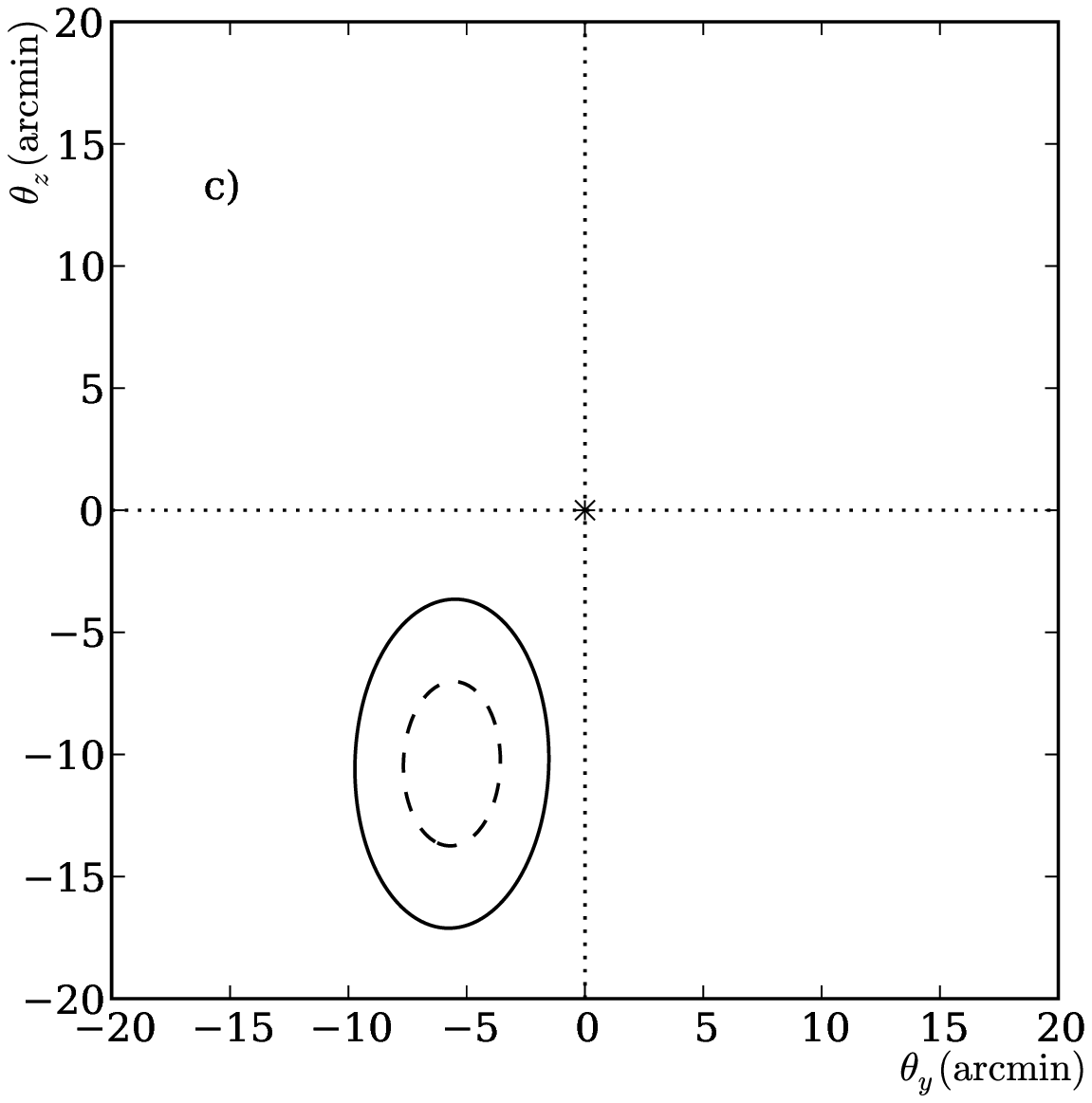}
   \caption{Projection of the likelihood surfaces into each of the planes a) $yx$, b) $zx$ and c) $zy$ for one day of data (August 3, 2008). Crosses correspond to the ideal location (perfect alignment), and contours are 68\% (dashed) and 95\% (solid) containment.}
   \label{fig:ellipse}
\end{figure}

The bright point sources for daily alignment check are selected based on the Test Statistic (TS), which is twice the difference of the log likelihood 
of a point source in a uniform background and the log likelihood of a purely uniform background for a 
particular energy band. The total TS is then the sum of the TS's for the energy bands above 500 MeV. Since 
the background fraction is already small, the assumption of uniformity is 
a good approximation to the distribution of the background events, and we select all sources 
with a total TS greater than 25. For the example shown in Figure~\ref{fig:ellipse}, 6 sources match this criterion, and these are used as seeds for 
the alignment procedure. The largest contribution comes from the Vela pulsar, which is dominated by photons between 1 GeV and its 10 GeV cutoff.
 
The likelihood is maximized with respect to the background for each energy band, since the shape of the PSF depends on energy. The total likelihood for a point source is then the product of the likelihood in each energy band.

Figure~\ref{fig:ellipse} shows typical projected error ellipses resulting from the likelihood fit.
The crosses correspond to the ideal location and the dashed and solid contour lines represent 1 and 2 sigma contours derived from the 
log likelihood, respectively. 
A typical day in orbit yields a precision of 1.5 $arcmin$ in $x$ and $y$, and 2.4 $arcmin$ in $z$, when using 174 photons collected from six astronomical sources. We expect, after the analysis of the first year data to reach the required value of 4 arcsec for the boresight alignment 
residuals.

\begin{figure}
  \center
   \includegraphics[width=16cm]{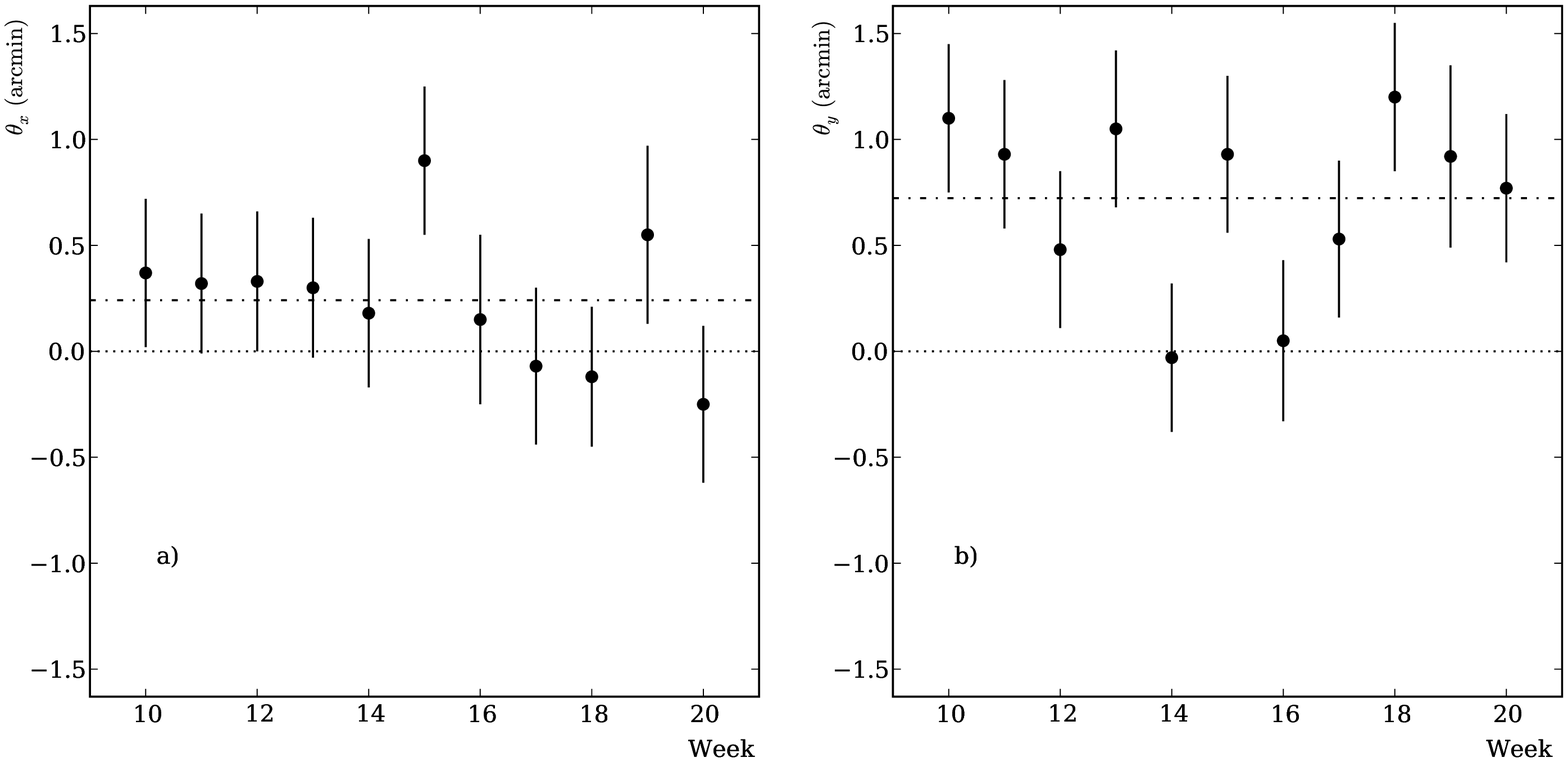}
   \includegraphics[width=7.6cm]{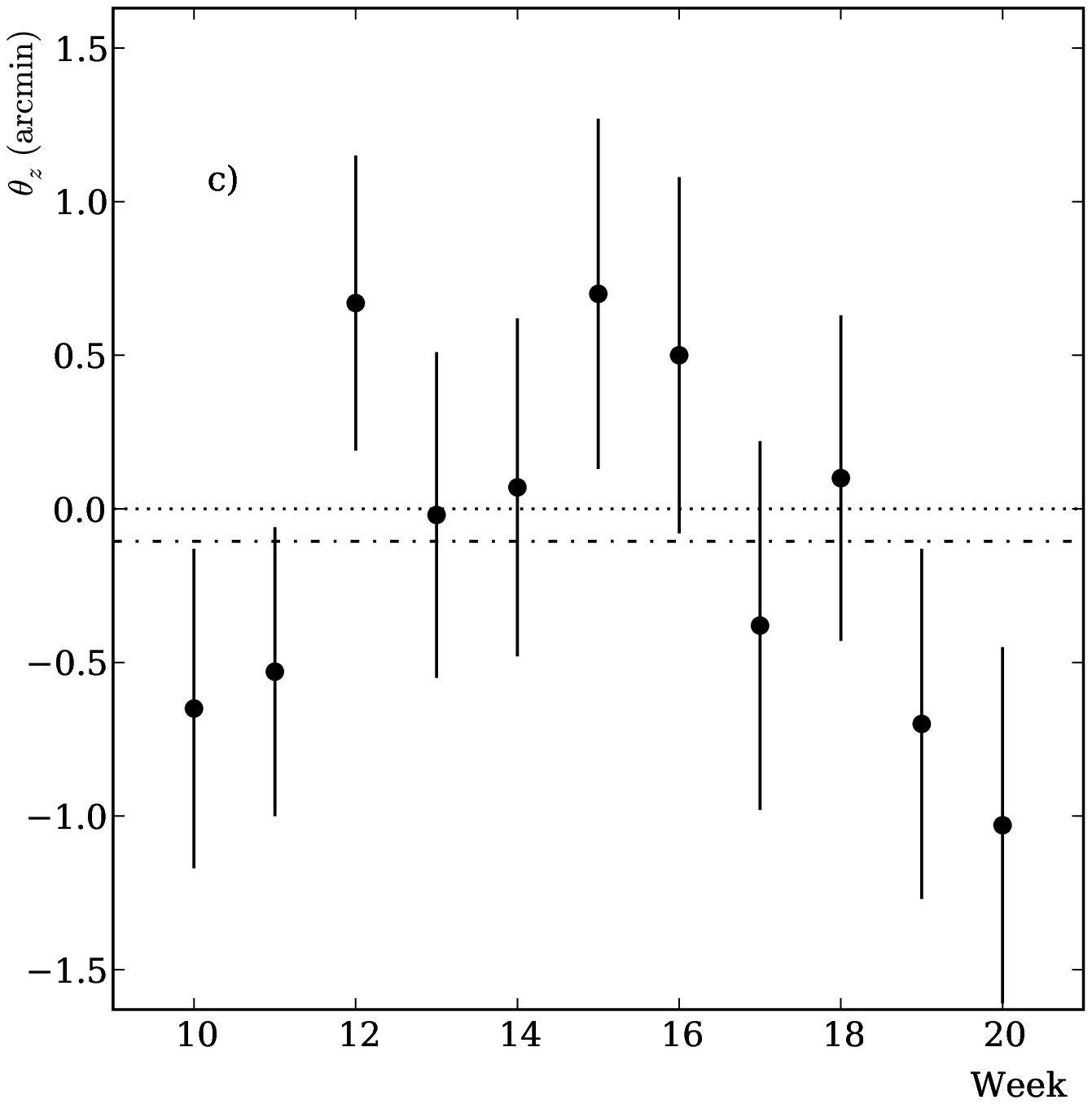}
   \caption{Results of the likelihood fit for weekly measurements over a period of 3 months for the rotation angles about the a) $x$, b) $y$ and c) $z$ axes. The cumulative mean is displayed as dash-dotted line, while the dashed line shows the reference at zero ($y$ axis scales are different). The start date for these measurements on Mission week 10 corresponds to August 10, 2008.}
   \label{fig:daily}
\end{figure}

As shown in Figure~\ref{fig:daily} we make independent measurements of the boresight alignment for each week, 
to monitor the stability. The figure shows the cumulative mean (dash-dotted line) for the rotation angles about the $x$, $y$ and $z$ axes for a period of 3 months. The cumulative mean is just the average of the parameters weighted by the errors. 
We use the same sources and data to optimize the combined likelihood with respect to the PSF parameters themselves, and feed this back into the analysis. Weekly measurements indicate a stability of 0.3 $arcmin$, constant with the statistical errors.
 
\begin{figure}
  \centering
   \includegraphics[width=14cm]{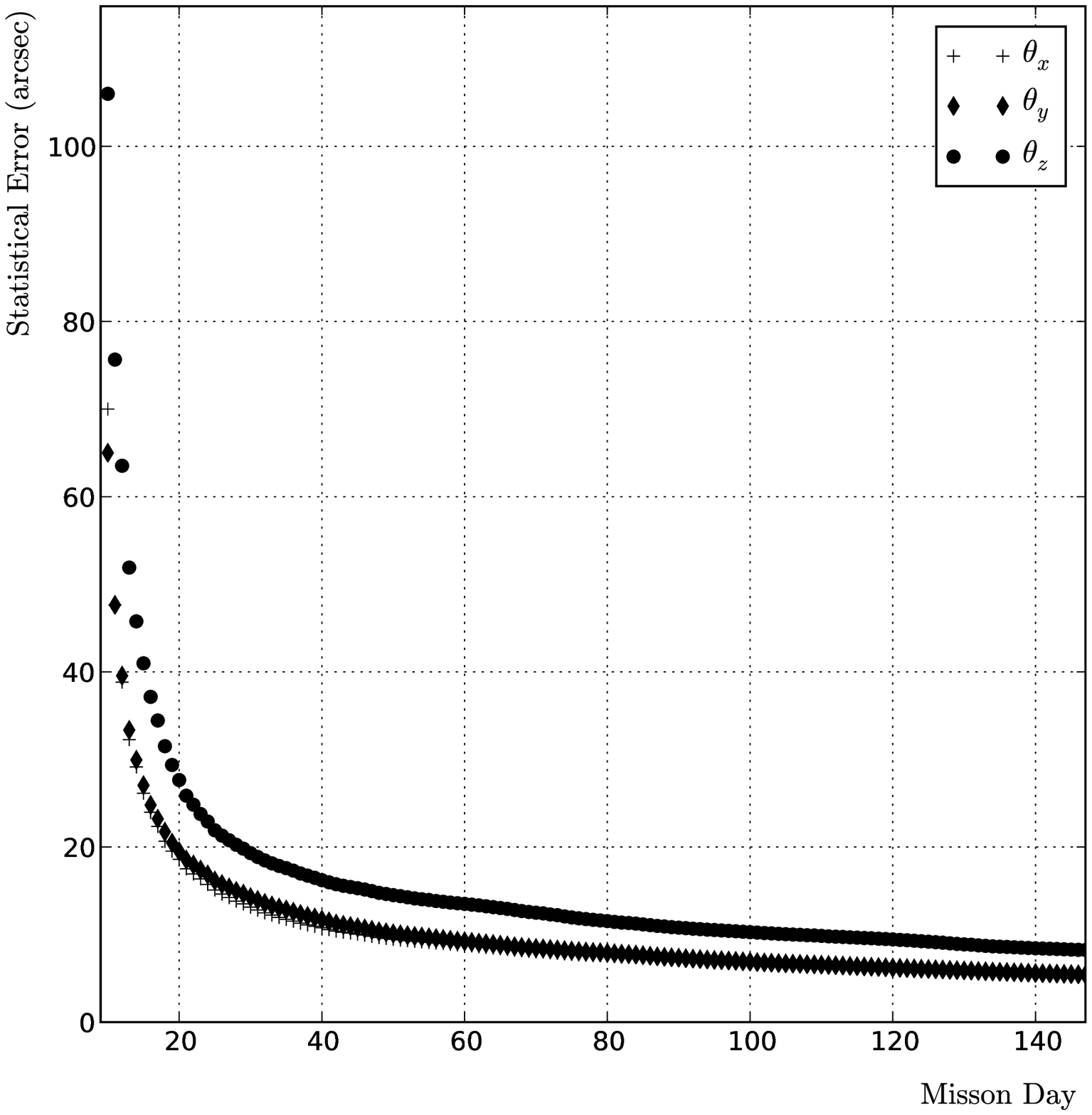}
   \caption{Data correspond to results of the likelihood fit for measurements over a period of 5 months for the rotation angles about the a) $x$, b) $y$ and c) $z$ axes.}
   \label{fig:months}
\end{figure}

We repeated the likelihood fits for a period of about 5 months. The results for the rotation angles are shown in 
Figure~\ref{fig:months}. We clearly see the improvement with more statistics. 
As expected, rotations about $x$ and $y$ axis lead to 
similar results (6 $arcsec$), while for rotations around $z$ we obtain values approaching 8 $arcsec$. These values should be added in quadrature to 
other contributions to obtain the systematic uncertainties for localizing point sources.

\section{Conclusions}
\label{sec:conclusion}

We have discussed the on-orbit calibrations for the {\em Fermi} LAT, which include synchronization of trigger signals, optimization of delays for latching data, determination of detector thresholds, gains and responses, evaluation of the perimeter of the South Atlantic Anomaly (SAA), measurements of live time and of absolute time and internal and spacecraft boresight alignments. The results summarized in Table~\ref{table:summary} were obtained using known astrophysical sources, galactic cosmic rays, and 
charge injection. There were only minor changes to calibration constants since launch and these quantities have been stable during the first eight months 
of operations.

The LAT with almost a million channels is a remarkably stable instrument and is expected to operate during the next few years with dedicated calibration runs 
of about three hours every three months. The frequency with which calibration updates are expected to occur varies from three to twelve months, but these do not
necessarily impact the LAT performance in any significant way. 
As a consequence, changes to high level datasets due to these calibrations will be infrequent. 
The results reported here have been used to calibrate the LAT datasets to be publicly released in August 2009.

\begin{table}
\begin{center}
\begin{tabular}{|c|c|c|c|} \hline\hline
Category           &     Title          & Summary & Sec.\\ \hline\hline
Trigger  &  Time coincidence window &  700 ns, settings within 50 ns        & \ref{sec:tsync} \\ \hline 
Trigger  &   Fast trigger delays &    synchronization within 50 ns   & \ref{sec:tsync} \\ \hline 
Trigger  &   Delays for latching data & optimized within 50 ns  & \ref{sec:tsync} \\ \hline 
ACD       &   Pedestal       &  width $<$ 4 ADC bins or $<$  0.01 MIPs  & \ref{sec:acdpeds} \\ \hline
ACD       &   Coherent noise     & removed down to 0.005 MIP   & \ref{sec:acdpeds} \\ \hline  
ACD       &   MIP peak     &   stability $<$10\%   & \ref{sec:acdmip} \\ \hline 
ACD       &   High range (CNO)   & width of carbon peak$\sim$20\% of peak   & \ref{sec:acdmip} \\ \hline 
ACD       &   Veto threshold &     turn-on at 0.4-0.5 MIP, set within $\pm$ 0.01 MIP  & \ref{sec:veto} \\ \hline
ACD       &   High level discriminator &  turn-on at 24-26 MIP, set within $\pm$ 1 MIP      & \ref{sec:veto} \\ \hline  
CAL       &   Pedestal      & RMS within 0.1-0.2 MeV   & \ref{sec:calpeds} \\ \hline 
CAL       &   Electronics linearity     &  corrected to $\le$1\% of the measured energy   & \ref{sec:callin} \\ \hline  
\multirow{2}{*} {CAL}       &   Energy scales     &  spread crystal-to-crystal $\le$1\%   & \ref{sec:callin} \\
         &       &  stability of peaks: MIP($\le$2\%), carbon($\le$1\%)  &  \\ \hline  
CAL       &   Light asymmetry &   2 mm (LEX1), 9 mm (HEX8) from 200-900 MeV & \ref{sec:calpos} \\ \hline 
CAL       &   Zero-suppression threshold & set at 2 MeV, $\sim$ 10 x electronics noise  & \ref{sec:calthresh} \\ \hline 
CAL       &   Low-energy threshold   &   set at 100 MeV ($\pm$ 1\%)    & \ref{sec:calthresh} \\ \hline 
CAL       &   High-energy threshold  &   set at 1 GeV ($\pm$ 2\%)    & \ref{sec:calthresh} \\ \hline 
CAL       &   Upper level discriminators  & set at 5\% below saturation level  & \ref{sec:calthresh} \\ \hline 
\multirow{2}{*}{TKR}       &   Noisy channels   & avg strip occ. (10$^{-5}$), add $\le$10\% to the TKR data & \ref{sec:tkrch} \\ 
       &      & electronic noise occupancy (10$^{-7}$), 0.04\% disabled &  \\ \hline  
\multirow{2}{*} {TKR}       &   Trigger threshold  & set at $\sim$0.28 MIP, spread channel-to-channel$\sim$5\%  & \ref{sec:tkrdac} \\ 
       &   & $\le$ 1 ADC shift since launch (99.96\% of channels)&  \\ \hline
TKR       &   Data latching threshold     & spread channel-to-channel$\sim$12\%    & \ref{sec:tkrdac} \\ \hline
TKR       &   ToT conversion parameters & fitted to $\sim$ 8\% (statistical error)   & \ref{sec:tkrtot} \\ \hline
TKR       &   MIP scale   &  RMS of correction factor (9\%)  & \ref{sec:tkrmip} \\ \hline 
SAA     &   SAA polygon &  in SAA for$\sim$13\% of the orbit time  & \ref{sec:saa} \\ \hline  
Timing    &   LAT timestamps &   $<$ 0.3 $\mu$s with respect to a reference GPS & \ref{sec:absolute} \\ \hline
Alignment     &   Intra tower &  for details see Tables~\ref{table:intrafinal} and~\ref{table:data}  & \ref{sec:intra} \\ \hline
Alignment     &   Inter tower &  for details see Tables~\ref{table:inter} and~\ref{table:data}  & \ref{sec:inter} \\ \hline  
Alignment     &   LAT boresight &  $\theta_x$ , $\theta_y$ $\le$ 6$^{''}$ and  $\theta_z$ $\le$ 8$^{''}$ (5 months)& \ref{sec:sc} \\ \hline 
\hline
\end{tabular}
\vspace*{+0.5cm}
\caption{Summary of the on-orbit Fermi LAT calibrations.}
\label{table:summary}
\end{center}
\end{table}

\section{Acknowledgments}

The {\em Fermi} LAT Collaboration acknowledges the generous ongoing support of a number of agencies and institutes that have supported 
both the development and the operation of the LAT as well as scientific data analysis. These include the National Aeronautics 
and Space Administration and the Department of Energy in the United States
, the Commissariat \`a l'Energie Atomique and the Centre National de la Recherche Scientifique / Institut National de Physique Nucl\'eaire et de Physique des Particules in France, the Agenzia Spaziale Italiana and the Istituto Nazionale di Fisica Nucleare in Italy and the Istituto Nazionale di Astrofisica, the Ministry of Education, Culture, Sports, Science and Technology (MEXT), High Energy Accelerator Research Organization (KEK) and Japan Aerospace Exploration Agency (JAXA) in Japan, and the K. A. Wallenberg Foundation and the Swedish National Space Board in Sweden.

Additional support for science analysis during the operations phase from the following agencies is also gratefully acknowledged: the Istituto Nazionale di Astrofisica in Italy and the K.~A.~Wallenberg Foundation in Sweden for providing a grant in support of a Royal Swedish Academy of Sciences Research fellowship for JC.


\begin{thebibliography}{99}

\bibitem{latpaper} {W. B. Atwood \it et al.}, in press {\em Astrophysical Journal}.  
\bibitem{gbmpaper} C. Meegan {\it et al}, in preparation. 
\bibitem{egret} D.J. Thompson  {\it et al.}, ApJ. Suppl. Vol {\bf 86} (1993) 629. 
\bibitem{Atwood:2007ra}
  W.~B.~Atwood {\it et al.},
  Astropart.\ Phys.\  {\bf 28}, (2007) 422.  
\bibitem{ssd}
T. Ohsugi {\it et al}, Nucl. Inst. and Meth. {\bf A 541} (2005) 29.
\bibitem{Johnson98}
R.~P.~Johnson, P.~Poplevin, H.~F.-W.~Sadrozinski, and E.~N.~Spencer, 
IEEE Trans. Nucl. Sci. {\bf 45}, (1998) 927.
\bibitem{Baldini:2006pv}
  L.~Baldini {\it et al.},
  IEEE Trans.\ Nucl.\ Sci.\  {\bf 53}, (2006) 466.
\bibitem{cal} J. E. Grove {\it et al.}, in preparation.  
\bibitem{acdover} A. Moiseev {\it et al.}, Astropart. Phys. \textbf{27}, (2007) 339.
\bibitem{acddet}  A. Moiseev {\it et al.}, Nucl. Inst. and Meth. {\bf A 583}, (2007) 372. 
\bibitem{perfpaper} A. Abdo {\it et al.}, in preparation.  
\bibitem{bt97}  W. Atwood {\it et al.}, Nucl. Inst. and Meth. {\bf A 446} (2000) 444.
\bibitem{bt99}  E. do Couto e Silva {\it et al.}, Nucl. Inst. and Meth. {\bf A 474}, (2001) 19. 
\bibitem{bt06} A. Abdo {\it et al.}, in preparation.  
\bibitem{tdaqpaper} A. Abdo {\it et al.}, in preparation.  
\bibitem{backsplash} A. Moiseev {\it et al.}, Astropart. Phys. \textbf{22}, (2004) 275.
\bibitem{lott2006} B. Lott {\it et al.}, Nucl. Inst. and Meth. \textbf{560}, (2006) 395.
\bibitem{ap8}  D. M. Sawyer and J. I. Vette, {\em AP-8 Trapped Proton Environment for Solar Maximum and Solar Minimum}, NSSDC/WDC-A-R\&S 76-06, 1976. 
\bibitem{psb97} D. Heynderickx, M. Kruglanski, V. Pierrard, J. Lemaire, M. D. Looper, and J. B. Blake, 
IEEE Trans. Nucl. Sci., {\bf 46}, (1999) 1475.
\bibitem{igrf} IGRF-10, IAGA Division V-MOD Geomagnetic Field Modeling, http://www.ngdc.noaa.gov/IAGA/vmod/igrf.html
\bibitem{pamela_icrc} M. Casolino, et al. {\em Observations of primary, trapped and quasi-trapped particles with the PAMELA experiment}. to appear in the proceedings of the 30$^{th}$ International Cosmic Ray Conference (ICRC 2007) 3-11 Jul 2007, Merida, Yucatan, Mexico. 
\bibitem{grb0801916C} A. Abdo {\it et al} submitted to {\em Science} (2008).
\bibitem{pulsartiming} D. A. Smith {\it et al.} A\&A {\bf 492} (2008) 923.
\bibitem{celeste} M. de Naurois, J. Holder {\it et al.} ApJ {\bf 566}, (2002) 343.
\bibitem{vela} A. Abdo {\it et al.}, in press {\em Astrophysical Journal} (2008).  
\bibitem{gpsStandard} Global Postioning System Standard Positioning Service Performance Standard', 4th edition (September 2008), http://pnt.gov/public/docs/2008-SPSPS.pdf, p. A-16, section A.4.8
\bibitem{0451} A. Abdo {\it et al.}, submitted to ApJ.  
\bibitem{2134} A. Abdo {\it et al.}, in preparation.  
\bibitem{like} A. Abdo {\it et al.}, in preparation.  
\end{thebibliography}
\end{document}